\documentclass[usenatbib]{mnras}

\usepackage{newtxtext,newtxmath}

\usepackage[T1]{fontenc}
\usepackage{ae,aecompl}

\usepackage{graphicx}	
\usepackage{amsmath}	
\usepackage{amssymb}	

\def\la{\raise.5ex\hbox{$&lt;$}\kern-.8em\lower 1mm\hbox{$\sim$}}
\def\ma{\raise.5ex\hbox{$&gt;$}\kern-.8em\lower 1mm\hbox{$\sim$}}

\def\kms{$\rm km\, s^{-1}$}
\def\cm3{$\rm cm^{-3}$}

\def\n0{$\rm n_{0}$}
\def\B0{$\rm B_{0}$}

\def\L12{L$_{12\mu m}$~}
\def\F12{F$_{12\mu m}$~}

\def\fe2{[Fe\,{\sc ii}]}
\def\h2{H$_{2}$}
\def\pp{$\pm$}
\def\mc{$\mu$m}
\def\lb{$\lbrack$}

\title[Optical/NIR stellar absorption and emission-line indices ]{Optical/NIR stellar absorption and emission-line indices from luminous infrared galaxies}

\author[Riffel et al.]{Rog\'erio Riffel$^1$\thanks{E-mail: riffel@ufrgs.br. Visiting Astronomer at the Infrared Telescope Facility, which is operated by the University of Hawaii
    under Cooperative Agreement no. NCC 5-538 with the National Aeronautics and Space Administration, Office of Space Science, Planetary Astronomy Program.},    
Alberto Rodr\'{\i}guez-Ardila$^{2,9}$, 
Michael S. Brotherton$^3$, \newauthor 
Reynier Peletier$^{4}$,
Alexandre Vazdekis$^{5}$,
Rogemar A. Riffel$^{6}$, 
Lucimara Pires Martins$^{7}$,   \newauthor 
Charles Bonatto$^{1}$,
Natacha Zanon Dametto$^{1}$,  
Luis Gabriel Dahmen-Hahn$^{1}$, \newauthor
Jessie Runnoe $^{3, 8}$,
 Miriani G. Pastoriza$^{1}$,
 Ana L. Chies-Santos $^{1}$, 
 Marina Trevisan$^{1}$ \\
$^{1}$Departamento de Astronomia, Universidade Federal do Rio Grande do Sul,  Av. Bento Gon\c calves 9500, 91501-970, Porto Alegre, RS, Brasil. \\
$^2$ Laborat\'{o}rio Nacional de Astrof\'{i}sica/MCT - Rua dos Estados Unidos 154, Bairro das Nac\~oes. CEP 37504-364, Itajub\'{a}, MG, Brasil\\
$^3$ Department of Physics and Astronomy, University of Wyoming, Laramie, WY 82071, USA\\
$^4$ Kapteyn Astronomical Institute, University of Groningen, Postbus 800, 9700 AV, Groningen, The Netherlands\\
$^5$ Instituto de Astro\'{\i}sica de Canarias, V\'{\i}a L\'actea, S/N, E-38205, La Laguna, Tenerife, Spain\\
$^6$ Departamento de F\'{\i}sica, CCNE, Universidade Federal de Santa Maria (UFSM), 97105-900 Santa Maria, RS, Brazil \\
$^7$ NAT - Universidade Cruzeiro do Sul, Rua Galv\~ao Bueno, 868, 01506-000, S\~ao Paulo, SP, Brazil.\\
$^8$ Department of Astronomy, University of Michigan, 1085 S. University Ave., Ann Arbor, MI 48109, USA\\
$^9$ Divis\~ao de Astrof\'{\i}sica, Instituto Nacional de Pesquisas Espaciais, 12227-010, S\~ao Jos\'e dos Campos, SP, Brazil\\
}

\date{Accepted XXX. Received YYY; in original form ZZZ}

\pubyear{2018}

\hypersetup{draft}
\begin{document}
\label{firstpage}
\pagerange{\pageref{firstpage}--\pageref{lastpage}}
\maketitle

\begin{abstract}

We analyze a set of optical-to-near-infrared long-slit nuclear spectra of 16 infrared-luminous spiral galaxies. All of the studied sources present \h2\ emission, which reflects the star-forming nature of our sample, and they clearly display H\,{\sc i} emission lines in the optical. Their continua contain many strong stellar absorption lines, with the most common features due to \ion{Ca}{i}, \ion{Ca}{ii}, \ion{Fe}{i}, \ion{Na}{i}, \ion{Mg}{i}, in addition to prominent absorption bands of TiO, VO, ZrO, CN and CO. We report a homogeneous set of equivalent width (EW) measurements for 45 indices, from optical to NIR species for the 16 star-forming galaxies as well as for 19 early type galaxies where we collected the data from the literature. This selected set of emission and absorption-feature measurements can be used to test predictions of the forthcoming generations of stellar population models. We find correlations among the different absorption features and propose here correlations between optical and NIR indices, as well as among different NIR indices, and compare them with model predictions. While for the optical absorption features the models consistently agree with the observations,the NIR indices are much harder to interpret. For early-type
spirals the measurements agree roughly with the models, while for  star-forming objects they fail to predict the strengths of these indices. 

\end{abstract}


\section{Introduction}

One challenge in modern astrophysics is to understand galaxy formation and evolution. Both processes are strongly related to the star-formation history (SFH) of galaxies. Thus, the detailed study of the different stellar populations found in galaxies is one of the most promising ways to shed some light on their evolutionary histories. So far, stellar population studies have been concentrated mainly in the optical spectral range \citep[e.g.][]{Bica+88,Worthey+94,Trager+00,Sanchez-Blazquez+06,Gonzalez-Delgado+15,Goddard+17,Martin-Navarro+18}. In the near infrared, (0.8-2.4\mc, NIR) even with some work dating back to the 1980s \citep[e.g.][]{Rieke+80}, stellar population studies have just started to become more common in the last  two decades \citep[][for example]{Origlia+93,Origlia+97,Riffel-Rogerio+07, Riffel-Rogerio+08,Riffel-Rogerio+09,Cesetti+09,Lyubenova+10,Chies-Santos+11a,Chies-Santos+11b,Riffel-Rogerio+11a,Kotilainen+12,Martins+13b,La_Barbera+13,Zibetti+13,Noel+13,Dametto+14,Riffel-Rogerio+15, Baldwin+17,Dametto+19,Alton+18,Dahmer-Hahn+18,Dahmer-Hahn+19,Francois+18}. Models have shown that the NIR spectral features provide very important insights, particularly into the stellar populations dominated by cold stars \citep[e.g.][]{Maraston05,Riffel-Rogerio+07,van-Dokkum+12,Conroy+12,Zibetti+13,Riffel-Rogerio+15,Rock+15,Rock+16}. For example, the stars in the thermally pulsing asymptotic giant branch (TP-AGB) phase may be responsible for nearly half of the luminosity in the K band for stellar populations with an ago of $\sim$ 1~Gyr \citep{Maraston98,Maraston05,Salaris+14}.

One common technique to study the unresolved stellar content of galaxies is the fitting of a combination of simple stellar populations (SSPs) to obtain the SFH. However, due to difficulties in theoretical treatment \citep{Maraston05,Marigo+08,Noel+13} and the lack of complete empirical stellar libraries in the NIR, \citep{Lancon+01,Chen+14,Riffel-Rogerio+15,Villaume+17} the available SSP models produce discrepant results \citep[e.g.][]{Baldwin+17}, thus making it very difficult to reliably analyse the SFH in the NIR.

On the other hand, the stellar content and chemical composition of the unresolved stellar populations of galaxies can also be obtained by the study of the observed absorption features present in their integrated spectra. So far, we still lack a comprehensive NIR dataset to compare with model predictions, required to make improvements to the models and 
to lead to a better understanding of the role played by the cooler stellar populations in the integrated spectra of galaxies. 

Among the best natural laboratories to study these kinds of stellar content are {\it infrared galaxies}, sources that emit more energy in the infrared ($\sim$5-500\mc) than at all the other wavelengths combined \citep{Sanders+96,Sanders+03}.  The relevance of studying these galaxies lies particularly in the fact that they are implicated in a variety of interesting astrophysical phenomena, including the formation of quasars and elliptical galaxies \citep[e.g.][]{Genzel+01,Veilleux+06,Wang+06}. When studying luminous infrared galaxies in the Local Universe, it is possible to obtain high-angular-resolution observations of these objects, thus allowing the investigation of their very central regions. Comparison of such objects with those at higher redshifts may help to understand the SFH over cosmic times.

With the above in mind, we obtained optical and NIR spectra of a sub-sample of galaxies selected from the {\it IRAS} Revised Bright Galaxy Sample present in the Local Universe. These galaxies are believed to be experiencing massive star formation, making them suitable for studying their most important spectral features that can be used as proxies to test and constrain stellar-population models. As part of a series of papers aimed at studying the stellar population and gas emission features, here we provide measurements for the most conspicuous emission and absorption features, and present new correlations between absorption features. 
The outline of the paper is as follows. In \S\ref{obs} we describe the observations and data reduction. The results are presented and discussed in \S\ref{resDisc}. Final remarks are made in \S\ref{finrem}.

\section{Observations and data reduction} \label{obs}

Our sample is composed of 16 Local Universe (v$\rm _r \lesssim$ 6400 km s$^{-1}$) galaxies that are very bright in the infrared (see Tab.~\ref{log}). They were selected from the {\it IRAS} Revised Bright Galaxy Sample,  which is regarded as a statistically complete sample of 629 galaxies, with 60 $\mu$m flux density $\gtrsim$ 5.24 Jy. Galaxies chosen for this study were those with log($L_{fir}/L_{\odot}$) $\gtrsim$ 10.10, accessible from the Infrared Telescope Facility (IRTF) and the Wyoming Infrared Observatory (WIRO, see below), and bright enough to reach a S/N $\sim$ 100 in the K-band within a reasonable on-source integration time.

\subsection{Near Infrared Data}
Cross-dispersed near-infrared (NIR) spectra in the range 0.8-2.4 $\mu$m were obtained on October 4, 6, and 7 in 2010 with the
SpeX spectrograph \citep{Rayner+03} attached to the NASA 3\,m IRTF telescope at the Mauna Kea observing site. 
The detector is a 1024$\times$1024 ALADDIN 3 InSb array with a spatial scale of 0.15$''$/pixel.   A 0.8$'' \times$15$''$ slit was used during the observations, giving a spectral resolution of R$\sim$ 1000 (or $\sigma =$ 127~\kms). Both the arc lamp spectra and the night-sky spectra are consistent with this value \citep{Riffel-Rogemar+13}. 
The observations were done by nodding in an Object-Sky-Object pattern with typical individual integration times of 120\,s
and total on-source integration times between 18 and 58 minutes. During the observations, A0\,V stars were observed near each target to provide telluric
standards at similar air masses. These stars were also used to flux calibrate the galaxy spectra by using black body functions to calibrate the observed spectra of the standard stars.  The seeing varied between 0.4$''$--0.7$''$ over the different nights and there were no obvious clouds.

We reduced the NIR observations following the standard data reduction procedures given by \citet{Riffel-Rogerio+06,Riffel-Rogerio+13}. In short, spectral extraction and wavelength calibration were performed using {\sc spextool}, software developed and provided by the SpeX team for the IRTF community \citep{Cushing+04}. The area of the integrated region is listed in Tab.~\ref{log}. Each extraction was centred at the peak of the continuum-light distribution for every object of the sample. No effort was made to extract spectra at positions different from the nuclear region, even though some objects show evidence of extended emission, as this goes beyond the scope of this analysis. 
Telluric absorption correction and flux calibration were applied to the individual 1-D spectra by means of the IDL 
routine {\it xtellcor} \citep{Vacca+03}.

\subsection{Optical Data}

For completeness, the same sample was also observed in the optical range on nearly the same dates as the NIR data were collected with the WIRO Long Slit Spectrograph.  The instrument is attached to the University of Wyoming's 2.3-meter telescope, located on Jelm Mountain at WIRO. The Cassegrain-mounted instrument uses a Marconi 2k$\times$2k CCD detector.  During our observations we used a 900 l/mm grating in first order to obtain spectra from approximately 4000-7000 \AA\ calibrated with a CuAr comparison lamp.  Given our 4-arc-second slit oriented North-South, the resolution was R$\sim$1200.  Due to the relatively large spatial extent of these low-redshift objects, we offset the telescope pointing by two arc-minutes to obtain sky spectra uncontaminated by galaxy light.  The seeing varied between 1-2 arc-seconds during the nights of observation.  We reduced the spectra using standard techniques in IRAF\footnote{IRAF is distributed by the National Optical Astronomy Observatories, which are operated by the Association of Universities for Research in Astronomy, Inc., under cooperative agreement with the National Science Foundation.}. Table~\ref{log} shows the observation log along with extraction apertures. The 1-D wavelength and flux-calibrated spectra were then corrected for redshift, determined from the average $z$ measured from the position of [S\,{\sc iii}] 0.953$\mu$m, Pa$\delta$, He\,{\sc i}~1.083$\mu$m, Pa$\beta$ and Br$\gamma$.

Examples of the final reduced spectra, from optical to NIR ($\sim$0.4\mc\ ---2.4\mc) are presented in Figure~\ref{spectra}, for the remaining galaxies see Appendix~\ref{FinRed} . For each galaxy we show the optical, $z$+$J$, $H$ and $K$ bands, from top to bottom, respectively. It is worth mentioning that the optical and NIR data do not share the same apertures, and the slit was not generally oriented at the same position angles. However, since we are interested in the nuclear region, the different slit orientations should not introduce large discrepancies in the measurements.
The ordinate axis represents the monochromatic flux in units of $\rm 10^{-15}\, erg \, cm^{-2} \, s^{-1}$~\AA$^{-1}$. The position of the most common and expected emission and absorption lines are indicated as dotted (red) and dashed (blue) lines, respectively.

\begin{table*}
\renewcommand{\tabcolsep}{0.60mm}
\centering
\caption{Near-Infrared observation log and basic sample properties.\label{log}} 
\begin{tabular}{lclcccccllcl}
\hline \hline
\noalign{\smallskip}
Source    &   $\alpha$	  &	$\delta$	&  ~~~~~~~~z  &   Obs.  Date &     Exp. Time	  & Airmass   & PA & Size & Activity          & log($\frac{L_{IR}}{L_{\odot}})^{\star}$& Morphology	\\ 
          &		  &		&	      & 	     &      (s) 	  &	      & (deg)	   &  (pc$\times$pc)      &	            & (13) \\
\noalign{\smallskip}
\hline
NGC 23    &  00h09m53.4s  &   +25d55m26s& 0.0157202   & 2010 10 07   &    29 $\times$ 120 &  1.04     & 330	   &  1348 $\times$ 270	   & SFG$\rm ^{1}$	  & 11.05 & SBa  \\ 
NGC 520   &  01h24m35.1s  &   +03d47m33s& 0.0080367   & 2010 10 04   &    16 $\times$ 120 &  1.04     & 300	   &  689 $\times$ 138 	   & SFG$\rm ^{2}$	  & 10.91 & S0    \\ 
NGC 660   &  01h43m02.4s  &   +13d38m42s& 0.0029152   & 2010 10 06   &    24 $\times$ 120 &  1.01     & 33	   &  237 $\times$ 50  	   & Sy2/HII$\rm ^{2,3,4}$	      & 10.49 & SBa pec  \\ 
NGC 1055  &  02h41m45.2s  &   +00d26m35s& 0.0036267   & 2010 10 04   &    16 $\times$ 120 &  1.07     & 285	   &  466 $\times$ 62  	   & LINER/HII$\rm ^{2,3,4}$	      & 10.09 & Sbc   \\ 
NGC 1134  &  02h53m41.3s  &   +13d00m51s& 0.0129803   & 2010 10 04   &    16 $\times$ 120 &  1.11     & 0	   &  1113 $\times$ 223	   & SFG$\rm ^{5}$	  & 10.83 & S? \\ 
NGC 1204  &  03h04m39.9s  &   -12d20m29s& 0.0154058   & 2010 10 07   &    16 $\times$ 120 &  1.23     & 66	   &  1321 $\times$ 264	   & LINER$\rm ^{6}$		  & 10.88 & S0/a \\ 
NGC 1222  &  03h08m56.7s  &   -02d57m19s& 0.0082097   & 2010 10 06   &    24 $\times$ 120 &  1.13     & 315	   &  598 $\times$ 141 	   & SFG$\rm ^{7}$	  & 10.60 & S0 pec \\ 
NGC 1266  &  03h16m00.7s  &   -02d25m38s& 0.0077032   & 2010 10 07   &    18 $\times$ 120 &  1.09     & 0	   &  661 $\times$ 132 	   & LINER$\rm ^{7}$		  & 10.46 & SB0 pec \\ 
UGC 2982  &  04h12m22.4s  &   +05d32m51s& 0.0177955   & 2010 10 04   &    9  $\times$ 120 &  1.11     & 295	   &  1526 $\times$ 305	   & SFG$\rm ^{8}$	  & 11.30 & SB \\ 
NGC 1797  &  05h07m44.9s  &   -08d01m09s& 0.0154111   & 2010 10 07   &    16 $\times$ 120 &  1.23     & 66	   &  1321 $\times$ 264	   & SFG$\rm ^{1}$	  & 11.00 & SBa \\ 
NGC 6814  &  19h42m40.6s  &   -10d19m25s& 0.0056730   & 2010 10 07   &    16 $\times$ 120 &  1.17     & 0	   &  486 $\times$ 97  	   & Sy~1$\rm ^{7}$		  & 10.25 & SBbc \\ 
NGC 6835  &  19h54m32.9s  &   -12d34m03s& 0.0057248   & 2010 10 06   &    22 $\times$ 120 &  1.21     & 70	   &  368 $\times$ 98  	   & SFG$\rm ^{9}$	  & 10.32 & SBa \\ 
UGC 12150 &  22h41m12.2s  &   +34d14m57s& 0.0214590   & 2010 10 04   &    15 $\times$ 120 &  1.08     & 37	   &  1656 $\times$ 368	   & LINER/HII$\rm ^{10}$	  & 11.29 & SB0/a \\ 
NGC 7465  &  23h02m01.0s  &   +15d57m53s& 0.0066328   & 2010 10 06   &    12 $\times$ 120 &  1.03     & 340	   &  569 $\times$ 114 	   & LINER/Sy~2$\rm ^{11}$	  & 10.10 & SB0 \\ 
NGC 7591  &  23h18m16.3s  &   +06d35m09s& 0.0165841   & 2010 10 07   &    16 $\times$ 120 &  1.03     & 0	   &  1422 $\times$ 284	   & LINER$\rm ^{7}$		  & 11.05 & SBbc \\ 
NGC 7678  &  23h28m27.9s  &   +22d25m16s& 0.0120136   & 2010 10 04   &    16 $\times$ 120 &  1.01     & 90	   &  927 $\times$ 206 	   & SFG$\rm ^{12}$		  & 10.77 & SBc \\ 
\noalign{\smallskip}
\hline
\end{tabular}
\begin{list}{Table Notes:}
\item SFG: Star-Forming Galaxies (Starburst or H{\sc ii} galaxies). LINER/HII were assumed to be pure LINERs in the text. The galaxies are listed in
order of right ascension, and the number of exposures refers to on-source integrations. The slit width is 0.8$''$. References - 1: \citet{Balzano+83}; 2: \citet{Ho+97b}; 3: \citet{Ho+97a}; 4: \citet{Filho+04}; 5: \citet{Condon+02}; 6: \citet{Sturm+06}; 7: \citet{Pareira-Santaella+10}; 8: \citet{Schmitt+06}; 9: Coziol+98; 10: \citet{Veilleux+95}; 11: \citet{Ferruit+00}; 12: \citet{Goncalves+98}; 13: \citet{Sanders+03}; 
\end{list}
\end{table*}

\begin{table}
\centering
\caption{Optical observation log. The slit was oriented North-South. \label{optlog}} 
\begin{tabular}{lclcl}
\hline \hline
\noalign{\smallskip}
Source    &    Obs.  Date &	Exp.      & Airmass  &  Size    \\ 
          &		  &	Time (s)  &	  &    (pc$\times$pc)  \\
\noalign{\smallskip}
\hline
NGC 23     &   2010 10 04  &   600  &	1.20  &  2359 $\times$ 1348 	\\ 
NGC 520    &   2010 10 03  &   600  &	1.28  &  4307 $\times$ 689  	\\ 
NGC 660    &   2010 10 04  &   600  &	1.18  &  437 $\times$ 250   	\\ 
NGC 1055   &   2010 10 04  &   600  &	1.36  &  544 $\times$ 311   	 \\ 
NGC 1134   &   2010 10 04  &   600  &	1.14  &  12243 $\times$ 1113     \\ 
NGC 1204   &   2010 10 04  &   600  &	1.68  &  2312 $\times$ 1321 	 \\ 
NGC 1222   &   2010 10 03  &   600  &	1.42  &  3344 $\times$ 704       \\ 
NGC 1266   &   2010 10 03  &   600  &	1.39  &  6440 $\times$ 661       \\ 
UGC 2982   &   2010 10 04  &   600  &	1.27  &  2670 $\times$ 1526      \\ 
NGC 1797   &   2010 10 02  &   600  &	1.56  &  11893 $\times$ 1321   \\ 
UGC 12150  &   2010 10 03  &   600  &	1.03  &  18401 $\times$ 1840	       \\ 
NGC 7465   &   2010 10 03  &   600  &	1.12  &  3270 $\times$ 569  	       \\ 
NGC 7591   &   2010 10 03  &   600  &	1.21  &  13154 $\times$ 1422     \\ 
NGC 7678   &   2010 10 04  &   600  &	1.18  &  1803 $\times$ 1030 	 \\ 
\noalign{\smallskip}
\hline
\end{tabular}
\begin{list}{Table Notes:}
\item The slit width is 4$''$.
\end{list}
\end{table}

\section{results }\label{resDisc}

\subsection{Emission-line spectra}\label{sec_emlines}
A visual inspection of the data reveals a wide diversity of emission-line strengths and species. The most common emission features detected are: H$\beta$, \lb\ion{O}{iii}] 4959, 5007\,\AA, \lb\ion{N}{ii}] 6548, 6583\,\AA, H$\alpha$, \lb\ion{S}{ii}] 6716,6730\,\AA, \lb\ion{S}{iii}] 9531\,\AA, Pa$\delta$, [C {\sc i}] 9824, 9850\,\AA, Pa$\beta$, \ion{He}{i} 10830\,\AA,  [\ion{P}{ii}] 11886\,\AA, \fe2\ 12570, 16436\,\AA,  Pa $\alpha$ \h2\ 19570\,\AA, \h2\ 21218\,\AA, and Br$\gamma$.  

Emission-line fluxes for each object of the sample were measured by fitting a Gaussian function to the observed profile and then integrating the flux under the curve. The LINER software \citep{Pogge93} was used for this purpose. No attempt to correct for stellar absorption was made before measuring the emission lines. This was done because NIR models with adequate spectral resolution (to allow the measurements of the weaker emission lines) are not available for the younger ages. \citet{Martins+13a} have shown that the underlying
stellar population has only a strong effect on the hydrogen recombination emission lines, with the largest differences in fluxes being about 25 per cent. This value is within the largest uncertainties on the fluxes values too.   For completeness, we have not subtracted the stellar features from the optical range too.

The results, including 3$\sigma$ uncertainties, are listed in Tables~\ref{flux} and \ref{flux2}.  For most of our targets, these measurements are made public for the first time. In addition, we computed the extinction coefficient, C$_{ext}$, for the NIR using the intrinsic value of 5.88 for the flux ratio of  Pa$\beta$/Br$\gamma$ \citep[][ using case B]{Hummer+87}. The \citet{Cardelli+89} extinction law was used, and the values obtained for the coefficients are listed in Tables~\ref{flux} and \ref{flux2}.

\begin{figure*}
\begin{minipage}[b]{0.5\linewidth}
\includegraphics[width=\textwidth]{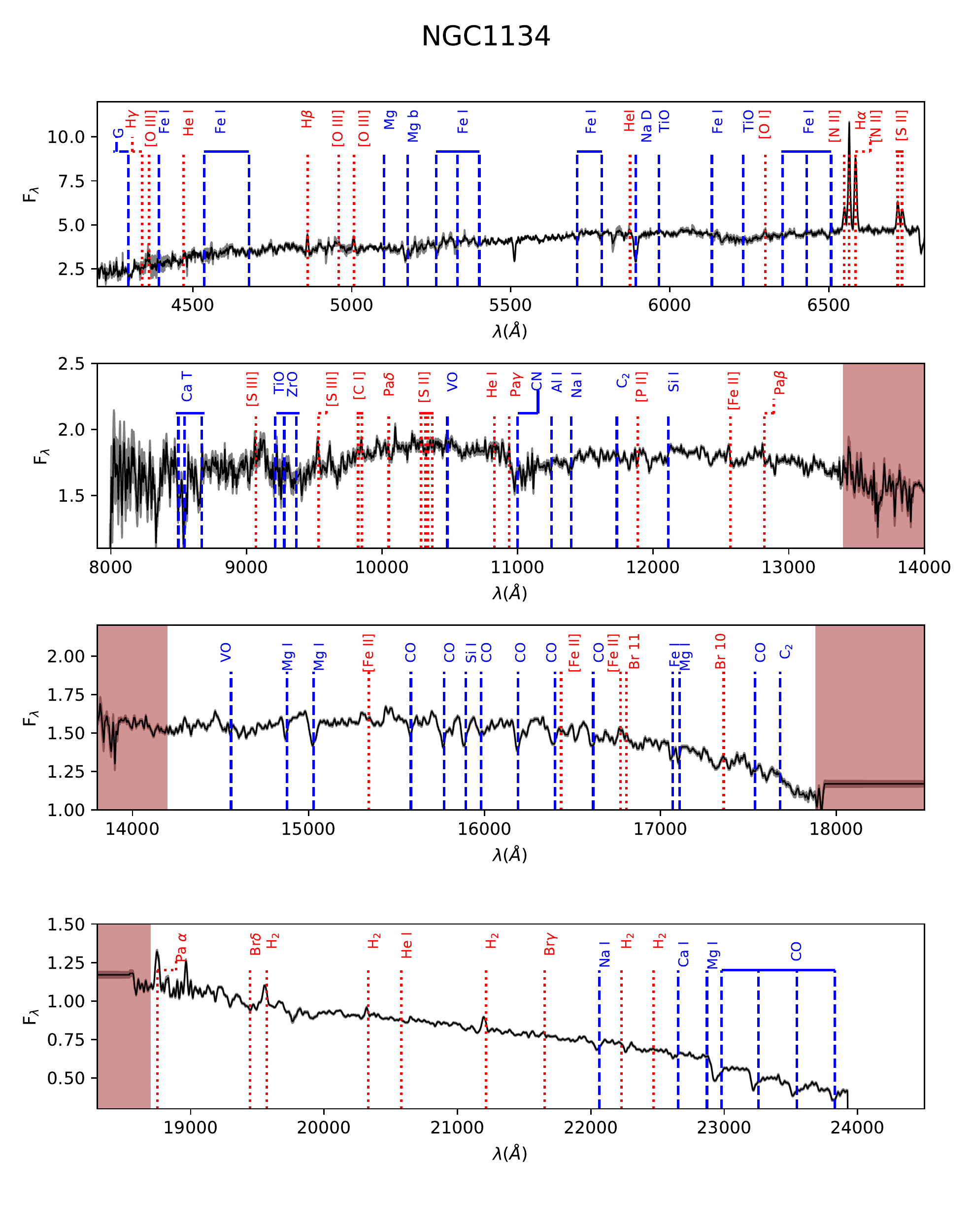}
\end{minipage}\hfill
\begin{minipage}[b]{0.5\linewidth}
\includegraphics[width=\textwidth]{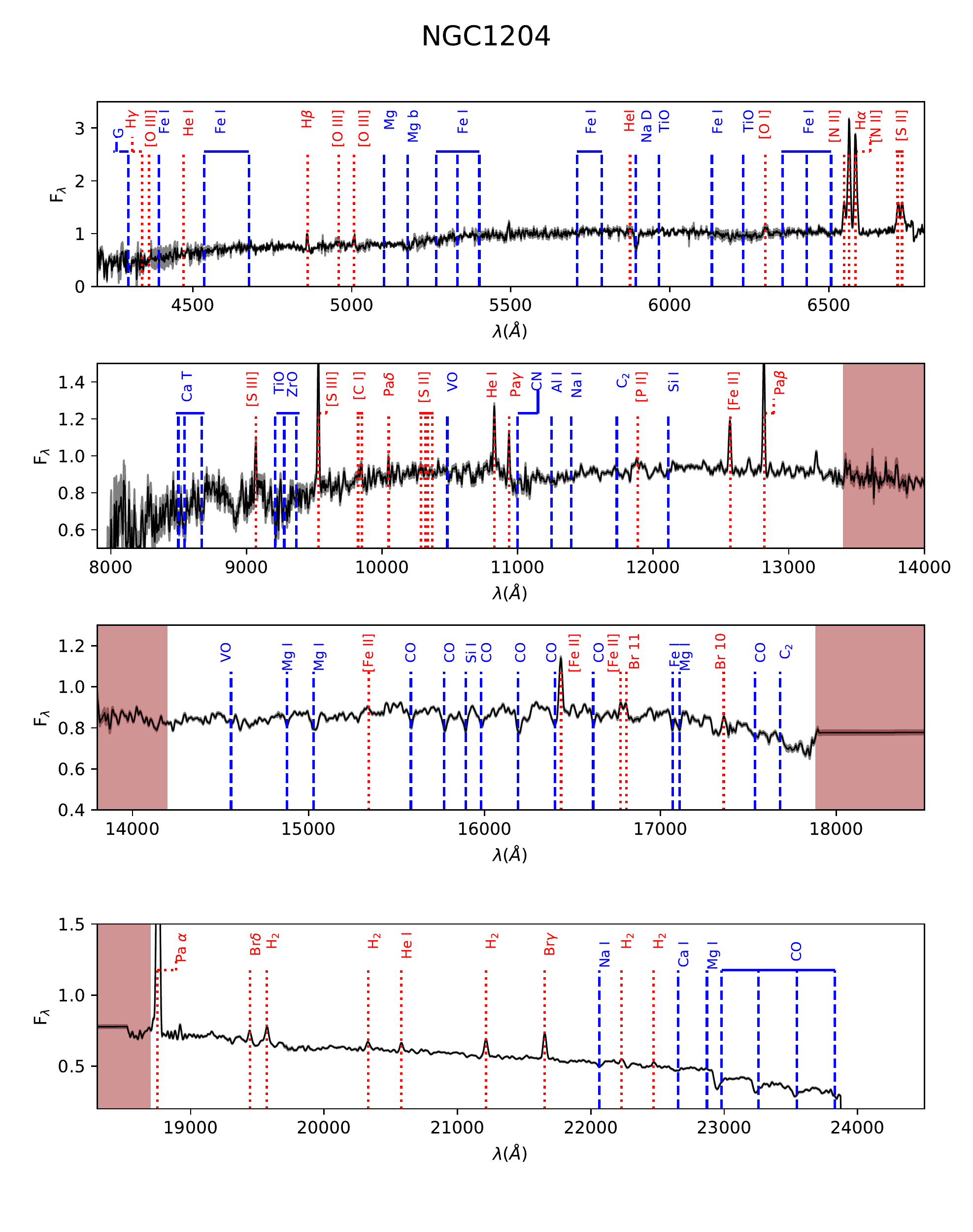}
\end{minipage}\hfill
\begin{minipage}[b]{0.5\linewidth}
\includegraphics[width=\textwidth]{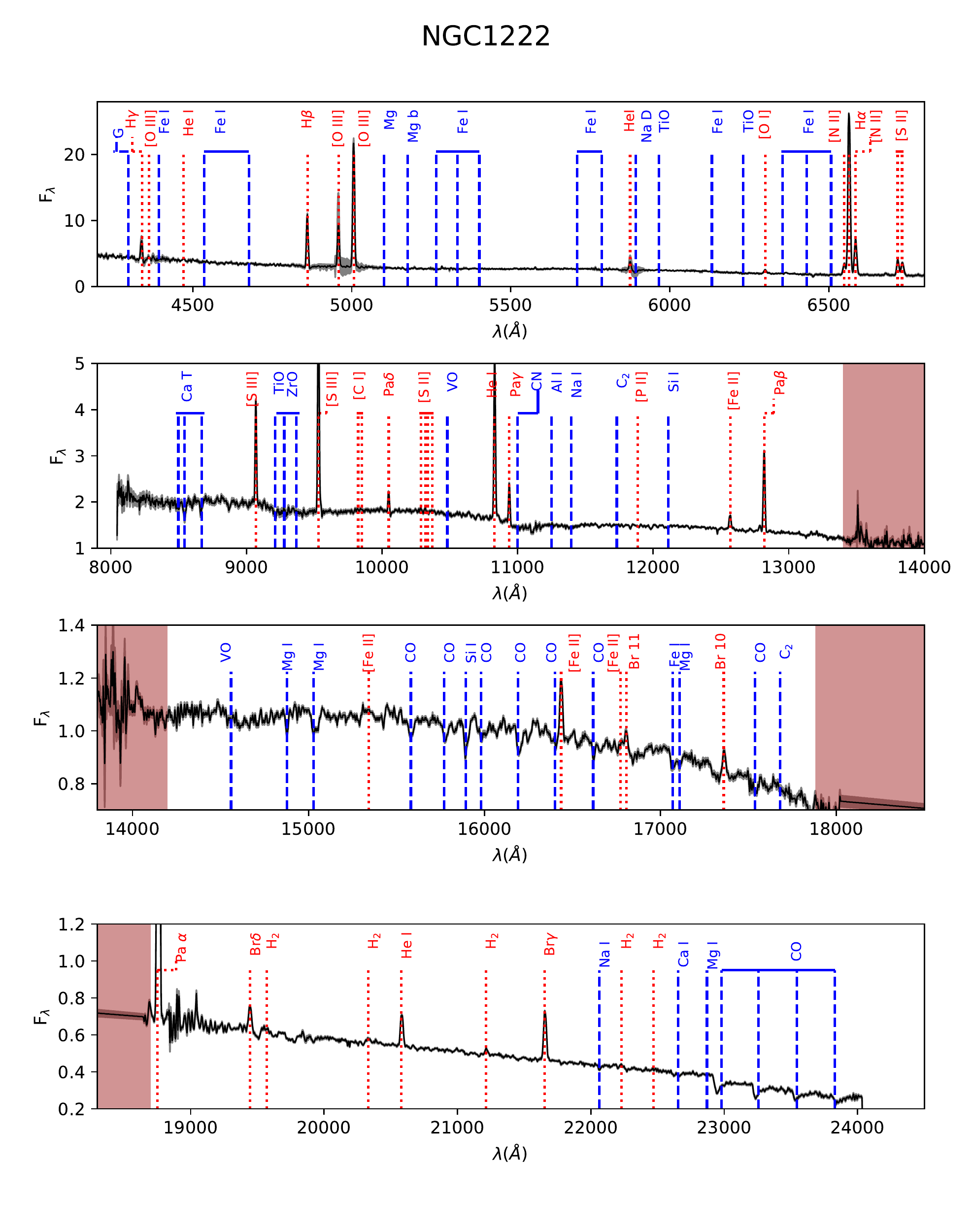}
\end{minipage}\hfill
\begin{minipage}[b]{0.5\linewidth}
\includegraphics[width=\textwidth]{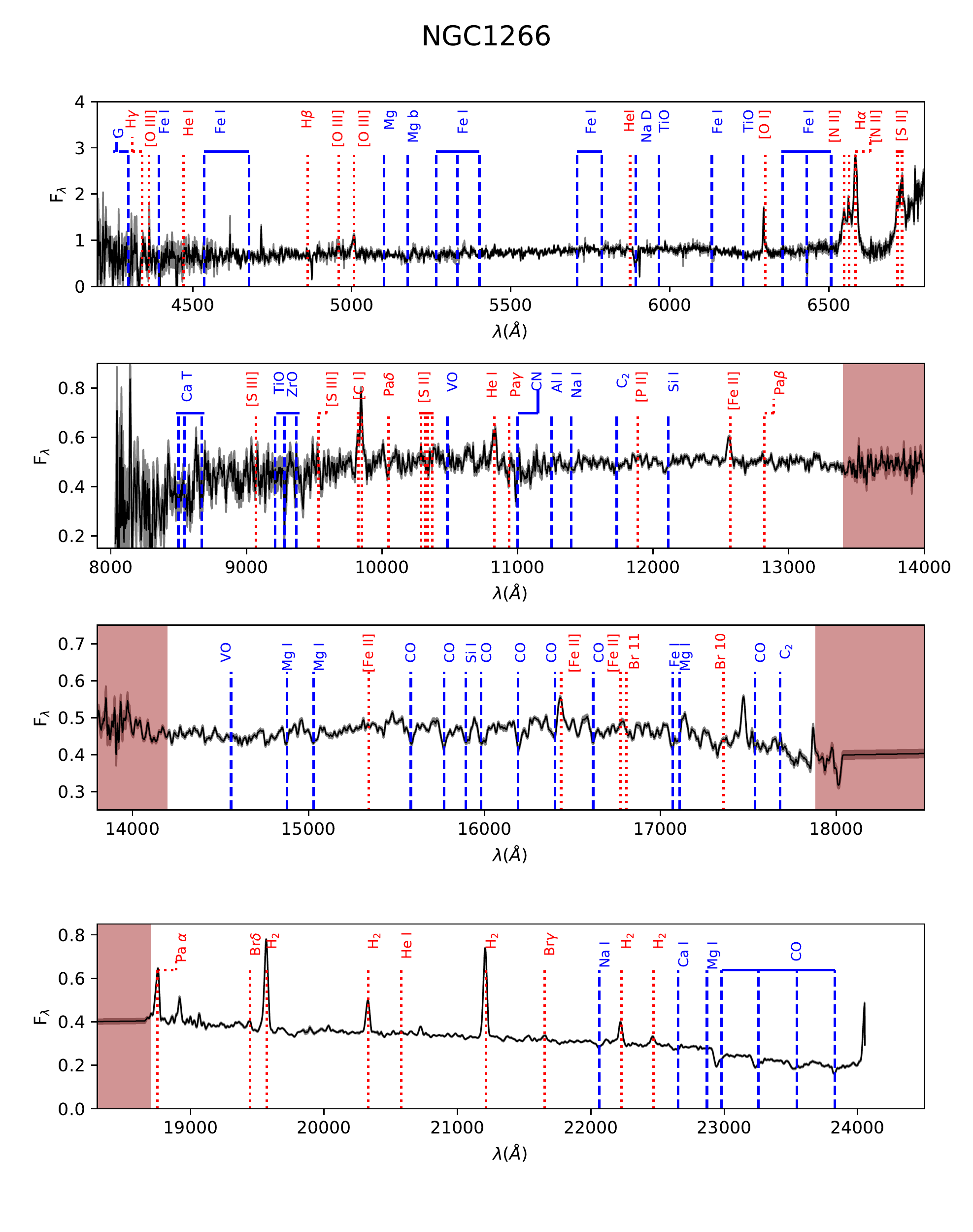}
\end{minipage}\hfill
\caption{Final reduced and redshift-corrected spectra for NGC~1134, NGC~1204, NGC~1222 and NGC~1266. For each galaxy we show from top to bottom the optical, $z+J$ , $H$ , and $K$ bands, respectively. The flux is in units of  $\rm 10^{-15}~ erg ~ cm^{-2} ~ s^{-1}$. The shaded grey area represents the uncertainties and the brown area indicates the poor transmission regions between different bands. The remaining spectra are shown in online material.}
\label{spectra}
\end{figure*}

\begin{figure}
\includegraphics[scale=0.5]{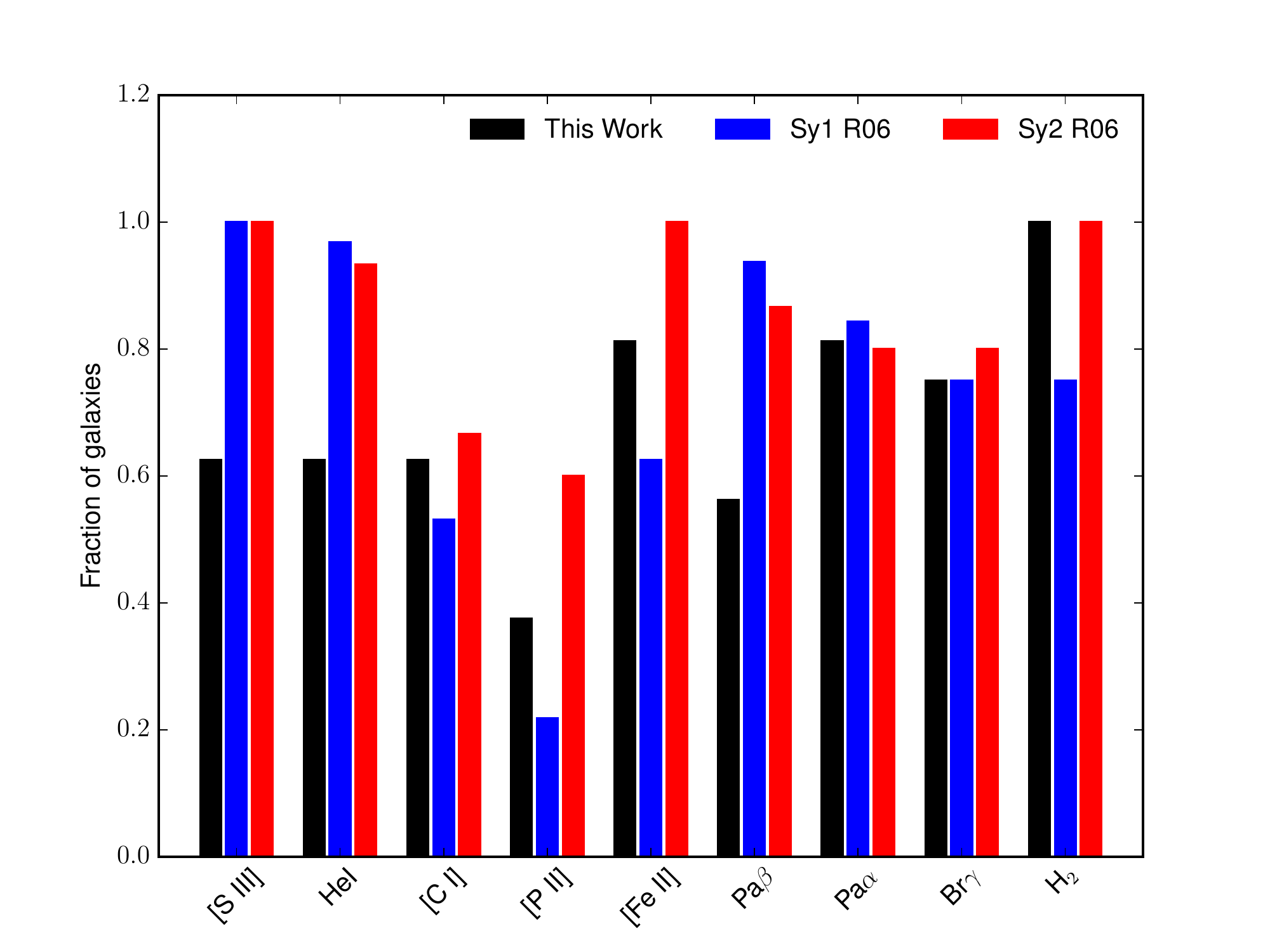}
\caption{Histogram showing statistics of the most common NIR emission lines.}
\label{freqhist}
\end{figure}

\begin{table*}
\renewcommand{\tabcolsep}{0.70mm}
\centering
\caption{Emission line fluxes in units of  $\rm 1x10^{-15}\,erg\,cm^{-2}\,s^{-1}$. \label{flux}} 
\begin{tabular}{lccccccccccccccc}
\hline \hline
\noalign{\smallskip}
Line    & Ion                  &  NGC23       &    NGC520     &    NGC660     &    NGC1055    &   NGC1134          &  NGC1204     &   NGC1222     &      NGC1266     \\
\noalign{\smallskip}
\hline                           
        & $C_{ext}$           &     --        &     --         & 3.42\pp0.13     &     --         &  --            & 2.95\pp0.14   & 1.06\pp0.05   &     6.74\pp0.73 	\\
4861    &H$\beta$             & 20.40\pp1.16  &       --       & 0.86\pp0.26     &      --        &  6.06\pp0.37   & 0.30\pp0.06   &  49.30\pp0.65  &     --        \\
4959    & [O {\sc iii}]       & 30.70\pp3.13  &       --       & 3.37\pp0.41     &      --        &  8.25\pp0.72   & 1.69\pp0.48   &  47.30\pp0.75  &     --        \\
5007    & [O {\sc iii}]       & 10.50\pp3.13  &   2.16\pp0.30  & 1.15\pp0.41     &      --        &  2.81\pp0.72   & 0.58\pp0.48   &  140.00\pp0.75 & 4.27\pp0.64   \\
6548    & [N {\sc ii}]        & 108.00\pp3.26 &   1.60\pp0.31  & 6.25\pp0.53     &  1.41\pp0.32   &  12.90\pp1.41  & 5.75\pp0.43   &  14.20\pp0.40  & 14.30\pp1.07  \\
6563    &H$\alpha$            & 83.80\pp3.89  &   8.68\pp0.44  & 17.20\pp0.46    &  4.97\pp0.34   &  40.50\pp0.95  & 20.50\pp0.42  &  222.00\pp0.43 & 11.40\pp0.76  \\
6583    & [N {\sc ii}]        & 27.90\pp3.89  &   6.96\pp0.49  & 18.60\pp0.48    &  4.56\pp0.39   &  38.30\pp1.21  & 18.60\pp0.42  &  49.70\pp0.45  & 30.60\pp0.87  \\
6716    & [S {\sc ii}]        & 46.50\pp1.92  &   3.29\pp0.46  & 5.15\pp0.48     &  1.25\pp0.29   &  14.60\pp1.23  & 5.76\pp1.23   &  20.70\pp0.48  & 18.70\pp1.16  \\
6730    & [S {\sc ii}]        & 34.70\pp1.92  &   3.07\pp0.55  & 4.88\pp0.61     &  1.31\pp0.44   &  12.90\pp1.50  & 5.69\pp1.23   &  19.90\pp0.57  & 21.10\pp1.16  \\
9069  & [S {\sc iii}]         & 14.80\pp3.96  &        --      &        --       &        --      &            --  & 10.30\pp1.40  &   32.80\pp1.59 &          --      \\
9531  & [S {\sc iii}]         & 15.70\pp3.96  &        --      &   23.90\pp0.94  &        --      &            --  & 11.90\pp0.54  &   72.60\pp1.49 &          --      \\
9824  & [C {\sc i}]           & 1.57\pp0.60   &        --      &   2.10\pp0.27   &        --      &            --  & 1.15\pp0.30   &   1.55\pp0.49  &          --      \\
9850  & [C {\sc i}]           & 5.15\pp0.60   &        --      &   2.53\pp0.27   &        --      &            --  & 1.88\pp0.30   &   0.97\pp0.49  &   9.21\pp1.06    \\
10049 &  Pa$\delta$           &      --       &        --      &   2.55\pp0.22   &        --      &            --  & 1.03\pp0.08   &   4.05\pp0.31  &          --      \\
10122 &  He {\sc ii}          &      --       &        --      &   2.78\pp0.22   &        --      &            --  & 1.34\pp0.08   &      --        &          --      \\
10830 & He {\sc i}            & 22.30\pp2.60  &        --      &   13.30\pp0.41  &        --      &            --  & 9.24\pp0.56   &   53.80\pp0.77 &   6.26\pp0.83    \\
10938 & Pa$\gamma$            & 5.65\pp1.58   &        --      &   8.57\pp0.31   &        --      &            --  & 2.78\pp0.27   &   12.90\pp0.76 &          --      \\
11470 &  [P {\sc ii}]         &      --       &        --      &   1.82\pp1.09   &        --      &            --  & 1.17\pp0.27   &      --        &          --      \\
11886 & [P {\sc ii}]          &      --       &        --      &   3.68\pp1.09   &        --      &            --  & 1.70\pp0.19   &      --        &          --      \\
12567 & [Fe {\sc ii}]         & 10.50\pp0.87  &        --      &   13.90\pp0.65  &        --      &            --  & 5.16\pp0.20   &   6.32\pp0.43  &   3.19\pp0.51    \\
12820 & Pa$\beta$             &      --       &        --      &   29.00\pp0.60  &        --      &            --  & 12.10\pp0.20   &   28.20\pp0.37 &  0.75\pp0.08      \\
12950 & [Fe {\sc ii}]         &      --       &        --      &   1.21\pp0.15   &        --      &            --  & 1.35\pp0.25   &      --        &          --      \\
13209 & [Fe {\sc ii}]         &      --       &        --      &   6.60\pp0.24   &        --      &            --  & 4.49\pp0.39   &      --        &          --      \\
15342 & [Fe {\sc ii}]         &      --       &        --      &        --       &        --      &            --  & 1.52\pp0.36   &      --        &          --      \\
16436 & [Fe {\sc ii}]         & 14.20\pp3.46  &   3.87\pp0.14  &   15.70\pp0.90  &        --      &            --  & 6.51\pp0.36   &   4.65\pp0.18  &   2.90\pp0.38    \\
16773 & [Fe {\sc ii}]+Br11    & 31.50\pp1.34  &        --      &        --       &        --      &            --  & 2.65\pp0.50   &      --        &          --      \\
17360 & Br10                  &      --       &        --      &        --       &        --      &            --  & 1.60\pp0.13   &      --        &          --      \\
18750 &  Pa$\alpha$           &      --       &   35.40\pp0.30 &   60.20\pp1.64  &        --      &            --  & 61.10\pp0.33  &   69.30\pp0.57 &   8.03\pp0.34    \\
19446 &  Br$\delta$           &      --       &   2.07\pp0.28  &   5.67\pp0.53   &        --      &            --  & 1.80\pp0.34   &   3.78\pp0.20  &          --      \\
19570 &  \h2\                 & 20.20\pp4.60  &   2.41\pp0.39  &   8.87\pp0.80   &   1.24\pp0.3   &            --  & 4.17\pp0.51   &   1.95\pp0.34  &   15.10\pp0.47   \\
20332 &  \h2\                 & 4.94\pp0.57   &   1.13\pp0.10  &   3.67\pp0.56   &        --      &  1.14\pp0.3    & 1.81\pp0.34   &   0.95\pp0.14  &   5.10\pp0.44    \\
20580 &  \h2\                 &      --       &   2.47\pp0.09  &   5.32\pp0.64   &        --      &            --  & 1.47\pp0.27   &   4.68\pp0.14  &          --      \\
21218 &  \h2\                 & 10.00\pp1.26  &   2.40\pp0.18  &   6.91\pp0.72   &   0.6\pp0.08   &  2.37\pp0.4    & 3.61\pp0.25   &   0.84\pp0.12  &   13.70\pp0.25   \\
21654 &  Br$\gamma$           &      --       &   6.67\pp0.19  &   16.60\pp0.71  &        --      &            --  & 5.88\pp0.27   &   6.98\pp0.07  &   1.40\pp0.33    \\
22230 &  \h2\                 & 4.53\pp2.97   &   0.67\pp0.19  &   2.01\pp1.20   &        --      &            --  & 0.87\pp0.09   &   0.49\pp0.19  &   3.26\pp0.12    \\
22470 &  \h2\                 & 1.47\pp0.37   &   0.83\pp0.12  &   1.18\pp0.12   &        --      &            --  & 0.72\pp0.10   &   0.29\pp0.04  &   1.51\pp0.14    \\
\noalign{\smallskip}
\hline
\end{tabular}
\end{table*}

\begin{table*}
\renewcommand{\tabcolsep}{0.70mm}
\centering
\caption{Continuation of Table~\ref{flux} \label{flux2}} 
\begin{tabular}{lccccccccccccccc}
\hline \hline
\noalign{\smallskip}
Line  & Ion                & UGC2982              &   NGC1797    &      NGC6814*        &   NGC6835       &   UGC12150   &     NGC7465    &    NGC7591     &    NGC7678          \\ 
\noalign{\smallskip}
\hline   
      & $C_{ext}$           &  --                  &  2.68\pp0.08  &          0.00         & 4.87\pp0.19      & 2.16\pp0.17   &  2.35\pp0.93    & 2.53\pp0.07     & 1.39\pp0.23  \\
4861  &H$\beta$             &  0.49\pp0.15         & 13.10\pp0.71  &             --        &     --           &     --        & 21.80\pp1.05    &     --          &  9.45\pp0.42   \\
4959  & [O {\sc iii}]       &  0.35\pp0.16         &     --        &             --        &     --           &     --        & 31.40\pp2.27    &     --          &      --        \\
5007  & [O {\sc iii}]       &  0.76\pp0.29         & 3.11\pp0.44   &             --        &     --           &     --        & 56.00\pp1.85    &     --          &  2.83\pp0.54   \\
6548  & [N {\sc ii}]        &  1.64\pp0.39         & 11.20\pp0.59  &             --        &     --           &     --        & 29.50\pp1.03    & 8.15\pp1.36     &  7.88\pp0.82   \\
6563  &H$\alpha$            &  13.90\pp0.40        & 88.70\pp0.74  &             --        &     --           &     --        & 122.00\pp0.89   & 21.10\pp1.00    &  53.40\pp1.00  \\
6583  & [N {\sc ii}]        &  5.62\pp0.39         & 46.40\pp0.76  &             --        &     --           &     --        & 71.10\pp0.98    & 18.60\pp1.04    &  27.60\pp1.00  \\
6716  & [S {\sc ii}]        &  2.77\pp0.45         & 10.80\pp0.41  &             --        &     --           &     --        & 39.70\pp1.17    & 4.87\pp1.94     &  9.23\pp0.97   \\
6730  & [S {\sc ii}]        &  2.66\pp0.75         & 10.30\pp0.46  &             --        &     --           &     --        & 34.10\pp1.22    & 3.46\pp1.94     &  10.30\pp1.37  \\
9069  & [S {\sc iii}]       &              --      &  5.80\pp0.59  &         23.70\pp0.49  &       --         &  5.16\pp0.27  &    18.00\pp1.43 &      --         &   6.59\pp0.64     \\
9531  & [S {\sc iii}]       &              --      &  12.10\pp0.59 &         55.50\pp0.58  &       --         &  5.22\pp0.20  &    38.20\pp0.78 &   10.00\pp0.32  &   15.60\pp0.64    \\
9824  & [C {\sc i}]         &              --      &  1.71\pp0.16  &            --         &       --         &       --      &    2.41\pp0.76  &   0.98\pp0.15   &   0.82\pp0.09     \\
9850  & [C {\sc i}]         &              --      &  1.67\pp0.16  &            --         &       --         &  2.49\pp0.17  &    3.32\pp0.76  &   2.58\pp0.15   &   1.74\pp0.09     \\
10049 &  Pa$\delta$         &              --      &       --      &            --         &       --         &       --      &      --         &   3.85\pp0.44   &      --           \\
10122 &  He {\sc ii}        &              --      &       --      &            --         &       --         &       --      &      --         &   3.28\pp0.23   &      --           \\
10830 & He {\sc i}          &              --      &  7.72\pp1.08  &            --         &       --         &  8.40\pp1.23  &    25.60\pp1.52 &   8.41\pp0.88   &   9.47\pp0.77     \\
10938 & Pa$\gamma$          &              --      &  3.98\pp1.08  &            --         &       --         &  2.57\pp0.57  &    8.07\pp1.20  &   3.02\pp0.47   &   4.08\pp0.55     \\
11470 &  [P {\sc ii}]       &              --      &  1.30\pp0.37  &            --         &       --         &  1.17\pp0.14  &      --         &   2.50\pp0.87   &      --           \\
11886 & [P {\sc ii}]        &              --      &  1.80\pp0.37  &         3.83\pp1.08   &           --     &  1.46\pp0.14  &           --    &   4.98\pp0.87   &      --           \\
12567 & [Fe {\sc ii}]       &              --      &  5.04\pp0.25  &         4.64\pp0.52   &  2.06\pp0.23     &  4.73\pp0.33  &    11.80\pp0.66 &   6.80\pp0.21   &   3.28\pp0.49     \\
12820 & Pa$\beta$           &              --      &  12.10\pp0.26 &        3.53\pp0.57    &  4.86\pp0.19     &  7.86\pp0.30  &    9.57\pp2.61  &   9.74\pp0.21   &   9.20\pp0.51     \\
12950 & [Fe {\sc ii}]       &              --      &  1.18\pp0.42  &            --         &       --         &       --      &      --         &      --         &     --            \\
13209 & [Fe {\sc ii}]       &              --      &  2.94\pp0.42  &            --         &       --         &       --      &    5.60\pp0.34  &      --         &     --            \\
15342 & [Fe {\sc ii}]       &              --      &       --      &            --         &       --         &       --      &      --         &      --         &     --            \\
16436 & [Fe {\sc ii}]       &              --      &  4.65\pp0.47  &         5.53\pp0.48   &  3.24\pp0.08     &  3.56\pp0.18  &    9.20\pp0.28  &   6.05\pp0.60   &   3.09\pp0.21     \\
16773 & [Fe {\sc ii}]+Br11  &              --      &       --      &            --         &       --         &       --      &      --         &      --         &     --            \\
17360 & Br10                &              --      &       --      &            --         &       --         &       --      &      --         &      --         &     --            \\
18750 &  Pa$\alpha$         &         4.70\pp0.18  &  60.40\pp0.46 &         82.50\pp3.49  &  32.60\pp0.29    &  34.30\pp0.17 &    31.60\pp3.36 &   36.70\pp0.67  &   23.30\pp0.32    \\
19446 &  Br$\delta$         &              --      &  2.12\pp0.39  &            --         &  2.36\pp0.22     &        --     &         --      &          --     &   1.05\pp0.10     \\
19570 &  \h2\               &         1.79\pp0.49  &  4.10\pp0.52  &         3.36\pp0.54   &  3.33\pp0.44     &  5.25\pp0.05  &    4.33\pp0.16  &   8.17\pp0.26   &   1.45\pp0.10     \\
20332 &  \h2\               &         0.61\pp0.08  &  1.48\pp0.30  &         1.25\pp0.20   &  0.86\pp0.05     &  1.65\pp0.07  &    2.36\pp0.39  &   2.46\pp0.20   &   2.03\pp0.20     \\
20580 &  \h2\               &              --      &  1.40\pp0.26  &            --         &  1.98\pp0.05     &        --     &    1.33\pp0.34  &   1.53\pp0.18   &   1.24\pp0.20     \\
21218 &  \h2\               &         0.60\pp0.02  &  3.21\pp0.28  &         2.14\pp0.20   &  1.34\pp0.21     &  4.00\pp0.13  &    3.92\pp0.27  &   4.80\pp0.36   &   0.88\pp0.14     \\
21654 &  Br$\gamma$         &         0.80\pp0.04  &  5.33\pp0.09  &         0.52\pp0.32   &  4.66\pp0.25     &  2.88\pp0.13  &    3.75\pp0.69  &   4.07\pp0.05   &   2.56\pp0.15     \\
22230 &  \h2\               &             --       &  1.20\pp0.17  &         1.05\pp0.23   &  0.65\pp0.18     &  1.70\pp0.85  &    1.71\pp0.10  &   2.25\pp0.25   &   0.71\pp0.03     \\
22470 &  \h2\               &            --        &  1.18\pp0.16  &         0.61\pp0.13   &           --     &  0.55\pp0.12  &    0.49\pp0.06  &   1.30\pp0.26   &   0.32\pp0.14     \\
\noalign{\smallskip}
\hline
\end{tabular}
\end{table*}

\subsection{The continuum spectra}

\begin{figure*}
\includegraphics[scale=.7]{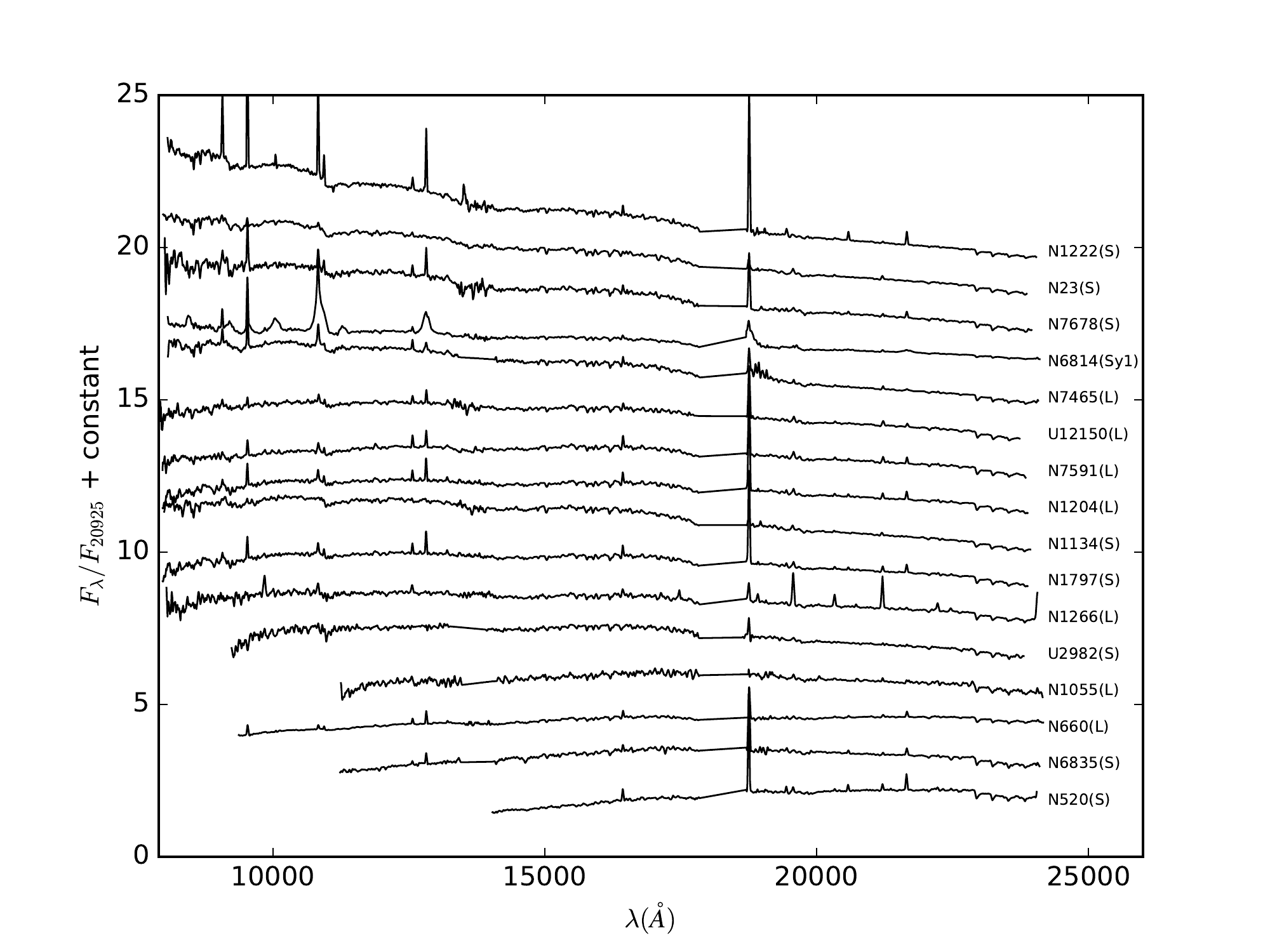}
\caption{Near-infrared normalized spectra ordered according to their shapes from steeper (top) to flatter (bottom). The data were normalized at 20925\AA. Activity types are listed (S = SFG and L=LINER).}
\includegraphics[scale=0.7]{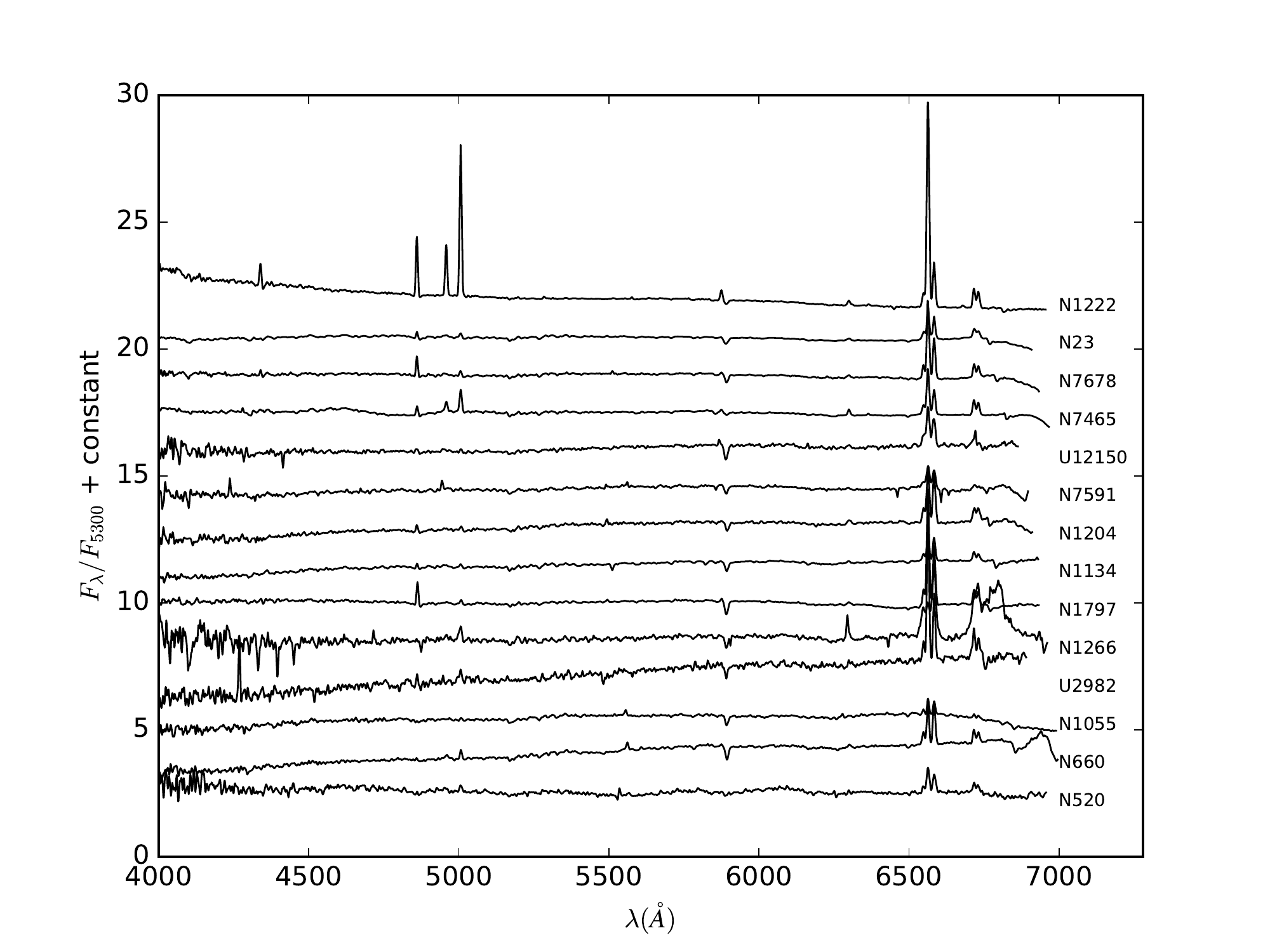}
\label{nircont}
\caption{Same order as Figure~\ref{nircont}, but for the optical spectral range. The data were normalized at 5300~\AA.}
\label{optcont}
\end{figure*}

The main goal of this section is to characterize the continuum emission observed in our sample and compare it to other data in the literature.  To help in the visual inspection\footnote{ Emission lines and equivalent widths of the absorption features were measured on the spectra previous to normalization.} of the individual spectra, we normalized the continuum
emission to unity in two regions free from emission/absorption features taken from \citet{Riffel-Rogerio+11c}. The NIR spectra were normalized at 20925~\AA\ and then sorted according to their continuum shapes. For a proper comparison with the optical portion of the spectrum, we normalized the optical spectra at 5300~\AA\ and plotted them in the same order as the NIR spectra (Figs.~\ref{nircont} and~\ref{optcont}).

A first-order inspection of Figure~\ref{nircont} allows us to infer that, contrary to what happens in Seyfert galaxies \citep{Riffel-Rogerio+06}, there seems to be no correlation between activity type (LINERs or SFGs) and continuum shape. In fact, these  very bright infrared galaxies present a continuum shape very similar to what is found in fainter H{\sc ii} sources and normal galaxies, as reported by \citet{Martins+13a}, which may indicate that the LINER spectrum of these galaxies is powered by starburst instead of a low-luminosity AGN. In addition, the continua of all the optical spectra look very similar.

A large diversity of atomic absorption lines and molecular bands is also apparent in the spectra. These features are seen from the very blue optical end to the red end of the observed NIR spectral region. The most common atomic absorption features are due to \ion{Ca}{i}, \ion{Ca}{ii}, \ion{Fe}{i}, \ion{Si}{i}, \ion{Na}{i} and \ion{Mg}{i}, besides the prominent absorption bands of CH, MgH, TiO, VO, ZrO and CO.  These features are identified in Fig.~\ref{spectra}.  It is clear in these figures that some of the most important features predicted for intermediate-age stellar populations, which are expected to be enhanced in the RGB and TP-AGB stellar phases \citep{Maraston05, Riffel-Rogerio+07, Riffel-Rogerio+15}, are detected in the spectra. Among these features are the ZrO/CN/VO at 9350~\AA, the 10560~\AA\ VO,  1.1~\mc\ CN and 1.6~\mc\ and 2.3~\mc\  CO bands.

\subsubsection{Towards a homogeneous NIR index definition}

The Equivalent Widths (EWs) of these features offer coarse but robust information about the stellar content of a galaxy spectrum, and therefore they can be used as powerful diagnostics of the stellar content of galaxies. Contrary to the optical range, where there exist indices defined in a homogeneous way by the Lick group \citep[see][and references]{Worthey+94}, in the NIR there is no such homogeneous set of definitions covering the full NIR wavelength range, and authors tend to use their own definitions \citep[e.g.][]{Riffel-Rogerio+07,Riffel-Rogerio+08,Silva+08,Marmol+09,Cesetti+09,Kotilainen+12,Riffel-Rogerio+11b,Riffel-Rogerio+15,Rock+17}, and therefore it is very difficult to compare results from different investigations. 

With this in mind, here we create a set of definitions for absorption features found in the NIR. We used two SSPs from the IRTF-based {\sc emiles} models \citep{Vazdekis+16,Rock+15,Rock+16}, with 1.0\,Gyr and 10\,Gyr, solar metallicity, calculated with the PADOVA evolutionary tracks and with $\sigma=228$\kms. We added up their light fractions (normalized to unity at $\lambda$=12230\AA) as follows: 

$ F_{\rm comb}=0.5 \frac{F_{\lambda}^{1Gyr}}{F_{\lambda=12230}^{1Gyr}} + 0.5\frac{F_{\lambda}^{10Gyr}}{F_{\lambda=12230}^{10Gyr}} $.

\begin{figure*}
\includegraphics[scale=0.8]{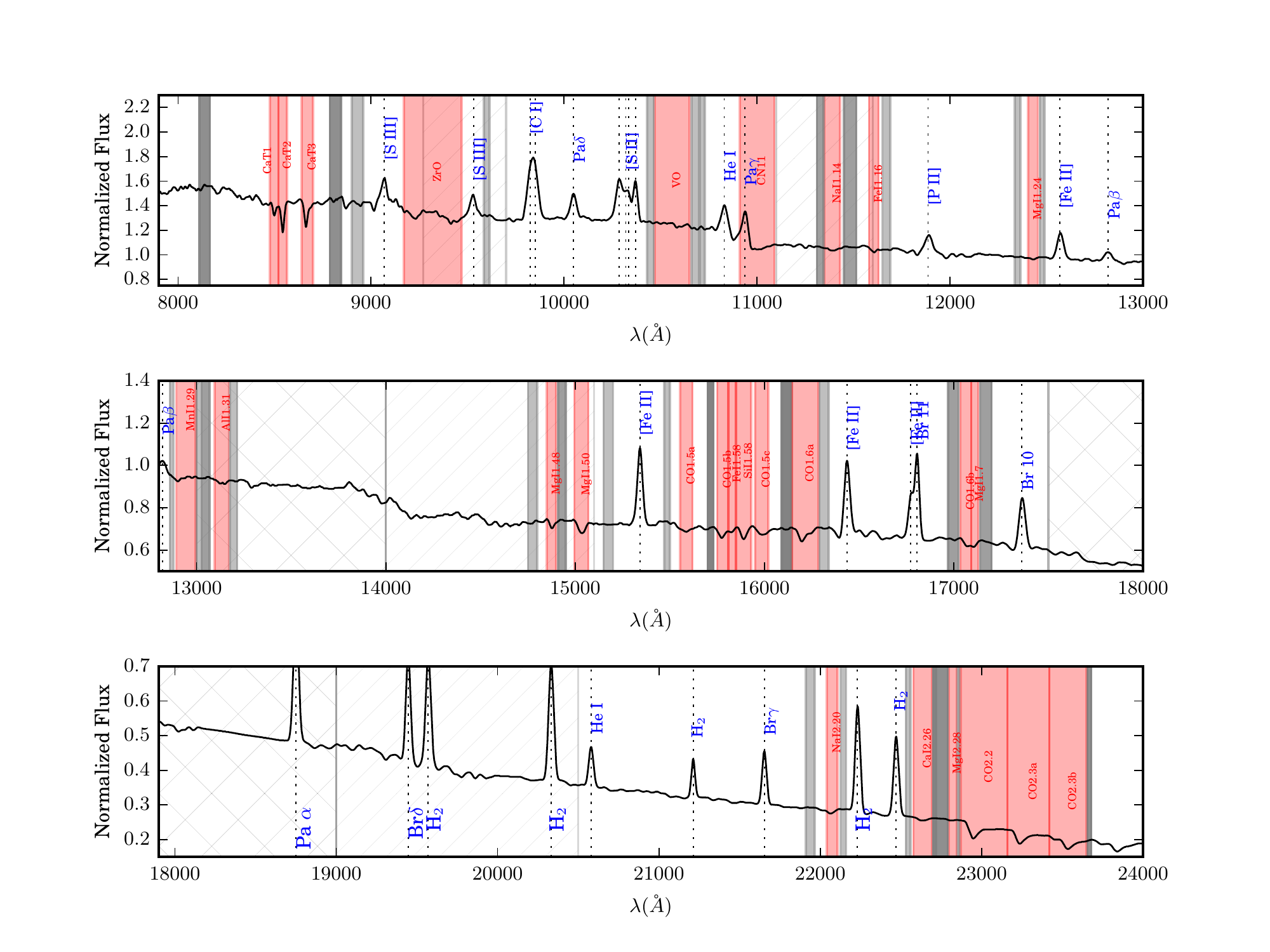}
\caption{Simulated spectrum showing the NIR indices definitions. Blue and red continuum band passes are in grey and line limits in red. Regions of strong (transmission $<2$0\%) telluric absorption are shaded with an ``X" pattern, while regions of moderate (transmission $<$80\%) telluric absorption are shaded with a line pattern. Emission lines and absorption features are labeled. See text for details.}
\label{indDefs}
\end{figure*}

\noindent To this resulting spectrum we added Gaussians to model emission-lines profiles. These lines are located at the wavelengths of the most common emission lines detected in galaxies in this spectral region (see Sec.~\ref{sec_emlines}) with Full Width at Half Maximums (FWHMs) characteristic of galaxies observed with SpeX with the configuration used here (25\AA~$\rm~\lesssim~FWHM~\lesssim$~40\AA) with arbitrary flux values. We employed the {\sc elprofile} routine of the {\sc ifscube} package\footnote{available at: https://bitbucket.org/danielrd6/ifscube.git} (Ruschel-Dutra, in preparation). Using this simulated spectrum we defined the line limits and continuum band passes as illustrated in  Fig.~\ref{indDefs} and listed in Table~\ref{ewdefs}.

We have measured the EWs for the most prominent absorption features using an updated {\sc python} version of the {\sc pacce} code \citep{Riffel+Vale11}.  In this code version, the EW uncertainties are assumed to be the standard deviation of 1000 EWs measurements of simulated spectra created by perturbing each flux point by its uncertainty through a Monte Carlo approach. The line definitions used are listed in Table~\ref{ewdefs}, and the measured values are in Tables~\ref{Ew1} and \ref{Ew2}. In order to have a sample of early type galaxies (ETG) to compare our results with, we have collected NIR spectra from the literature and measured the EW of the absorption features with the same definitions used for our sample. Tables \ref{Ewbald1} and \ref{Ewbald2} present the measurements for the sample of galaxies presented in \citet{Baldwin+17}. For four of the galaxies we were able to find Sloan Digital Sky Survey data used to measure the optical EW, while for the remaining objects we collected the values of Fe5015, Mg$_b$ and Fe5270  from \citet{McDermid+15}. We also measured the values from the spectra presented by \citet{Dahmer-Hahn+18}, which values are listed in Tab.~\ref{Ewlgdh}.

\section{discussion}

\subsection{Emission Lines}

In order to compare the frequency of occurrence of the emission lines in our sample with what is seen in Seyfert galaxies, we show a histogram in Fig.~\ref{freqhist} where the lines found here are compared to those of \citet{Riffel-Rogerio+06}. What clearly emerges from this figure is that  \lb\ion{S}{iii}],  \ion{He}{i}, and Pa$\beta$ lines are less frequent in our sample (occurring in $\sim$60\% of the sources) than in Seyferts (present in almost all of the objects). On the other hand, we find a higher frequency of occurrence of lines of [C {\sc i}] ($\sim$ 65\%), [\ion{P}{ii}] ($\sim$ 40\%) and \fe2\ ($\sim$ 65\%) than in Sy~1 objects, and a similar rate as in Sy~2s. The remaining emission lines occur with similar frequencies in the present sample and in Seyferts \citep[see also][]{Lamperti+17}. Lines that are less frequent in the present sample compared to AGNs are located in regions with strong stellar features. Thus, it is possible that the absence of these features is because they are intrinsically weaker than in AGNs and/or diluted by the broad absorption features that dominate the $z+J$ band. Note though that for three objects (NGC\,1055, NGC\,6835 and NGC\,520, see Fig~\ref{nircont}), our spectral range excludes the \lb\ion{S}{iii}],  \ion{He}{i}, and  [C {\sc i}] emission lines. If present in these spectra, they would show up in $\sim$80\% of our sample.  

\begin{figure}
\includegraphics[scale=0.45]{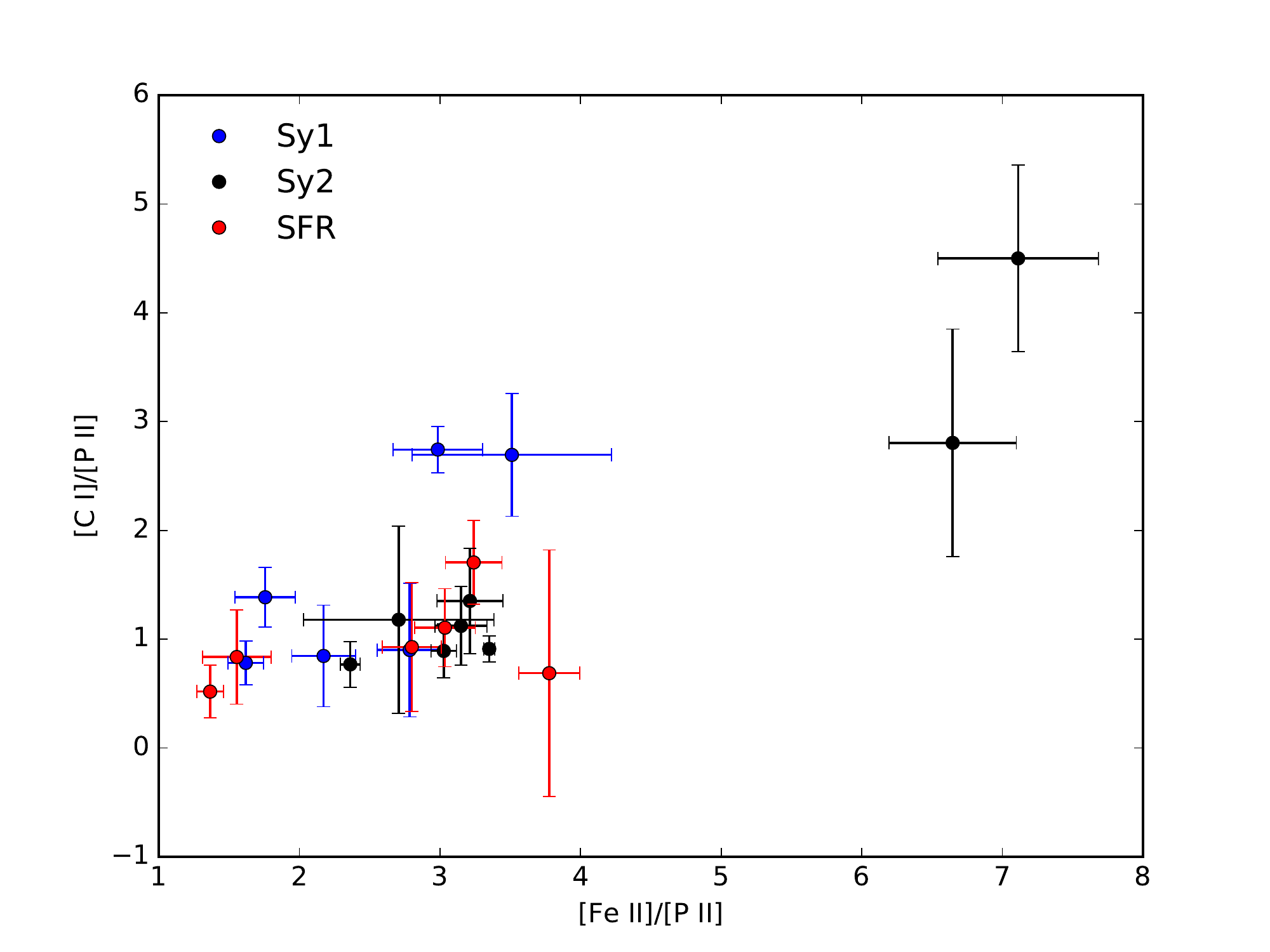}
\label{pii_cor}
\caption{The correlation between the emission-line ratios of [\ion{C}{i}]\,$\lambda$9850\,\AA   and \fe2\,$\lambda$12570\,\AA} relative to  [\ion{P}{ii}]\,$\lambda$11886\,\AA.
\end{figure}

It is worth mentioning that the kinematics of the [S {\sc iii}], \fe2\  and \h2\ lines as well as the  excitation mechanisms of the \fe2\ and \h2\ lines of the galaxies of this sample were explored in \citet{Riffel-Rogerio+13}. However, the low-ionization forbidden lines of [\ion{C}{i}] (i.e. $\lambda$\,9850\,\AA) and [\ion{P}{ii}] (i.e. $\lambda$\,11886\,\AA), also
detected in our sample, were not yet analyzed. Although the [\ion{P}{ii}] line is stronger compared to \fe2 $\lambda$\,12570\,\AA in Sy~2s \citep{Riffel-Rogerio+06} than in the other types of galaxies, the detection of [\ion{P}{ii}] lines is surprising here. This is because at Solar metallicity, Phosphorus is about 1000 times less abundant than Carbon  \citep{Ferguson+97} and 100 times less abundant than Iron \citep{Oliva+01}.  Hence, if the P/C abundance is near to solar, the [\ion{P}{ii}] lines should not be  present, unless other strong abundant elements are much more optically thick than they appear. 
A similar problem is found in some quasars for which broad absorption lines of 
\ion{P}{v}~$\lambda\lambda$1118,1128~\AA\ are detected and extreme abundances ratios for 
P/C are found \citep{Hamann+98,Hamann+01,Borguet+12}. According to \citet{Oliva+01} for a solar Fe/P $\sim$ 100 abundance ratio, one expects that  $\frac{[\ion{Fe}{ii}]}{[\ion{P}{ii}]} = 50$, similar to what is expected for supernova remnants.  The NIR [\ion{P}{ii}] emission lines may probably help to set some constraints on the abundance of Phosphorus in galaxies.

As discussed in \citet{Oliva+01} bright \fe2\ lines can only be formed in regions where hydrogen is partially ionized. Such regions of hot, partially ionized gas can only be produced in an efficient way by shocks and/or photoionisation by soft X-rays. According to these authors, [\ion{Fe}{ii}]/[\ion{P}{ii}] can be used to distinguish between shocks (ratio $\gtrsim$20) and photoionisation (ratio $\lesssim$ 2). In order to test this hypothesis, we plotted in   Fig.~\ref{pii_cor}  [\ion{Fe}{ii}]/[\ion{P}{ii}] $\times$  [\ion{C}{i}]/[\ion{P}{ii}] for our sample as well as the Seyfert galaxies of \citet{Riffel-Rogerio+06}. 
As can be seen in this figure, there is a good correlation and no clear separation between the SFGs and the Seyferts, suggesting that the dominant excitation mechanism is 
the same for the three ions.
Furthermore, due to the low values derived for the [\ion{Fe}{ii}]/[\ion{P}{ii}] ratio, that
excitation mechanism might be expected to be photoionisation based on the arguments of \citet{Oliva+01}. To test this, we have computed photoionisation models using  {\sc cloudy/C17.01}\footnote{Available at https://www.nublado.org.} \citep{Ferland+17} updated with the next release of collisional strengths for [P II]  \citep[taken from][]{Tayal04} as well as with new transition probabilities\footnote{They are a 
combination of data taken from the MCHF/MCDHF Database at http://nlte.nist.gov/MCHF/ and data from the NIST Atomic Spectra  Database at https://www.nist.gov/pml/atomic-spectra-database} (private communication), however these models are not able to reproduce the observed line ratios, underestimating both (the models values for both ratios are nearly zero). This may be due to the fact that these lines are not in fact excited by photoionisation, but mostly driven by shocks.

\subsection{Absorption Features}
It is crucial to be able to derive ages and chemical composition in order to understand the dominant underlying unresolved stellar content of galaxies \citep{Rock+17}. So far, the NIR is lacking a clear procedure based on  absorption-line strengths.  The obvious choice to do this kind of study is using stellar clusters as probes, instead of the use of more complex star-forming objects. However, while observations of the integrated spectra of stellar clusters in the optical region have been available for almost 30 years \citep[e.g.][]{Bica+88} in the NIR such observations are very difficult since the light emitted by the stars of the clusters in the NIR bands is dominated by a few very bright stellar phases making it difficult to get reliable integrated spectra of such objects in the NIR \citep[e.g][]{Lyubenova+10,Riffel-Rogerio+11a}.

In order to have a more homogeneous data-set, in addition to the data-set we present here representing complex SFHs of SFGs (\S~\ref{obs}), we collected spectra of nearby ETGs (which tend to have less complex SFHs than our sample) observed similarly as those in the present work. Our final data set representing the older stellar population is composed of 12 ETG selected in order to span a wide range of ages (1-15 Gyr) at approximately solar metallicity and observed by \citet{Baldwin+17} using Gemini/GNIRS in the cross-dispersed mode ($\sim 0.8 - 2.5\mu m; R\sim 1700$; $\sigma \sim$75 \kms) plus 6 ETG selected from the Calar Alto Legacy Integral Field Area Survey \citep[CALIFA][]{Sanchez+16} and observed by \citet{Dahmer-Hahn+18} using the TripleSpec spectrograph attached to the Astrophysical Research Consortium (ARC) 3.5-meter telescope ($\sim 0.95 - 2.45\mu m; R\sim 2000$; $\sigma \sim$64 \kms). In addition to these NIR spectra, we also collected, when available, the optical spectra of the sources. In the case of \citet{Baldwin+17} galaxies, the optical spectra where taken from the Sloan Digital Sky Survey \citep{Ahn+14}, while for the sample of \citet{Dahmer-Hahn+18} we took the data from the  \citep[CALIFA][]{Sanchez+16}. The optical and NIR indices were measured by us using the definitions of Tab.~\ref{ewdefs} and are listed as online material in Tabs. \ref{Ewbald1}, \ref{Ewbald2} and \ref{Ewlgdh}.

\subsubsection{Previous NIR index - index correlations }

\begin{figure}
\includegraphics[scale=.45]{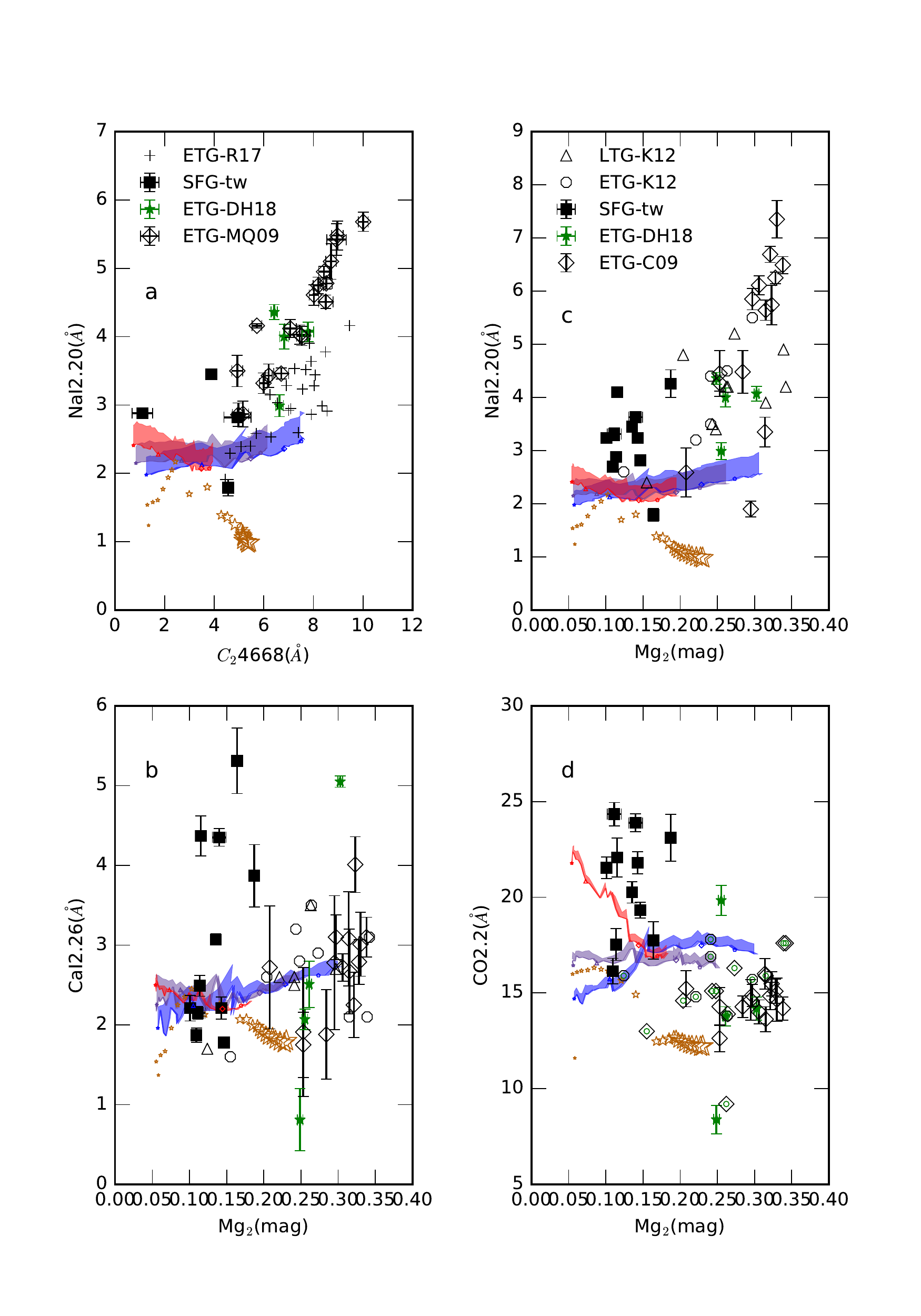}
\caption{Index-index diagrams. Filled boxes are from this work. Plus markers indicate early-type objects indices presented in \citet[][panel a, R17]{Rock+17}. Open diamonds are indices measured in early-type galaxies taken from \citet[panel a, ETG-MQ09][]{Marmol+09} and \citet[panels b, c and d, ETG-C09][]{Cesetti+09}. Triangles and octagons represent, respectively, the late and early-type galaxies studied by \citet[][LTG-K12 and ETG-K12]{Kotilainen+12}. Filled green stars represent our new measurements for the early-type galaxies of \citet[][ETG-DH18]{Dahmer-Hahn+18} in all panels. Note that the literature data may have different definitions between them selves  and with the measurements we present here. Open brown stars represent \citet{Maraston+11} solar metalicity, Pickles-based models, with the size of the points scaling with ages (smallest points for 300Myr and largest for 15Gyr). The shaded areas represent IRTF-based {\sc emiles} models  \citep{Vazdekis+16,Rock+15,Rock+16} with red, gray and blue indicating [Fe/H]=-0.40, [Fe/H]=0.00 and [Fe/H]=0.22, respectively. The shaded area represent models with a spectral resolution of $\sigma$=60 $\rm km/s$ (the lowest available) to $\sigma$=228 $\rm km/s$. The age range used is between 0.3\,Gyr and 15.0\,Gyr, with arrows, triangles, diamonds and pentagons representing 0.3, 1, 5 and 10~Gyr, respectively. The E-MILES models with ages smaller than 1\,Gyr should be taken with caution. For more details see text. }
\label{index_index}
\end{figure}

Due to the lack of adequate data sets to test predictions of NIR data, compared to the optical \citep[see][for example]{Thomas+03}, there are only a few studies trying to understand the behaviour of NIR $\times$ NIR indices.   For instance, \citet{Marmol+09} studied a sample of early type galaxies and found a strong correlation between $C_2$4668 and NaI2.20 indices. In Fig.~\ref{index_index}a we show the  \citet{Marmol+09} measurements (open diamonds) and the literature compilation presented by \citet[plus symbols,][]{Rock+17} together with our data (squares). Even though we only measured both indices for 4 sources, this correlation seems to still hold for SFGs, which populate the lower left end of the correlation (Fig.~\ref{index_index}a).

Using a similar approach, \citet{Cesetti+09} reported a trend of correlation of the optical Mg$_2$ band with NIR indexes, such as NaI2.20, CaI2.26 and CO2.2 for early type galaxies. In Fig~\ref{index_index}b,c and d we plotted our sample (filled squares), together with those of \citet[][open diamonds]{Cesetti+09} and \citet[open triangles][]{Kotilainen+12} for early type sources. Additionally we also added the inactive spirals (octagons, LTG-K12) of \citet[][]{Kotilainen+12}. From Fig.~\ref{index_index}d we have excluded the two Seyfert galaxies (NGC\,660 and NGC\,6814) since the CO band can be very diluted in these kind of sources \citep{Riffel-Rogerio+09,Burtscher+15}. 

From Fig.~\ref{index_index} it is clear that the trend seems to hold for NaI2.20 $\times$  Mg$_2$, while for CaI2.26 $\times$  Mg$_2$ there is no clear correlation, and in the case of CO2.2 $\times$  Mg$_2$ instead of a positive correlation there seems to be an inverse correlation. Additionally there seems to be a segregation between early and late-type galaxies in this plot (panel d). This indicates that CO is enhanced in younger stellar populations, in agreement with the predictions of the \citet{Maraston05} models as shown in  \citet{Riffel-Rogerio+07}.

To help in the interpretation of these results, on these index-index diagrams we have over-plotted the new optical-to-NIR  IRTF-based  stellar population synthesis models of the E-MILES team \citep{Vazdekis+12,Vazdekis+16,Rock+16}. The models employed are those computed using the PADOVA isochrones \citep{Girardi+00}, with ages in the range 0.3\,Gyr $< t < $ 15.0\,Gyr and metallicities within [Fe/H]~=~-0.40, [Fe/H]~=~0.00 and [Fe/H]~=~0.22 with two different spectral resolutions ($\sigma$=60\kms\ and $\sigma$=228\kms, the shaded area represents the differences caused by $\sigma$). We also plotted TP-AGB heavy \citep[see][for a comparison between TP-AGB heavy and light models]{Zibetti+13}, Pickles-based  models of \citet[][M11 hereafter]{Maraston+11}, which do have the same prescription than \citet{Maraston05} models but with a higher spectral resolution (R = 500) than the 2005 models, therefore making them more suitable for our comparisons. However, it is important to have in mind that M11 models do have a poorer spectral resolution than our data, the effects on the indices strengths by degrading the resolution to M11 models is within the uncertainties of our measurements. These models are shown as open brown stars and are only available for solar metalicity. What emerges from this exercise is that the models in general are not able to predict the NIR indices and that  there is a segregation between early (open diamonds and plus markers) and late-type (filled squares and octagons) galaxies in these diagrams. The upper panels show significantly larger NaI2.20 index values than predicted by the models with standard IMF. Both the optical Mg and C dominated indices are stronger than the models for the most massive galaxies (i.e. the ones with the largest index values).  In the case of the NaI2.20 index, \citet{Rock+17} concluded that for early-type sources the large values obtained for this index are due to a combination of a bottom-heavy initial mass function and the [Na/Fe] abundances.  On the other hand, \citet{Alton+18} found that their sample of massive ETGs is consistent with having a Milky Way-like IMF, or at most a modestly bottom-heavy IMF, and suggested that their extreme abundance values for Na, in the cores of massive ETGs, may be explained by the metallicity-dependent nucleosynthetic yield of Na. 

The lower panels of Fig.~\ref{index_index} show that the ETGs are in better agreement with the predicted values. However, about half of our SFGs sample show stronger CaI2.26 and CO2.2 values than predicted by the models.  From these plots, we also can infer that the TP-AGB phase does not change substantially the CO index, once the solar metalicity M11 models are in agreement with the E-MILES ones for the younger ages ($t \lesssim$ 1 Gyr), with a large discrepancy for the older ages. Besides age, metallicity appears as an additional discriminator for the measured strengths of CO bands, with low metalicity ([Fe/H] = -0.40) and intermediate ages ($\sim$350Myr) showing the largest values for the CO2.2 index.  This is in agreement with the previous findings of \citet{Kotilainen+12}, who found that the evolved red stars completely dominate the NIR spectra, and that in this age range, the hot, young stars contribution to the EWs is virtually nonexistent. So far, to fully access these younger stellar content of the galaxies it is necessary to fit the full spectrum, taking the continuum into account \citep[see][for example]{Baldwin+17,Dahmer-Hahn+18}. However, this is beyond the scope of the present paper and will be the subject of a  future investigation (Riffel et al., {\it in preparation}). On the other hand, the lower values of the CO index presented by the ETGs are also not explained by the models, with M11 models underestimating and E-MILES models overestimating the main locus occupied by these sources.

\subsubsection{New index - index correlations}

\begin{figure*}
 \includegraphics[scale=0.9]{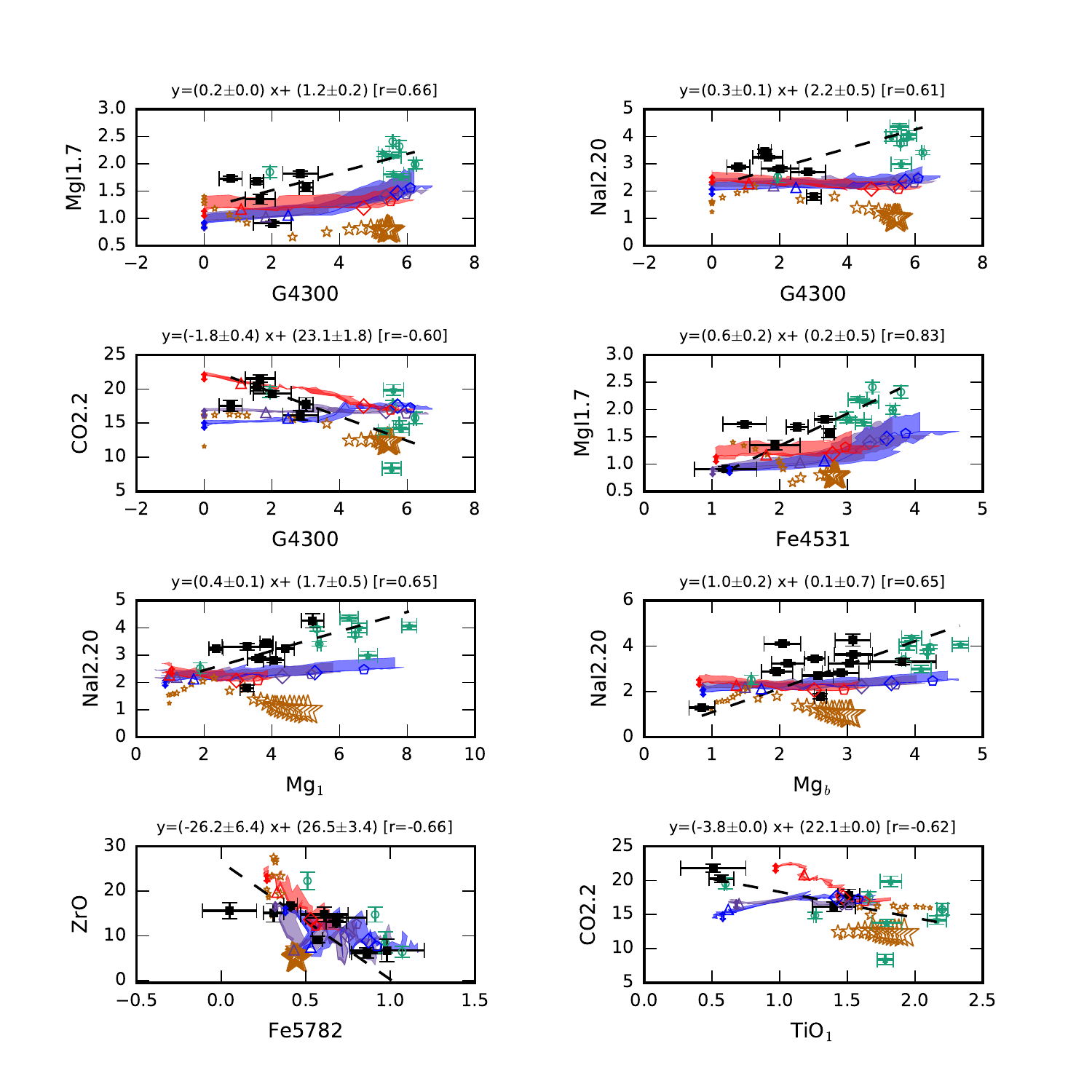}
 \caption{Index-index correlations. Black squares are the data points of the present work. Green diamonds are from \citet{Baldwin+17} and filled green stars are from \citet{Dahmer-Hahn+18}. Open brown stars represent \citet{Maraston+11} solar metalicity, Pickles-based models, with the size of the points scaling with ages (smallest points for 300Myr and largest for 15Gyr).  The shaded areas represent represent IRTF-based {\sc emiles} models  \citep{Vazdekis+16,Rock+15,Rock+16} with red, gray and blue representing [Fe/H]=-0.40, [Fe/H]=0.00 and [Fe/H]=0.22, respectively. The shaded area represents models with a spectral resolution of $\sigma$=60 $\rm km/s$ (the lowest available) to $\sigma$=228 $\rm km/s$. The age range used is between 0.3\,Gyr and 15.0\,Gyr, with arrows, triangles, diamonds and pentagons representing 0.3, 1, 5 and 10~Gyr, respectively. The models with ages smaller than 1\,Gyr should be taken with caution. For more details see text.  }
 \label{ewplt1}
 \end{figure*}

\begin{figure*}
\includegraphics[scale=0.9]{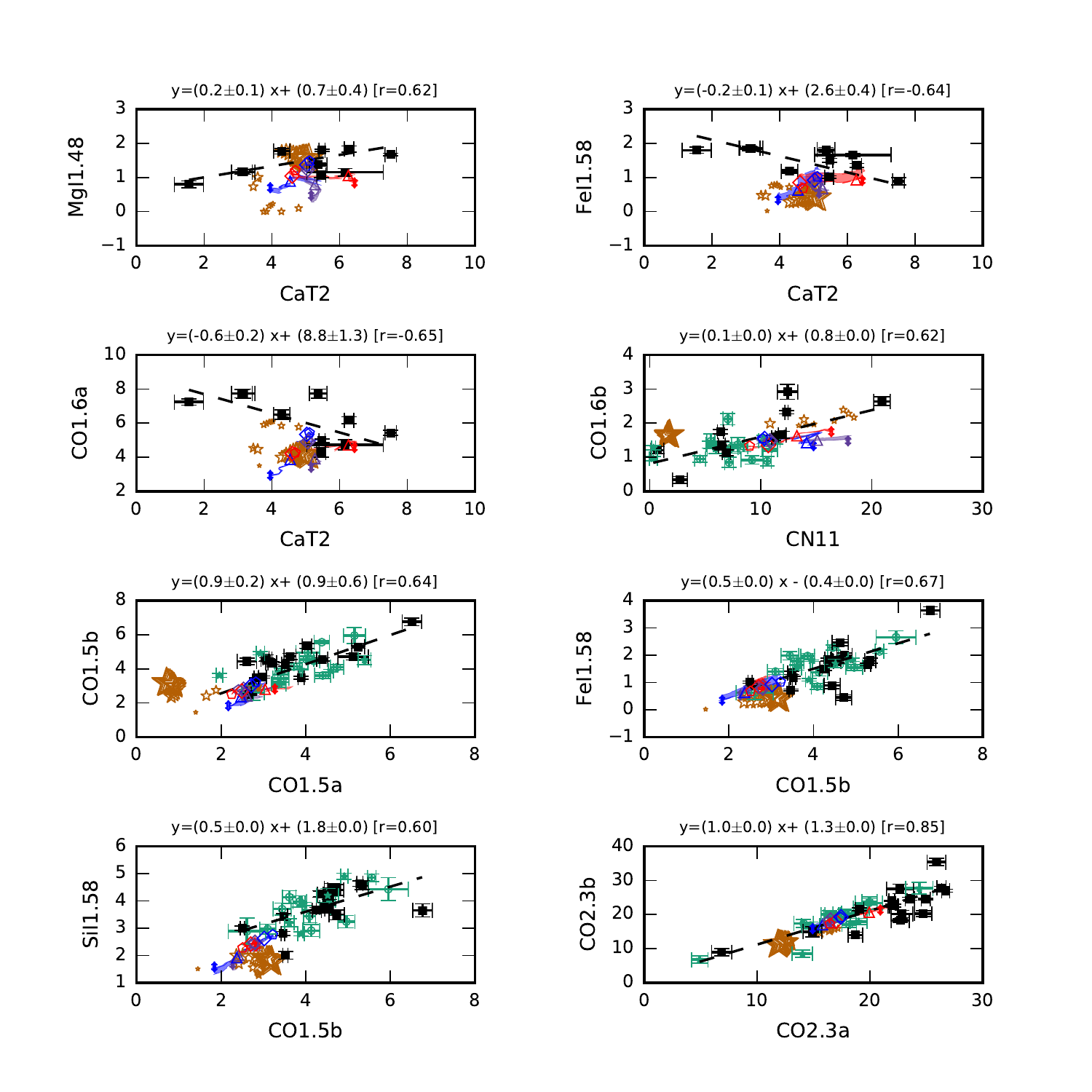}
\caption{Same as Fig.~\ref{ewplt1} but for NIR $x$ NIR indices.}
\label{ewplt2}
\end{figure*}

Because we measured a large set of lines for our sample, we have tried to find new correlations among the different absorption features by plotting all the EWs listed in Tab.~\ref{Ew1} and \ref{Ew2}, as well as literature data (Tabs.~\ref{Ewbald1} to \ref{Ewlgdh}) against each other. From these, we removed the correlations already discussed above (Fig.~\ref{index_index}) as well as the optical $\times$ optical indices correlations since these are well studied \footnote{The NaI2.20 and CO2.2 are well studied, however, we decided to keep them here for diagrams distinct from those presented in Fig.~\ref{index_index} because correlations with other lines may help to shed some light in the understanding of the mechanisms driving these lines.}. Since the CaT lines are correlated \citep[e.g.][]{Cenarro+01}, we only used CaT2 in our search for correlations.  The final set of optical {\it versus} NIR and NIR {\it versus} NIR indices the correlations are shown in Figs.~\ref{ewplt1} and \ref{ewplt2}, together with a linear regression using the orthogonal distance regression (ODR) method that takes errors both in the x and y variables into account \citep{Boggs+90}.  We note that when it was not possible to measure one of the indices used in the correlations, we have removed the galaxy from the plots and regression. In addition, we only considered the cases were both indices were measured at least for 6 sources.  To help to understand these plots we have over-plotted the same model set as discussed above.

What emerges from Fig.~\ref{ewplt1} is that both model sets are able to predict well all the measured values for the optical indices. In the NIR, however, the models fail in their predictions, except for CO2.2 and ZrO, with E-MILES making better predictions of strengths than M11, especially in the case of atomic absorption features. In addition, there is a clear separation of the ETGs and SFGs on the G4300$\times$MgI1.7, G4300$\times$NaI2.20, Fe4531$\times$MgI1.7 and Mg$_1 \times$NaI2.20  diagrams with ETGs in general showing higher values for both optical and NIR indices.  A less evident separation of ETGs and SFGs is observed on the G4300$\times$CO2.2 and Mg$_b\times$NaI2.20 diagrams, while no separation is observed for the  Fe5782$\times$ZrO and  TiO$_1\times$ CO2.2 diagrams. 

The optical indices (G4300, Fe4531 and Mg$_1$) are not very sensitive to the $\alpha$/Fe ratio  while G4300 is mainly sensitive to the C and O abundances \citep{Thomas+03}. This may indicate that the MgI1.7 and NaI2.20 indices are also sensitive to C and/or O abundances.  This is also in agreement with the findings of \citet{Rock+17} who suggested that [C/Fe] enhancement might contribute to the values observed for NaI2.20 in ETGs. However, the good correlation of NaI with Mg$_b$ may also indicate that this index is $\alpha/Fe$ dependent, since Mg$_b$ is sensitive to changes in the $\alpha/Fe$ ratio \citep{Thomas+03}. red The CO2.2 index values are well described by the model predictions for both SFGs and ETGs, with an age-metallicity dependence for the SFGs and no evidence of strong changes on their strengths caused by the amount of TP-AGB stars (see above). This is additionally supported by the CO2.2 $\times$ TiO$_1$ diagram, where M11 models, independent of age, do populate the locus filled by the ETGs, while E-MILES models do not reproduce the larger TiO and smallest CO strengths. The CO and TiO$_1$ correlation is not unexpected since these absorptions depend on O being available. The models do show that ZrO is more metallicity dependent while TiO$_1$ seems to be age dependent. In addition, some ETG show TiO$_1$ values larger than the models (specially E-MILES models), which can be interpreted as an IMF effect \citep[see][]{La_Barbera+13}. In the case of the Mg-dominated indices (in the NIR and optical) the large values for these indices can be associated with the most massive ETGs, and can be explained by an [Mg/Fe] enhancement \citep[e.g.][]{Worthey+92,Martin-Navarro+18}.

\begin{figure*}
 \includegraphics[scale=0.9]{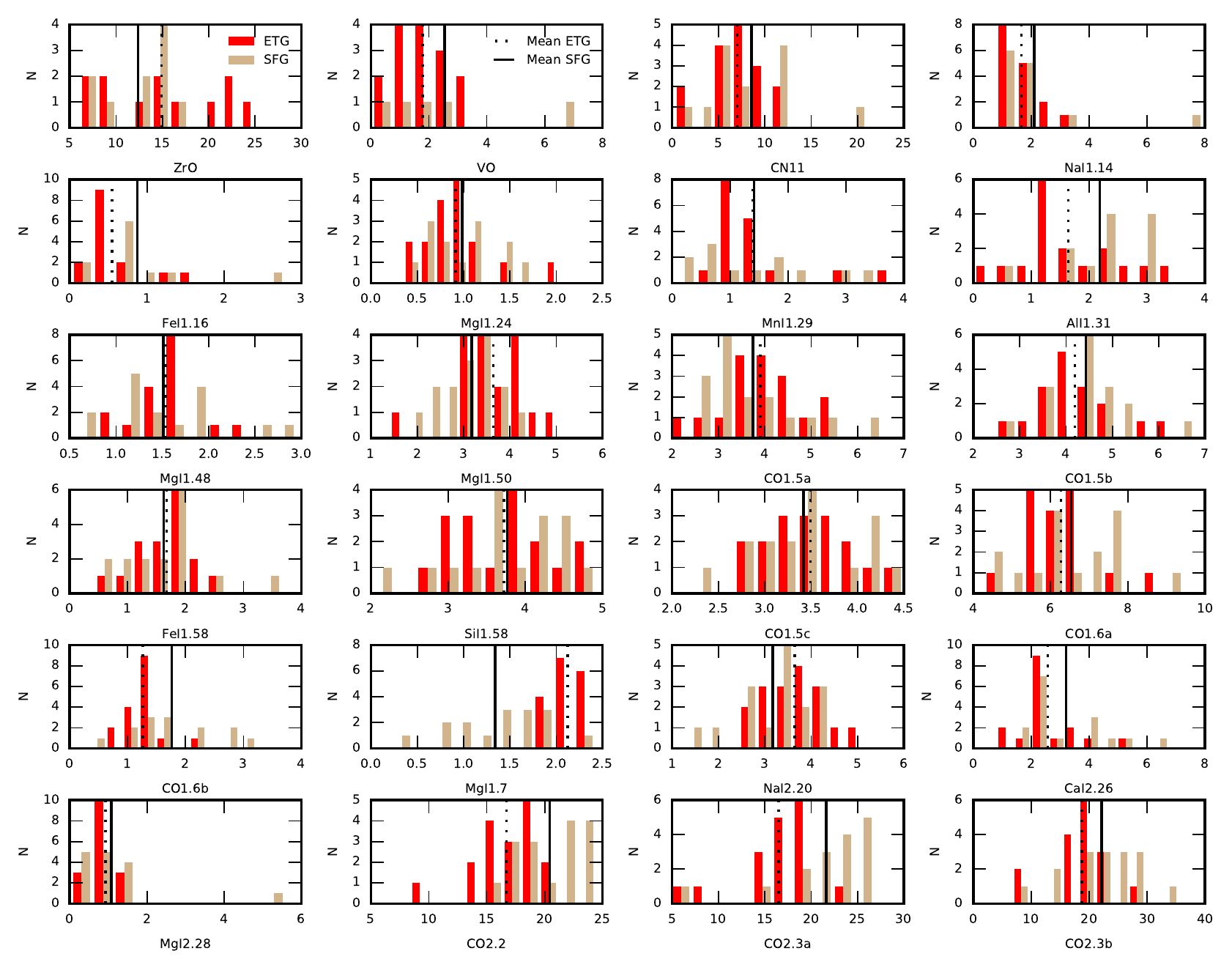}
 \caption{ Comparison of NIR indices strengths between SFG and ETG.}
 \label{EwHist}
 \end{figure*}

The correlations found from this exercise for the NIR indices are shown in Fig.~\ref{ewplt2}. One particularly relevant correlation is CO1.6b$\times$CN11, as the CN11 index is believed to be heavily dominated by the AGB evolutionary phase and particularly by C stars \citep{Maraston05}. Almost 50\% of our SFG do show CN11$\gtrsim$10\AA, with a mean value $\sim$20\% larger than in ETG (see Fig.~\ref{EwHist}) and are consistent with the intermediate age (0.3 - 2\,Gyr) models. M11 models do cover better the space of values of the measurements, but all the older ages M11 models (t$\gtrsim$ 3\,Gyr) do predict more or less constant values for CN11 (the same hapens for CO1.6b).
The ETGs are more or less matched by SSP models with old ages and no indication of an intermediate age population is required to explain the absorption features of these sources, once, their strengths in some cases are smaller than those of the older E-MILES SSPs.  The CO2.2 and CO2.3a,b (also CO1.5a and CO1.5b) indices are to some extent described by the models, with larger values predicted for intermediate age SSPs.The remaining strengths are not predicted by the models and no clear separation is found for SFGs and ETGs.

\begin{figure*}
 \includegraphics[scale=0.9]{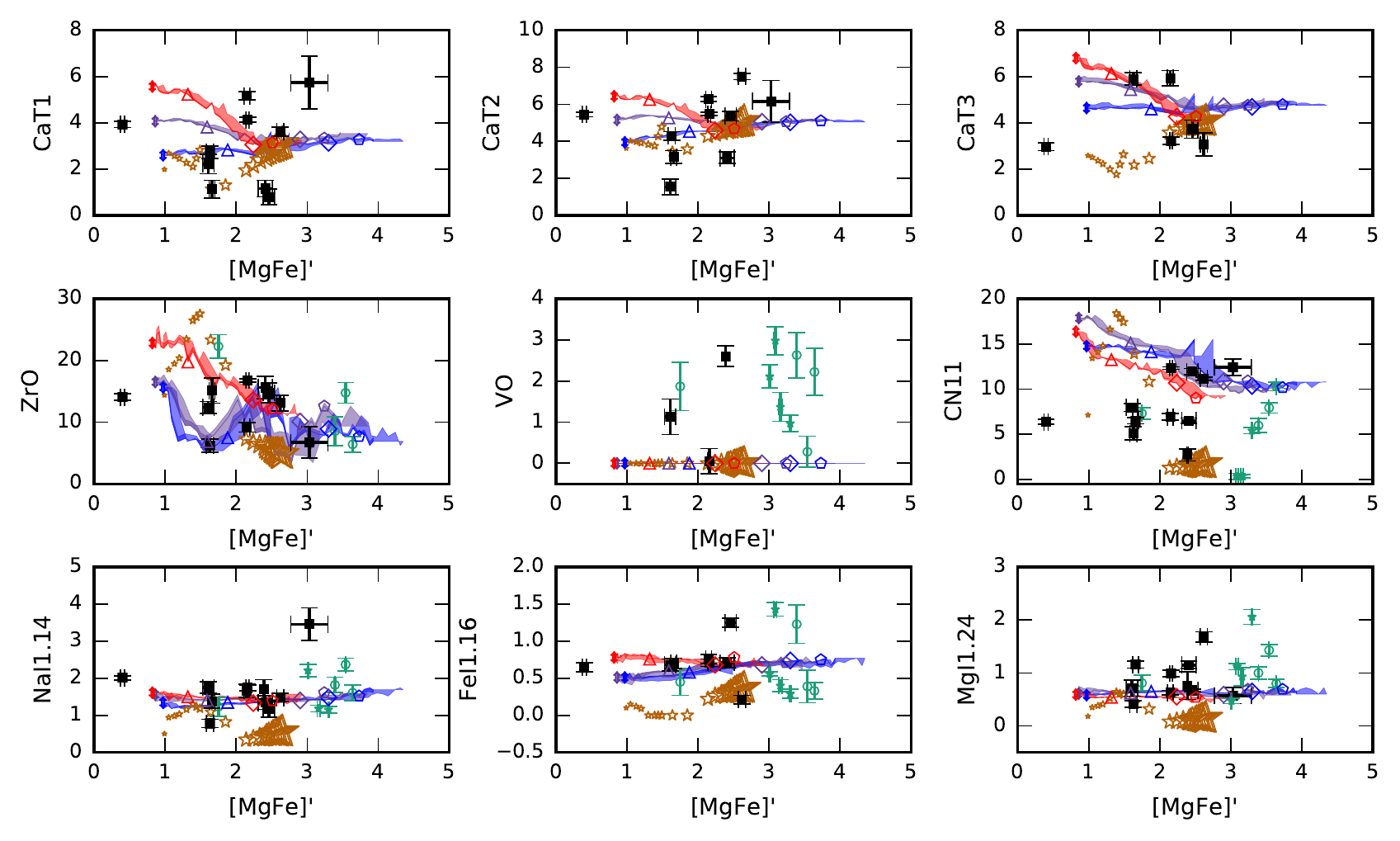}
 \caption{ Comparison of NIR indices with [MgFe]'. The models are the same as Fig.~\ref{ewplt1}.}
 \label{Fe}
 \includegraphics[scale=0.9]{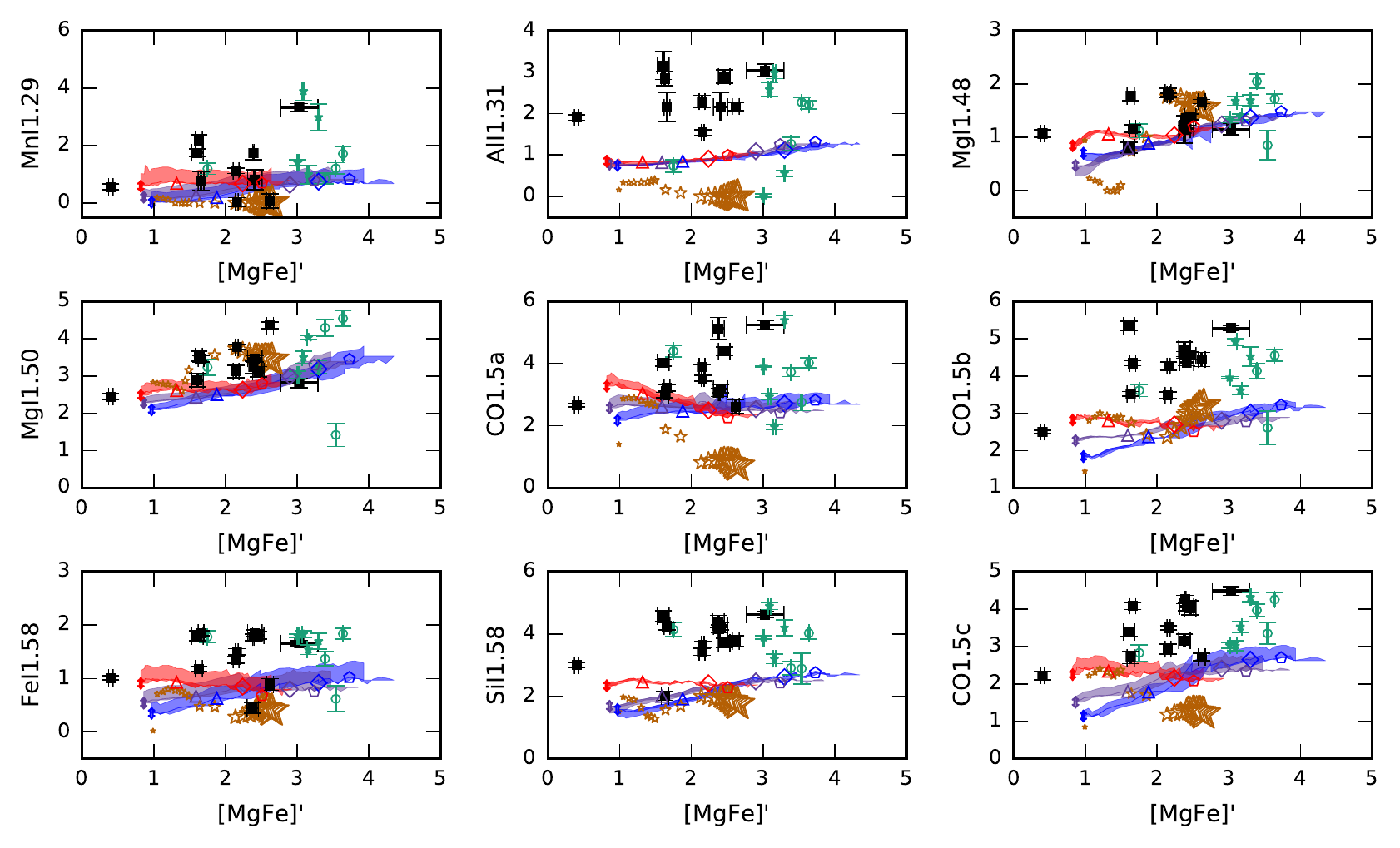}
 \caption{ Comparison of NIR indices with [MgFe]'. The models are the same as Fig.~\ref{ewplt1}.}
 \label{Fe1}
 \end{figure*}

\begin{figure*}
 \includegraphics[scale=0.9]{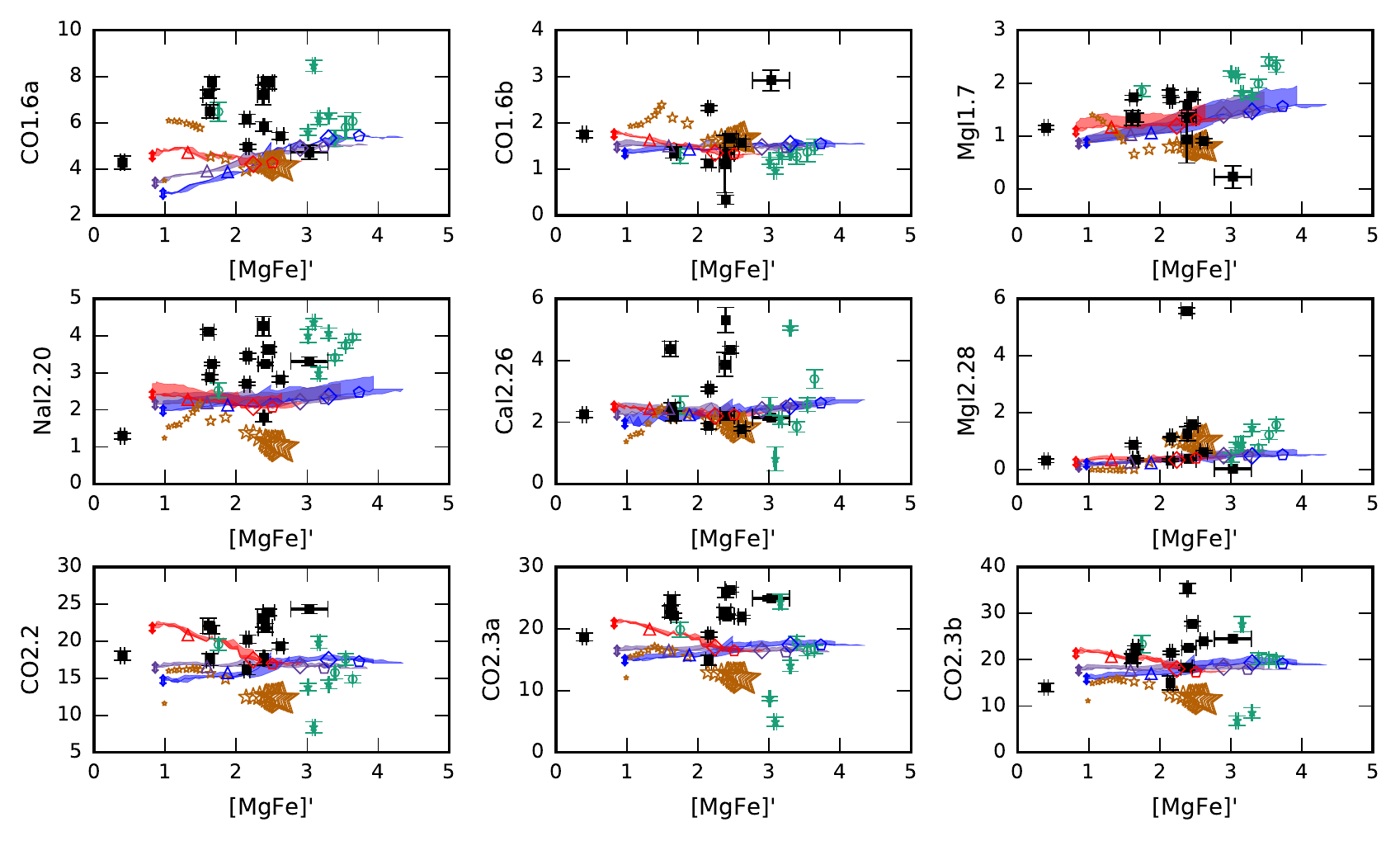}
 \caption{ Comparison of NIR indices with [MgFe]'. The models are the same as Fig.~\ref{ewplt1}.}
 \label{Fe2}
 \end{figure*}

With the aim of understanding the behaviour of the NIR indices, we plotted them against the [MgFe]$'$ index of \citet{Thomas+03} defined as: 
\begin{equation}
    [MgFe]' \equiv \sqrt{Mgb(0.72 \times Fe5270+028 \times Fe5335)}
\end{equation}
which, for this sample with a small range in metallicity, is basically an age-indicator and is completely independent of the $\alpha/Fe$ ratio. Assuming that the ETGs used here are objects with relatively normal old stellar populations, which is a valid assumption since their full spectra can be well fitted with SSP models \citep[see][for details]{Baldwin+17,Dahmer-Hahn+18}. This can also be seen in Figs.~\ref{Fe} to \ref{Fe2}, where the ETGs do in general show less scatter in the [MgFe]$'$ index than the SFGs. This indicates a more complex SFH for the latter, most likely with a strong contribution from intermediate ($\sim$1 Gyr) stellar populations. In order to test the effect of a more complex SFH on the NIR strengths, we show in Fig.~\ref{EwHist} histograms comparing the strenght distributions between SFG and ETG. Except for a few indices (ZrO, MgI1.48,  MgI1.50, CO1.5a, FeI1.58, CO1.5c, MgI1.7, NaI2.20), the mean value for SFG is larger than that for ETG. This more complex SFH can also explain why the CN and CO bands are in general stronger for the SFGs than the ETGs. These bands are enhanced by the short-lived younger red giant branch (RGB) and thermally pulsing asymptotic giant branch (TP-AGB) stars \citep{Maraston05,Riffel-Rogerio+07,Riffel-Rogerio+15}. According to \citet{Maraston98}, these stars can be responsible for up to 70 per cent of the total flux in the NIR. However, for the case of the NaI2.20 index, \citep{Rock+17} constructed models using enhanced contribution from AGB stars and found that these stars have only a very limited effect on the model predictions and do not improve significantly the fit of the model NaI2.20 indices. They also show that small fractions (3 per cent) do have a similar impact on NaI2.20 than those with larger amounts of these stars. This result is consistent with our findings that NaI2.20 index has a mean value $\sim$20\% larger in ETG than in SFG.

In general, the NIR line strengths are not well reproduced by any set of models, suggesting that the SFH of the galaxies cannot be recovered when only using NIR indices. Our results are in agreement with the finding of \citet{Baldwin+17} who have studied the SFH of a sample of ETG by fitting different SSP models and found that the SFH vary dramatically among the different EPS models when fitting NIR data, with higher spectral resolution models producing more consistent results. They also found  variations in ages in the NIR tend to be small, and largely encoded in the shape of the continuum.  This was also noticed in \citet{Riffel-Rogerio+15} who suggested that TP-AGB stars contribute noticeably to a mean stacked NIR spectrum made up with mostly late type galaxies hosting a low luminosity AGN, from the Palomar survey \citep{Mason+15}. This result was obtained by fitting a mix of individual IRTF stars to the mean galaxy spectrum. Nevertheless, in this same work we have shown that other evolved stars (red giants, C-R and E-AGB stars) can reproduce most of the absorption features detected, without having to resort to stars in the TP-AGB phase.

\begin{table*}
\renewcommand{\tabcolsep}{0.70mm}
\centering
\caption{Line limits and continuum bandpasses. \label{ewdefs}}
\begin{scriptsize}
\begin{tabular}{lllllll}
\hline\hline
\noalign{\smallskip}
Centre & Main Absorber        & Index Name    & Line Limits   & Blue continuum  & Red Continuum  & Reference \\
   (\AA)     &       & & (\AA)         &  (\AA) & (\AA) \\ 
\noalign{\smallskip}
\hline \noalign{\smallskip}
4228.5     &	Ca {\sc i}	   &   Ca4227   & 4222.250 -- 4234.750	   &  4211.000 -- 4219.750    &  4241.000 -- 4251.000  &   \citet{Worthey+94} \\ 
4298.875   &	CH (G-Band)	   &   G4300    & 4281.375 -- 4316.375	   &  4266.375 -- 4282.625    &  4318.875 -- 4335.125  &   \citet{Worthey+94}  \\ 
4394.75    &	Fe {\sc i}	   &   Fe4383   & 4369.125 -- 4420.375	   &  4359.125 -- 4370.375    &  4442.875 -- 4455.375  &   \citet{Worthey+94}  \\ 
4463.375   &	Ca {\sc i}	   &   Ca4455   & 4452.125 -- 4474.625	   &  4445.875 -- 4454.625    &  4477.125 -- 4492.125  &   \citet{Worthey+94}  \\ 
4536.75    &	Fe {\sc i}	   &   Fe4531   & 4514.250 -- 4559.250	   &  4504.250 -- 4514.250    &  4560.500 -- 4579.250   &   \citet{Worthey+94} \\ 
4677.125   &	C$_2$		   &   Fe4668   & 4634.000 -- 4720.250	   &  4611.500 -- 4630.250    &  4742.750 -- 4756.500   &   \citet{Worthey+94} \\ 
5015.875   &	Fe {\sc i}	   &   Fe5015   & 4977.750 -- 5054.000	   &  4946.500 -- 4977.750    &  5054.000 -- 5065.250   &   \citet{Worthey+94} \\ 
5101.625   &	MgH		   &   Mg1      & 5069.125 -- 5134.125	   &  4895.125 -- 4957.625    &  5301.125 -- 5366.125   &   \citet{Worthey+94} \\ 
5175.375   &	MgH		   &   Mg2      & 5154.125 -- 5196.625	   &  4895.125 -- 4957.625    &  5301.125 -- 5366.125    &   \citet{Worthey+94}\\ 
5176.375   &	Mg b		   &   Mgb      & 5160.125 -- 5192.625	   &  5142.625 -- 5161.375    &  5191.375 -- 5206.375  &   \citet{Worthey+94}  \\ 
5265.65    &	Fe {\sc i}	   &   Fe5270   & 5245.650 -- 5285.650	   &  5233.150 -- 5248.150    &  5285.650 -- 5318.150  &   \citet{Worthey+94}  \\ 
5332.125   &	Fe {\sc i}	   &   Fe5335   & 5312.125 -- 5352.125	   &  5304.625 -- 5315.875    &  5353.375 -- 5363.375  &   \citet{Worthey+94}  \\ 
5401.25    &	Fe {\sc i}	   &   Fe5406   & 5387.500 -- 5415.000	   &  5376.250 -- 5387.500    &  5415.000 -- 5425.000  &   \citet{Worthey+94}  \\ 
5708.5     &	Fe {\sc i}	   &   Fe5709   & 5696.625 -- 5720.375	   &  5672.875 -- 5696.625    &  5722.875 -- 5736.625  &   \citet{Worthey+94}  \\ 
5786.625   &	Fe {\sc i}	   &   Fe5782   & 5776.625 -- 5796.625	   &  5765.375 -- 5775.375    &  5797.875 -- 5811.625 &   \citet{Worthey+94}   \\ 
5893.125   &	Na {\sc i}	   &   NaD      & 5876.875 -- 5909.375	   &  5860.625 -- 5875.625    &  5922.125 -- 5948.125  &   \citet{Worthey+94}  \\ 
5965.375   &	TiO		   &   TiO1     & 5936.625 -- 5994.125	   &  5816.625 -- 5849.125    &  6038.625 -- 6103.625  &   \citet{Worthey+94}  \\ 
6230.875   &	TiO		   &   TiO2     & 6189.625 -- 6272.125	   &  6066.625 -- 6141.625    &  6372.625 -- 6415.125  &   \citet{Worthey+94}  \\ 
8498.0     &	Ca {\sc ii}	   &   CaT1     & 8476.000 -- 8520.000	   &  8110.000 -- 8165.000    &  8786.000 -- 8844.000 & \citet{Bica+87} ($\dagger$)  \\ 
8542.0     &	Ca {\sc ii}	   &   CaT2     & 8520.000 -- 8564.000	   &  8110.000 -- 8165.000    &  8786.000 -- 8844.000  & \citet{Bica+87} ($\dagger$)  \\ 
8670.0     &	Ca {\sc ii}	   &   CaT3     & 8640.000 -- 8700.000	   &  8110.000 -- 8165.000    &  8786.000 -- 8844.000  & \citet{Bica+87} ($\dagger$)  \\ 
9320.0     &	ZrO/TiO/CN     &   ZrO      & 9170.000 -- 9470.000	   &  8900.000 -- 8960.000    &  9585.000 -- 9615.000   & New Definition ($\alpha$) \\ 
10560.0    &	VO		   &   VO       & 10470.000 -- 10650.000   &  10430.000 -- 10465.000  &  10660.000 -- 10700.000 & New Definition ($\alpha$) \\  
11000.0    &	CN		   &   CN11     & 10910.000 -- 11090.000   &  10705.000 -- 10730.000  &  11310.000 -- 11345.000 & New Definition ($\beta$) \\  
11390.0    &   Na {\sc i}          &   NaI1.14  & 11350.000 -- 11430.000   &  11310.000 -- 11345.000  &  11450.000 -- 11515.000 & New Definition ($\beta$)  \\ 
11605.0    &   Fe {\sc i}          &   FeI1.16  & 11580.000 -- 11630.000   &  11450.000 -- 11515.000  &  11650.000 -- 11690.000 &\citet{Roeck+15} \\ 
12430.0    &   Mg {\sc i}          &   MgI1.24  & 12405.000 -- 12455.000   &  12335.000 -- 12365.000  &  12465.000 -- 12490.000 & \citet{Roeck+15} \\ 
12944.0    &   Mn {\sc i}          &   MnI1.29  & 12893.000 -- 12995.000   &  12858.000 -- 12878.000  &  13026.000 -- 13068.000 & New Definition \\ 
13132.5    &   Al {\sc i}          &  AlI1.31  &  13095.000 -- 13170.000  &   13000.000 -- 13070.000 &   13175.000 -- 13215.000 & \citet{Roeck+15}\\  
14875.0    &   Mg {\sc i}          &   MgI1.48  & 14850.000 -- 14900.000   &  14750.000 -- 14800.000  &  14910.000 -- 14950.000  & New Definition \\ 
15032.5    &   Mg {\sc i}          &   MgI1.50  & 14995.000 -- 15070.000   &  14910.000 -- 14950.000  &  15150.000 -- 15200.000  & New Definition\\ 
15587.5    &   CO+Mg {\sc i}       &   CO1.5a   & 15555.000 -- 15620.000   &  15470.000 -- 15500.000  &  15700.000 -- 15730.000  & New Definition ($\epsilon$)\\ 
15780.0    &   CO+Mg {\sc i}       &   CO1.5b   & 15750.000 -- 15810.000   &  15700.000 -- 15730.000  &  16095.000 -- 16145.000  & New Definition ($\epsilon$)\\ 
15830.0    &   Fe {\sc i}          &   FeI1.58  & 15810.000 -- 15850.000   &  15700.000 -- 15730.000  &  16090.000 -- 16140.000  & New Definition\\ 
15890.0    &Si {\sc i} + Mg {\sc i}&   SiI1.58  & 15850.000 -- 15930.000   &  15700.000 -- 15730.000  &  16090.000 -- 16140.000  & New Definition\\ 
15985.0    &   CO+Si {\sc i}       &   CO1.5c   & 15950.000 -- 16020.000   &  15700.000 -- 15730.000  &  16090.000 -- 16140.000  & New Definition ($\epsilon$)\\ 
16215.0    &CO+Si {\sc i}+Ca {\sc i}&   CO1.6a   &16145.000 -- 16285.000    & 16090.000 -- 16140.000   & 16290.000 -- 16340.000  & New Definition ($\epsilon$) \\
17064.0    &   CO + Fe {\sc i}      &   CO1.6b   &17035.000 -- 17093.000    & 16970.000 -- 17025.000   & 17140.000 -- 17200.000  & New Definition ($\epsilon$) \\
17111.5    &   Mg {\sc i}          &   MgI1.7   & 17093.000 -- 17130.000   &  16970.000 -- 17025.000  &  17140.000 -- 17200.000  & \citet{Roeck+15}\\ 
22073.5    &   Na {\sc i}          &   NaI2.20  & 22040.000 -- 22107.000   &  21910.000 -- 21966.000  &  22125.000 -- 22160.000  & \citet{Frogel+01}\\ 
22634.5    &   Ca {\sc i}          &   CaI2.26  & 22577.000 -- 22692.000   &  22530.000 -- 22560.000  &  22700.000 -- 22720.000 &  \citet{Frogel+01} ($\gamma$) \\ 
22820.0    &   Mg {\sc i}          &   MgI2.28  & 22795.000 -- 22845.000   &  22700.000 -- 22720.000  &  22850.000 -- 22865.000 & New Definition ($\delta$)   \\ 
23015.0    &   CO                  &   CO2.2    & 22870.000 -- 23160.000   &  22700.000 -- 22790.000  &  23655.000 -- 23680.000 & New Definition ($\epsilon$)\\ 
23290.0    &   CO                  &   CO2.3a   & 23160.000 -- 23420.000   &  22700.000 -- 22790.000  &  23655.000 -- 23680.000 & New Definition ($\epsilon$)\\ 
23535.0    &   CO                  &   CO2.3b   & 23420.000 -- 23650.000   &  22700.000 -- 22790.000  &  23655.000 -- 23680.000 & New Definition ($\epsilon$)\\ 
\noalign{\smallskip}
\hline
\end{tabular}
\end{scriptsize}
\begin{list}{Table Notes:}
\item The optical indices are those of the LICK observatory \citep[][and references]{Worthey+94}. The CaT indices are those of  \citet{Bica+87} with a change in the blue continuum band passes in order to fit in our spectral region; $\alpha$ Based on \citet{Riffel-Rogerio+15} with small changes on the line limits; $\beta$ New continuum limits with central bandpasses from \citet{Roeck+15}; $\epsilon$ adapted from \citet{Riffel-Rogerio+07} with fixed continuum band passes, with better identifications of the main absorbers as well as better constraints of the line limits;  $\gamma$ We made a small change on the blue continuum band pass to remove possible \h2\ emission lines.;  $\delta$ Adapted from \citet{Silva+08} in order to better accomodate the continuum regions for the CO lines. 
\end{list}
\end{table*}

\begin{table*}
\renewcommand{\tabcolsep}{0.70mm}
\centering
\caption{ Absorption feature Equivalent Widths (in \AA) .\label{Ew1}} 
\begin{tabular}{llccccccccccccccccccc}
\hline\hline
\noalign{\smallskip}
Line           & NGC23            & NGC520          & NGC660          & NGC1055          & NGC1134        & NGC1204          & NGC1222        & NGC1266  \\
\noalign{\smallskip}
\hline \noalign{\smallskip}
Ca4227         &   0.36$\pm$0.06  &   --	    &  0.53$\pm$0.61  &   --		&  1.14$\pm$0.16  &   --	    &	--	      &   --	           \\
G4300          &   1.56$\pm$0.19  &   --	    &  5.18$\pm$1.62  &   --		&  2.01$\pm$0.56  &   --	    &	--	      &   --	           \\
Fe4383         &   1.67$\pm$0.22  &   --	    &  7.23$\pm$0.68  &   --		&  3.84$\pm$1.05  &   --	    &	--	      &  7.53$\pm$2.22     \\
Ca4455         &   0.49$\pm$0.1   &  3.22$\pm$0.47  &  2.01$\pm$0.3   &   --		&  0.28$\pm$0.22  &   --	    &	--	      &   --	           \\
Fe4531         &   2.26$\pm$0.16  &   --	    &	--	      &   --		&  1.2$\pm$0.46   &   --	    &	--	      &   --	           \\
$C_2$4668      &   3.88$\pm$0.2   &   --	    &	--	      &   --		&  4.94$\pm$0.55  &   --	    &	--	      &   --	           \\
Fe5015         &    --  	  &   --	    &	--	      &   --		&   --  	  &   --	    &	--	      &   --	           \\
Mg$_1$         &   3.84$\pm$0.19  &   --	    &  5.26$\pm$0.3   &  5.2$\pm$0.33	&  4.06$\pm$0.31  &  4.39$\pm$0.27  &	--	      &  3.27$\pm$0.74     \\
Mg$_2$         &   4.98$\pm$0.13  &   --	    &  6.46$\pm$0.21  &  6.73$\pm$0.2	&  5.36$\pm$0.18  &  5.24$\pm$0.18  &	--	      &  4.14$\pm$0.37     \\
Mg$_b$         &   2.52$\pm$0.15  &   --	    &  2.71$\pm$0.25  &  3.08$\pm$0.26  &  2.9$\pm$0.27   &  3.03$\pm$0.31  &  0.85$\pm$0.19  &  3.81$\pm$0.5      \\
Fe5270         &   1.91$\pm$0.12  &   --	    &  2.53$\pm$0.35  &  1.92$\pm$0.32  &  2.44$\pm$0.2   &  2.09$\pm$0.3   &	--	      &  2.43$\pm$0.45     \\
Fe5335         &   1.73$\pm$0.12  &   --	    &  2.02$\pm$0.24  &  1.64$\pm$0.34  &  2.17$\pm$0.2   &  1.48$\pm$0.33  &  0.66$\pm$0.28  &  2.38$\pm$0.67     \\
Fe5406         &   0.95$\pm$0.04  &   --	    &  1.0$\pm$0.15   &  0.57$\pm$0.27  &  1.25$\pm$0.14  &  0.68$\pm$0.3   &  0.09$\pm$0.1   &  2.33$\pm$0.32     \\
Fe5709         &   0.64$\pm$0.04  &   --	    &  0.77$\pm$0.1   &  0.72$\pm$0.13  &  0.79$\pm$0.07  &  0.75$\pm$0.14  &  0.45$\pm$0.04  &  0.74$\pm$0.4      \\
Fe5782         &   0.41$\pm$0.04  &   --	    &  1.11$\pm$0.08  &   --		&  0.68$\pm$0.06  &  0.05$\pm$0.16  &  0.68$\pm$0.18  &  0.98$\pm$0.22     \\
NaD            &   4.25$\pm$0.08  &  1.91$\pm$0.25  &  5.07$\pm$0.19  &  4.02$\pm$0.27  &  4.71$\pm$0.12  &  3.56$\pm$0.21  &	--	      &  6.37$\pm$0.23     \\
TiO$_1$        &   0.57$\pm$0.09  &   --	    &	--	      &   --		&   --  	  &  0.51$\pm$0.24  &	--	      &   --	           \\
TiO$_2$        &   3.82$\pm$0.11  &   --	    &  5.79$\pm$0.2   &  7.96$\pm$0.28  &  6.13$\pm$0.2   &  6.07$\pm$0.27  &  3.87$\pm$0.19  &  8.34$\pm$0.51     \\
CaT1           &   4.13$\pm$0.11  &   --	    &	--	      &   --		&  3.62$\pm$0.20  &  1.16$\pm$0.36  &  3.93$\pm$0.13  &  5.74$\pm$1.14     \\
CaT2           &   5.48$\pm$0.09  &   --	    &	--	      &   --		&  7.51$\pm$0.17  &  3.11$\pm$0.31  &  5.46$\pm$0.13  &  6.16$\pm$1.13     \\
CaT3           &   3.22$\pm$0.16  &   --	    &	--	      &   --		&  3.07$\pm$0.50  &   --	    &  2.97$\pm$0.17  &   --	           \\
ZrO            &   16.76$\pm$0.27 &   --	    &	--	      &   --		&  13.09$\pm$1.29 &  15.60$\pm$1.78 &  14.06$\pm$0.56 &  6.70$\pm$2.53     \\
VO             &   0.05$\pm$0.31  &   --	    &  1.41$\pm$0.63  &   --		&   --  	  &   --	    &	--	      &   --	           \\
CN11           &   12.32$\pm$0.14 &   --	    &  3.76$\pm$0.28  &   --		&  11.15$\pm$0.17 &  6.48$\pm$0.31  &  6.37$\pm$0.27  &  12.41$\pm$0.92    \\
NaI1.14        &   1.74$\pm$0.08  &   --	    &  2.43$\pm$0.18  &   --		&  1.48$\pm$0.11  &  1.34$\pm$0.18  &  2.01$\pm$0.06  &  3.46$\pm$0.44     \\
FeI1.16        &   0.71$\pm$0.05  &   --	    &  0.44$\pm$0.07  &   --		&  0.21$\pm$0.06  &  0.71$\pm$0.07  &  0.65$\pm$0.06  &   --	           \\
MgI1.24        &   0.99$\pm$0.06  &   --	    &  0.74$\pm$0.05  &   --		&  1.68$\pm$0.10  &  1.15$\pm$0.07  &	--	      &  0.57$\pm$0.15     \\
MnI1.29        &   0.03$\pm$0.15  &   --	    &  0.28$\pm$0.14  &   --		&  0.06$\pm$0.25  &  0.80$\pm$0.35  &  0.55$\pm$0.10  &  3.32$\pm$0.14     \\
AlI1.31        &   1.54$\pm$0.07  &   --	    &  1.93$\pm$0.57  &   --		&  2.17$\pm$0.10  &  2.16$\pm$0.34  &  1.90$\pm$0.07  &  3.04$\pm$0.15     \\
MgI1.48        &   1.80$\pm$0.03  &  2.94$\pm$0.17  &  1.96$\pm$0.03  &  1.14$\pm$0.25  &  1.67$\pm$0.04  &  1.16$\pm$0.06  &  1.07$\pm$0.07  &  1.15$\pm$0.11     \\
MgI1.50        &   3.77$\pm$0.07  &  3.28$\pm$0.30  &  2.43$\pm$0.10  &   --		&  4.35$\pm$0.09  &  3.46$\pm$0.08  &  2.44$\pm$0.08  &  2.81$\pm$0.13     \\
CO1.5a         &   3.52$\pm$0.09  &  6.51$\pm$0.23  &  4.33$\pm$0.10  &  5.12$\pm$0.36  &  2.61$\pm$0.23  &  3.18$\pm$0.13  &  2.66$\pm$0.05  &  5.24$\pm$0.15     \\
CO1.5b         &   4.26$\pm$0.11  &  6.76$\pm$0.23  &  4.94$\pm$0.10  &  4.71$\pm$0.19  &  4.44$\pm$0.19  &  4.35$\pm$0.08  &  2.50$\pm$0.05  &  5.28$\pm$0.07     \\
FeI1.58        &   1.50$\pm$0.06  &  3.64$\pm$0.13  &  2.11$\pm$0.06  &  0.45$\pm$0.10  &  0.89$\pm$0.11  &  1.85$\pm$0.06  &  1.01$\pm$0.04  &  1.66$\pm$0.05     \\
SiI1.58        &   3.66$\pm$0.10  &  3.65$\pm$0.24  &  4.03$\pm$0.13  &  4.40$\pm$0.20  &  3.77$\pm$0.18  &  4.25$\pm$0.11  &  3.00$\pm$0.08  &  4.63$\pm$0.10     \\
CO1.5c         &   3.50$\pm$0.06  &  3.39$\pm$0.20  &  3.83$\pm$0.11  &  3.16$\pm$0.17  &  2.72$\pm$0.10  &  4.07$\pm$0.11  &  2.22$\pm$0.12  &  4.49$\pm$0.12     \\
CO1.6a         &   4.97$\pm$0.11  &  7.60$\pm$0.39  &  6.70$\pm$0.15  &  7.21$\pm$0.44  &  5.43$\pm$0.16  &  7.73$\pm$0.25  &  4.29$\pm$0.29  &  4.73$\pm$0.30     \\
CO1.6b         &   2.32$\pm$0.05  &  3.21$\pm$0.19  &  0.79$\pm$0.07  &  1.11$\pm$0.62  &  1.57$\pm$0.08  &  1.34$\pm$0.13  &  1.75$\pm$0.07  &  2.92$\pm$0.22     \\
MgI1.7         &   1.68$\pm$0.05  &  0.76$\pm$0.21  &  1.37$\pm$0.04  &  0.94$\pm$0.45  &  0.91$\pm$0.04  &  1.35$\pm$0.08  &  1.16$\pm$0.04  &  0.23$\pm$0.21     \\
NaI2.20        &   3.45$\pm$0.08  &  2.51$\pm$0.10  &  2.88$\pm$0.07  &  4.26$\pm$0.26  &  2.82$\pm$0.08  &  3.24$\pm$0.04  &  1.29$\pm$0.08  &  3.31$\pm$0.12     \\
CaI2.26        &   3.07$\pm$0.07  &  4.07$\pm$0.12  &  2.40$\pm$0.18  &  3.87$\pm$0.39  &  1.78$\pm$0.07  &  2.21$\pm$0.14  &  2.25$\pm$0.10  &  2.15$\pm$0.08     \\
MgI2.28        &   1.14$\pm$0.02  &  1.20$\pm$0.09  &  0.60$\pm$0.04  &  5.57$\pm$0.10  &  0.62$\pm$0.04  &  0.37$\pm$0.04  &  0.31$\pm$0.06  &  0.03$\pm$0.01     \\
CO2.2          &   20.26$\pm$0.56 &  23.16$\pm$0.36 &  12.23$\pm$0.39 &  23.11$\pm$1.22 &  19.33$\pm$0.41 &  21.80$\pm$0.58 &  18.03$\pm$0.61 &  24.34$\pm$0.61    \\
CO2.3a         &   19.07$\pm$0.38 &  26.71$\pm$0.35 &  12.45$\pm$0.66 &  22.69$\pm$0.80 &  21.93$\pm$0.26 &  22.02$\pm$0.43 &  18.71$\pm$0.64 &  24.94$\pm$0.40    \\
CO2.3b         &   21.53$\pm$0.37 &  27.00$\pm$0.42 &  12.50$\pm$0.78 &  18.28$\pm$0.56 &  24.03$\pm$0.13 &  22.57$\pm$0.44 &  13.99$\pm$0.85 &  24.52$\pm$0.26    \\
\noalign{\smallskip}
\hline
\end{tabular}
\end{table*}

\begin{table*}
\renewcommand{\tabcolsep}{0.70mm}
\centering
\caption{ Absorption feature Equivalent Widths (in \AA). \label{Ew2}} 
\begin{tabular}{llccccccccccccccccccc}
\hline\hline
\noalign{\smallskip}
Line &           UGC2982          & NGC1797         & NGC6814         & NGC6835         & UGC12150        & NGC7465         & NGC7591         & NGC7678 \\
\noalign{\smallskip}
\hline \noalign{\smallskip}
Ca4227         &    --  	  &  0.91$\pm$0.12  &	--	      &   --		&   --  	  &  0.47$\pm$0.1   &	--	      &   --	          \\
G4300          &    --  	  &  1.65$\pm$0.44  &	--	      &   --		&   --  	  &  2.84$\pm$0.52  &	--	      &  0.78$\pm$0.34    \\
Fe4383         &    --  	  &   --	    &	--	      &   --		&   --  	  &  2.77$\pm$0.27  &	--	      &  2.24$\pm$0.4     \\
Ca4455         &    --  	  &   --	    &	--	      &   --		&   --  	  &  0.53$\pm$0.16  &	--	      &  1.11$\pm$0.12    \\
Fe4531         &    --  	  &  1.93$\pm$0.37  &	--	      &   --		&   --  	  &  2.67$\pm$0.16  &	--	      &  1.48$\pm$0.32    \\
$C_2$4668      &    --  	  &   --	    &	--	      &   --		&   --  	  &   --	    &	--	      &  1.1$\pm$0.42     \\
Fe5015         &    --  	  &   --	    &	--	      &   --		&   --  	  &   --	    &	--	      &   --	          \\
Mg$_1$         &    --  	  &  2.35$\pm$0.21  &	--	      &   --		&   --  	  &   --	    &	--	      &  3.61$\pm$0.33    \\
Mg$_2$         &    --  	  &  3.77$\pm$0.13  &	--	      &   --		&  4.28$\pm$0.26  &  4.07$\pm$0.24  &  5.14$\pm$0.36  &  4.23$\pm$0.16    \\
Mg$_b$         &    --  	  &  2.12$\pm$0.24  &	--	      &   --		&  2.04$\pm$0.27  &  2.56$\pm$0.23  &  3.09$\pm$0.28  &  1.96$\pm$0.23    \\
Fe5270         &    --  	  &  1.32$\pm$0.19  &	--	      &   --		&  1.34$\pm$0.28  &  1.78$\pm$0.17  &  1.82$\pm$0.29  &  1.31$\pm$0.16    \\
Fe5335         &    --  	  &  1.22$\pm$0.13  &	--	      &   --		&  1.11$\pm$0.37  &  1.85$\pm$0.15  &  2.32$\pm$0.3   &  1.5$\pm$0.29     \\
Fe5406         &    --  	  &  0.87$\pm$0.11  &	--	      &   --		&  0.44$\pm$0.2   &  0.78$\pm$0.15  &  1.31$\pm$0.21  &  1.04$\pm$0.12    \\
Fe5709         &    --  	  &  0.35$\pm$0.06  &	--	      &   --		&  0.64$\pm$0.1   &  0.7$\pm$0.06   &  1.14$\pm$0.17  &  0.78$\pm$0.11    \\
Fe5782         &    --  	  &  0.31$\pm$0.06  &	--	      &   --		&   --  	  &  0.57$\pm$0.03  &  0.61$\pm$0.14  &  0.86$\pm$0.09    \\
NaD            &   4.78$\pm$0.37  &  6.27$\pm$0.19  &	--	      &   --		&  7.47$\pm$0.52  &  1.67$\pm$0.27  &  3.86$\pm$0.19  &  4.17$\pm$0.25    \\
TiO$_1$        &    --  	  &   --	    &	--	      &   --		&   --  	  &  1.4$\pm$0.16   &	--	      &   --	          \\
TiO$_2$        &    --  	  &  2.08$\pm$0.23  &	--	      &   --		&  4.88$\pm$0.52  &  4.63$\pm$0.15  &  3.4$\pm$0.25   &  4.6$\pm$0.21     \\
CaT1           &    --  	  &  1.13$\pm$0.39  &	--	      &   --		&  2.22$\pm$0.42  &  5.17$\pm$0.17  &  0.80$\pm$0.33  &  2.73$\pm$0.26    \\
CaT2           &    --  	  &  3.16$\pm$0.34  &  0.36$\pm$0.19  &   --		&  1.55$\pm$0.43  &  6.28$\pm$0.14  &  5.37$\pm$0.26  &  4.29$\pm$0.24    \\
CaT3           &    --  	  &   --	    &  3.35$\pm$0.06  &   --		&   --  	  &  5.93$\pm$0.33  &  3.73$\pm$0.39  &  5.90$\pm$0.26    \\
ZrO            &    --  	  &  15.11$\pm$2.04 &  1.20$\pm$1.74  &   --		&  12.38$\pm$0.98 &  9.16$\pm$0.75  &  14.76$\pm$1.61 &  6.18$\pm$1.10    \\
VO             &   7.10$\pm$1.35  &   --	    &  3.89$\pm$0.43  &   --		&  1.13$\pm$0.43  &   --	    &	--	      &   --	          \\
CN11           &   20.92$\pm$0.74 &  6.49$\pm$0.33  &	--	      &   --		&  7.95$\pm$0.42  &  6.91$\pm$0.37  &  11.93$\pm$0.34 &  5.08$\pm$0.72    \\
NaI1.14        &    --  	  &  1.38$\pm$0.19  &  3.99$\pm$0.13  &  7.92$\pm$0.83  &  1.75$\pm$0.15  &  1.66$\pm$0.18  &  1.17$\pm$0.22  &  0.78$\pm$0.11    \\
FeI1.16        &   0.17$\pm$0.20  &  0.70$\pm$0.07  &  0.53$\pm$0.06  &  2.78$\pm$0.11  &  0.68$\pm$0.08  &  0.76$\pm$0.06  &  1.25$\pm$0.06  &   --	          \\
MgI1.24        &   1.40$\pm$0.06  &  1.16$\pm$0.07  &  0.76$\pm$0.05  &  1.49$\pm$0.06  &  0.71$\pm$0.15  &  0.63$\pm$0.04  &  0.67$\pm$0.02  &  0.41$\pm$0.06    \\
MnI1.29        &   3.11$\pm$0.62  &  0.77$\pm$0.33  &  10.00$\pm$0.43 &  1.53$\pm$0.11  &  1.74$\pm$0.14  &  1.12$\pm$0.10  &	--	      &  2.18$\pm$0.20    \\
AlI1.31        &   0.46$\pm$0.26  &  2.15$\pm$0.35  &  1.33$\pm$0.11  &   --		&  3.16$\pm$0.34  &  2.29$\pm$0.15  &  2.89$\pm$0.16  &  2.84$\pm$0.17    \\
MgI1.48        &   0.57$\pm$0.10  &  1.16$\pm$0.07  &  0.94$\pm$0.03  &  1.83$\pm$0.14  &  0.80$\pm$0.10  &  1.83$\pm$0.09  &  1.38$\pm$0.05  &  1.77$\pm$0.08    \\
MgI1.50        &   2.41$\pm$0.10  &  3.46$\pm$0.08  &  2.07$\pm$0.07  &  1.82$\pm$0.25  &  2.88$\pm$0.18  &  3.12$\pm$0.17  &  3.11$\pm$0.14  &  3.53$\pm$0.14    \\
CO1.5a         &   2.83$\pm$0.05  &  3.21$\pm$0.10  &  2.53$\pm$0.07  &  3.06$\pm$0.07  &  4.03$\pm$0.11  &  3.89$\pm$0.07  &  4.39$\pm$0.13  &  2.98$\pm$0.09    \\
CO1.5b         &   3.46$\pm$0.04  &  4.34$\pm$0.07  &  3.10$\pm$0.05  &  4.63$\pm$0.19  &  5.34$\pm$0.13  &  3.48$\pm$0.09  &  4.55$\pm$0.10  &  3.52$\pm$0.06    \\
FeI1.58        &   0.73$\pm$0.03  &  1.85$\pm$0.05  &  0.77$\pm$0.03  &  2.46$\pm$0.11  &  1.80$\pm$0.09  &  1.35$\pm$0.06  &  1.81$\pm$0.07  &  1.18$\pm$0.05    \\
SiI1.58        &   2.81$\pm$0.07  &  4.25$\pm$0.12  &  2.57$\pm$0.07  &  4.45$\pm$0.19  &  4.58$\pm$0.17  &  3.43$\pm$0.11  &  3.71$\pm$0.15  &  2.01$\pm$0.14    \\
CO1.5c         &   2.99$\pm$0.07  &  4.09$\pm$0.10  &  2.51$\pm$0.06  &  3.46$\pm$0.18  &  3.39$\pm$0.13  &  2.93$\pm$0.13  &  4.03$\pm$0.19  &  2.70$\pm$0.16    \\
CO1.6a         &   6.02$\pm$0.19  &  7.73$\pm$0.27  &  4.26$\pm$0.20  &  9.42$\pm$0.45  &  7.24$\pm$0.18  &  6.17$\pm$0.20  &  7.73$\pm$0.26  &  6.50$\pm$0.28    \\
CO1.6b         &   2.64$\pm$0.13  &  1.36$\pm$0.12  &  0.92$\pm$0.05  &  2.07$\pm$0.28  &   --  	  &  1.13$\pm$0.09  &  1.66$\pm$0.10  &   --	          \\
MgI1.7         &   0.85$\pm$0.11  &  1.35$\pm$0.09  &  1.11$\pm$0.03  &  2.21$\pm$0.29  &  1.34$\pm$0.14  &  1.82$\pm$0.05  &  1.76$\pm$0.07  &  1.73$\pm$0.04    \\
NaI2.20        &   4.15$\pm$0.15  &  3.24$\pm$0.05  &  1.54$\pm$0.03  &  3.41$\pm$0.04  &  4.10$\pm$0.07  &  2.70$\pm$0.04  &  3.63$\pm$0.08  &  2.88$\pm$0.07    \\
CaI2.26        &   6.74$\pm$0.13  &  2.21$\pm$0.16  &  1.04$\pm$0.04  &  2.43$\pm$0.04  &  4.37$\pm$0.25  &  1.87$\pm$0.09  &  4.35$\pm$0.11  &  2.49$\pm$0.13    \\
MgI2.28        &   0.71$\pm$0.18  &  0.36$\pm$0.04  &  0.05$\pm$0.04  &  1.05$\pm$0.04  &   --  	  &  0.33$\pm$0.05  &  1.58$\pm$0.05  &  0.86$\pm$0.06    \\
CO2.2          &   17.96$\pm$0.81 &  21.54$\pm$0.56 &  6.24$\pm$0.12  &  22.26$\pm$0.54 &  22.08$\pm$1.02 &  16.14$\pm$0.67 &  23.89$\pm$0.47 &  17.53$\pm$0.83   \\
CO2.3a         &   22.66$\pm$1.19 &  22.13$\pm$0.39 &  2.96$\pm$0.16  &  23.54$\pm$0.41 &  22.84$\pm$0.95 &  14.90$\pm$0.82 &  26.34$\pm$0.40 &  24.72$\pm$0.77   \\
CO2.3b         &   27.53$\pm$1.39 &  22.58$\pm$0.44 &  4.36$\pm$0.20  &  24.65$\pm$0.41 &  20.18$\pm$1.07 &  14.97$\pm$1.47 &  27.79$\pm$0.50 &  20.27$\pm$0.89  \\
\noalign{\smallskip}
\hline
\end{tabular}
\end{table*}

\section{Final Remarks}\label{finrem}

We analysed long-slit spectra spanning optical to near-infrared wavelengths of 16 infrared-luminous star-forming galaxies with the aim of offering  the community a set of emission and absorption feature measurements that can be used to test the predictions of the forthcoming generations of stellar population models. The optical and NIR spectra were obtained at WIRO and at SpeX/IRTF, respectively. In addition to these, we collected literature spectra of early-type galaxies and performed the equivalent width measurements using a new homogeneous set of continuum and band pass definitions.  The main findings can be summarized as follows: 

\begin{itemize}

\item All our sources display \h2\ emission, characteristic  of the star-forming nature of our sample. In the optical they clearly display H\,{\sc i} emission lines. However, NGC\,1055 and NGC\,1134 show a NIR spectrum free of H\,{\sc i} emission lines. We interpret this latter result as the result of the low sensitivity of the NIR detector in this wavelength interval, thus the expected Br$\gamma$ fluxes are below the detection limit. 

\item The continua are dominated by stellar absorption features. The most common features are due to \ion{Ca}{i}, \ion{Ca}{ii}, \ion{Fe}{i}, \ion{Na}{i}, \ion{Mg}{i}, plus prominent absorption bands of: TiO, VO, ZrO and CO. In most cases (70\%) ,the stellar continua also show evidence of dust extinction. 

\item We present new definitions of continuum and line band passes for the NIR absorption lines. These definitions were made taking into account the position of the most common emission lines detected in this wavelength range. 

\item We report EW measurements for 45 indices, including both optical and NIR features. We also present measurements for most of these indices in spectra of ETGs taken from literature. 
To the best of our knowledge, they represent the most complete set of EW measurements reported in the literature to date, and can be used to test the predictions of stellar population models from the optical to the NIR. 

\item We looked for correlations among the different absorption features, presenting as the most robust ones those with a Pearson correlation coefficient r$>$0.6. 
In addition to the already-known correlations in the optical region, we propose here correlations between optical and NIR indices, as well as correlations between different NIR indices, and compare them with model predictions.

\item While for the optical absorption features the new generation of models, with scaled-solar abundance ratios and standard IMF, share the same locus as the observed data points, they fail to predict the strengths of most of the NIR indices for the SFGs, while in the case of the early-type sources they roughly reproduce the observations. This may indicate more complex SFHs for the SFGs, which we interpreted as a strong contribution from the younger stellar populations, thus explaining the fact that the CN and CO bands are in general larger for the SFGs than the ETGs. These bands are enhanced in stars in the TP-AGB phase, however, they seems to have a limited impact on the indices of ETGs.

\end{itemize}

\section*{Acknowledgements}

We are grateful to the referee for insight that has improved the quality of this paper.
RR and RAR thank CNPq, CAPES and FAPERGS for financial support for this project and to Luis Colina for helpful discussions on this project.  The authors are also grateful to Richard McDermid and Christina Baldwin for kindly offering information on the ETGs used in this work, as well as, to Jari Kotilainen for sharing information on his sample of objects. We also thank Peter van Hoof for providing the new collisional strengths for [\ion{P}{ii}] lines, which will be part of the new {\sc cloudy} release,  as well as for useful discussions.  ARA thanks CNPq for financial support. RFP acknowledges financial support from the European Union's Horizon 2020 research and innovation program under the Marie Sklodowska-Curie grant agreement No. 721463 to the SUNDIAL ITN network. AV acknowledges support from grant AYA2016-77237-C3-1-P from the Spanish Ministry of Economy and Competitiveness (MINECO).




\bibliographystyle{mnras}
\bibliography{references.bib}




\appendix
\section{Final reduced spectra}\label{FinRed}

 Final reduced and redshift-corrected spectra  for the remaining sample. Available as online material.

 \begin{figure*}
 \begin{minipage}[b]{0.5\linewidth}
 \includegraphics[width=\textwidth]{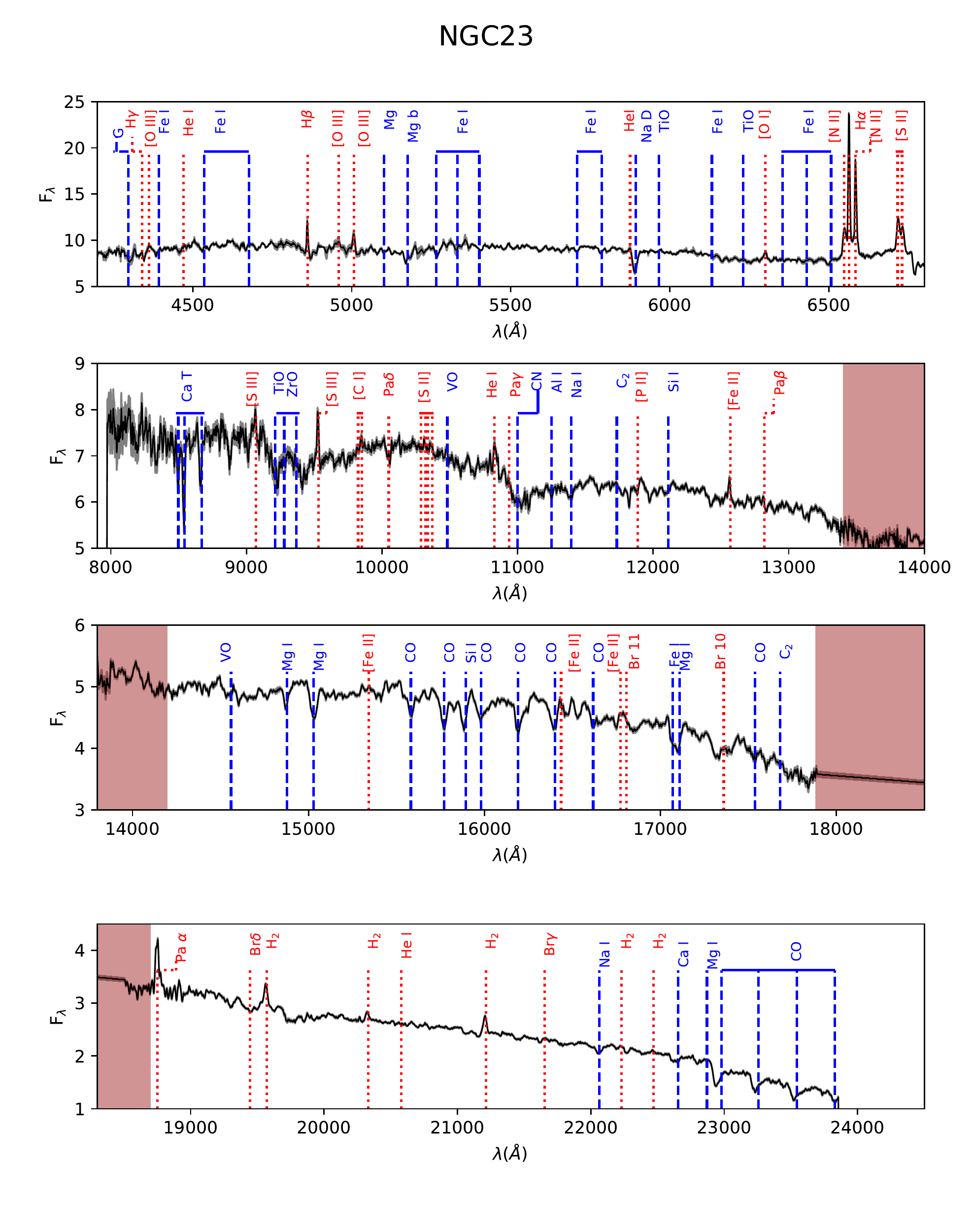}
 \end{minipage}\hfill
 \begin{minipage}[b]{0.5\linewidth}
 \includegraphics[width=\textwidth]{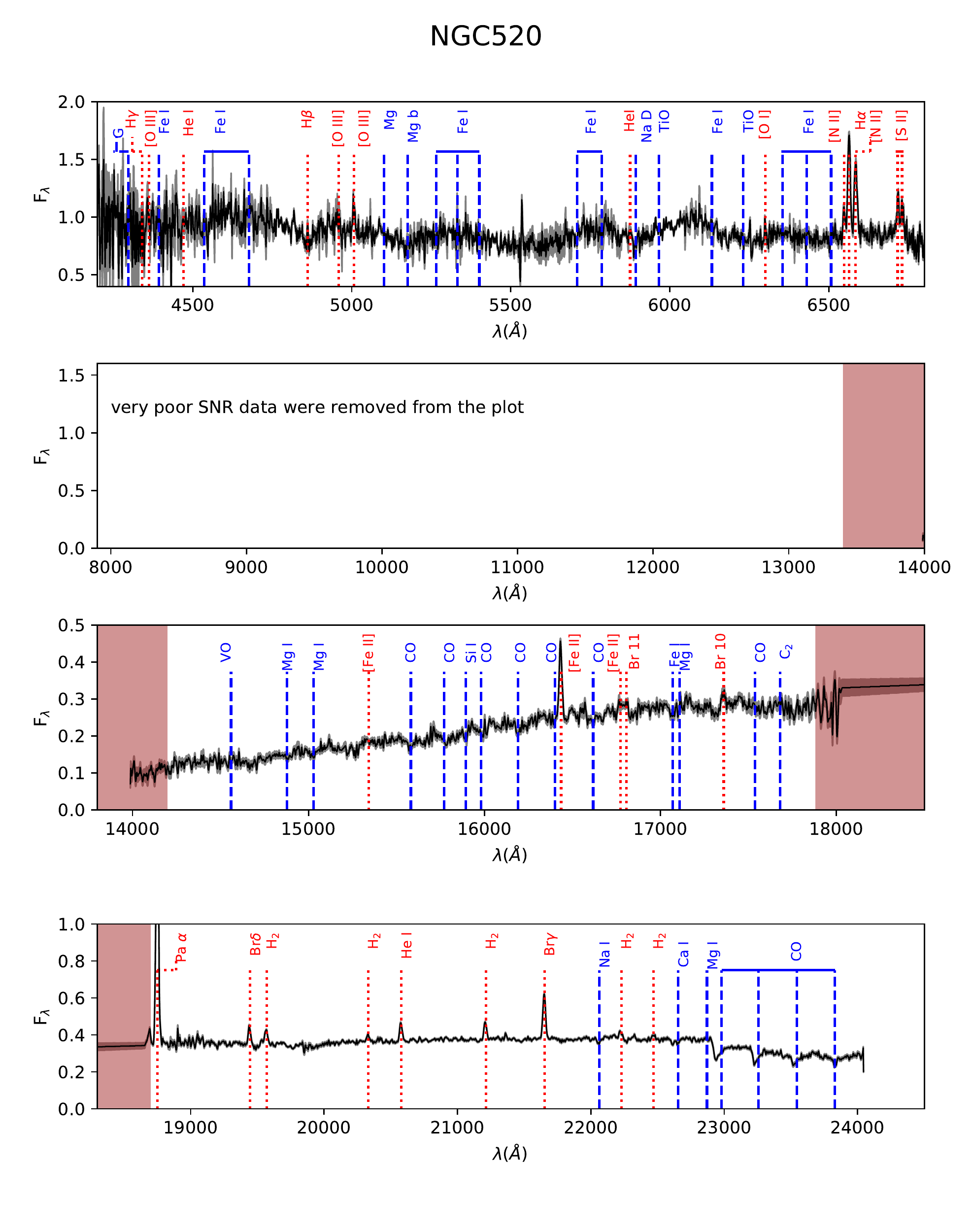}
 \end{minipage}\hfill
 \begin{minipage}[b]{0.5\linewidth}
 \includegraphics[width=\textwidth]{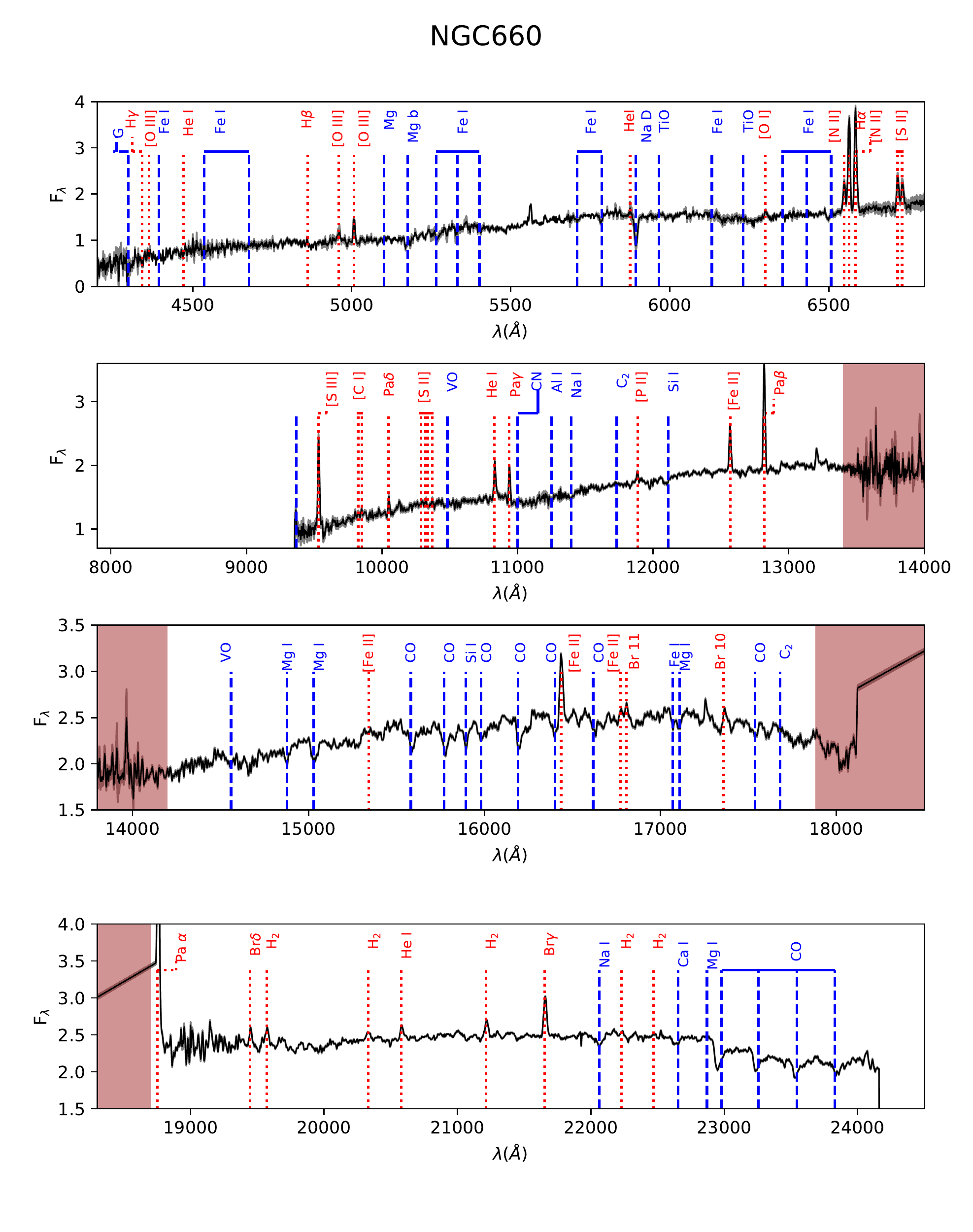}
 \end{minipage}\hfill
 \begin{minipage}[b]{0.5\linewidth}
 \includegraphics[width=\textwidth]{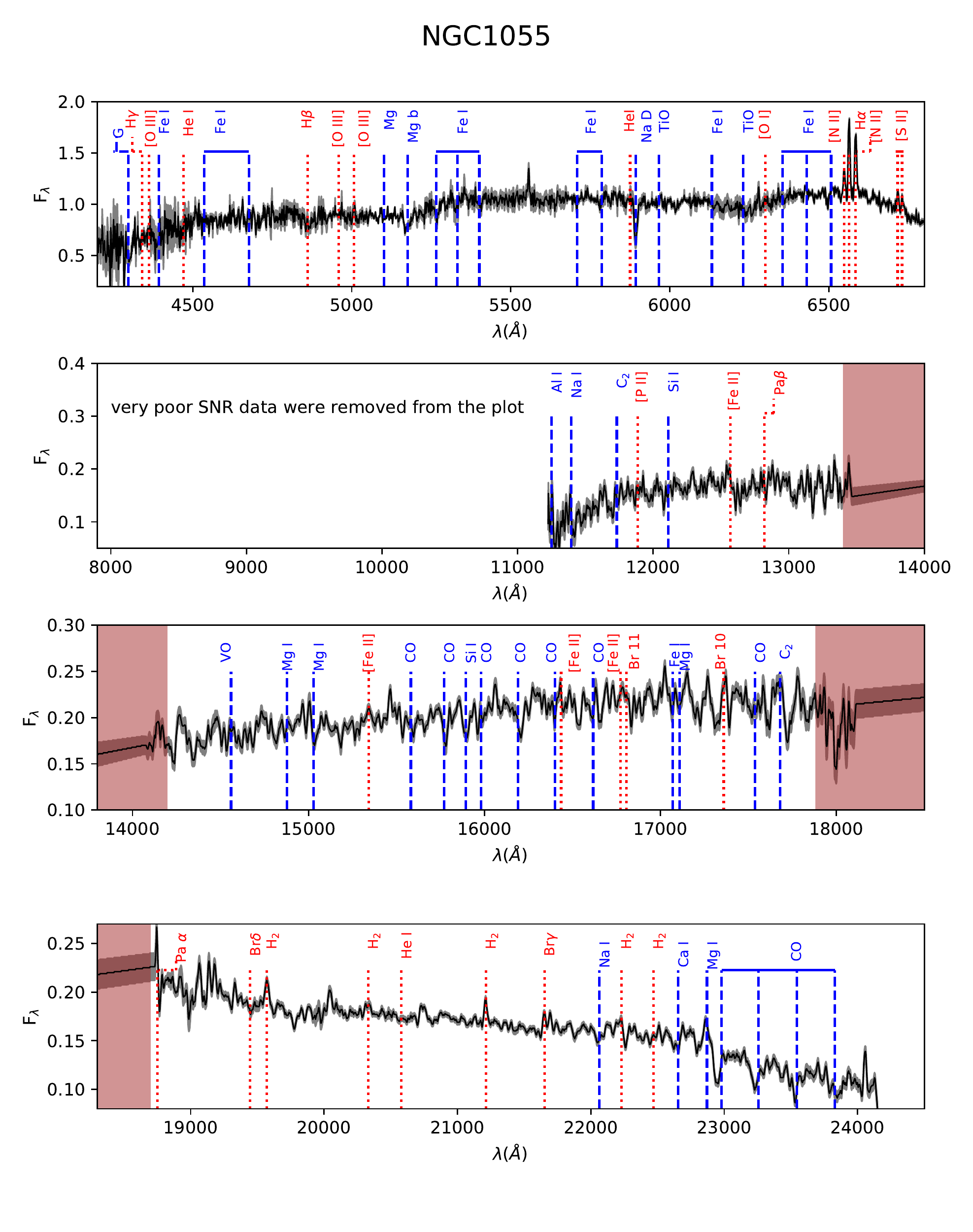}
 \end{minipage}\hfill
 \caption{Final reduced spectra in the Earth's velocity frame, The sources are labeled. For each galaxy we show from top to bottom the optical, $z+J$ , $H $, and $K$ bands, respectively. The flux is in units of  $\rm 10^{-15}~ erg ~ cm^{-2} ~ s^{-1}$. The shaded grey area represents the uncertainties and the brown area shows the poor transmission region between different bands.}
 \label{spectraapend1}
 \end{figure*}

 \begin{figure*}
 \begin{minipage}[b]{0.5\linewidth}
 \includegraphics[width=\textwidth]{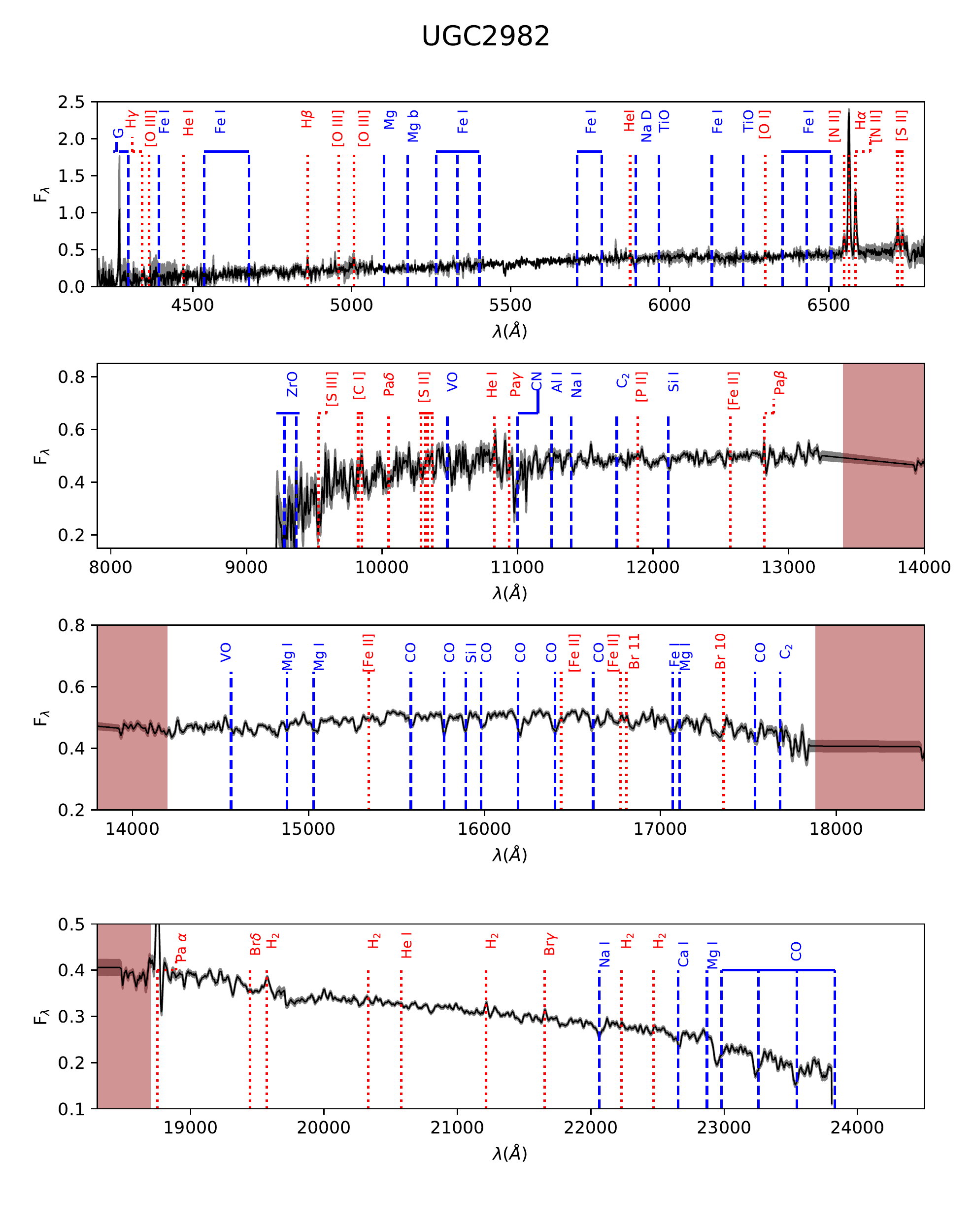}
 \end{minipage}\hfill
 \begin{minipage}[b]{0.5\linewidth}
 \includegraphics[width=\textwidth]{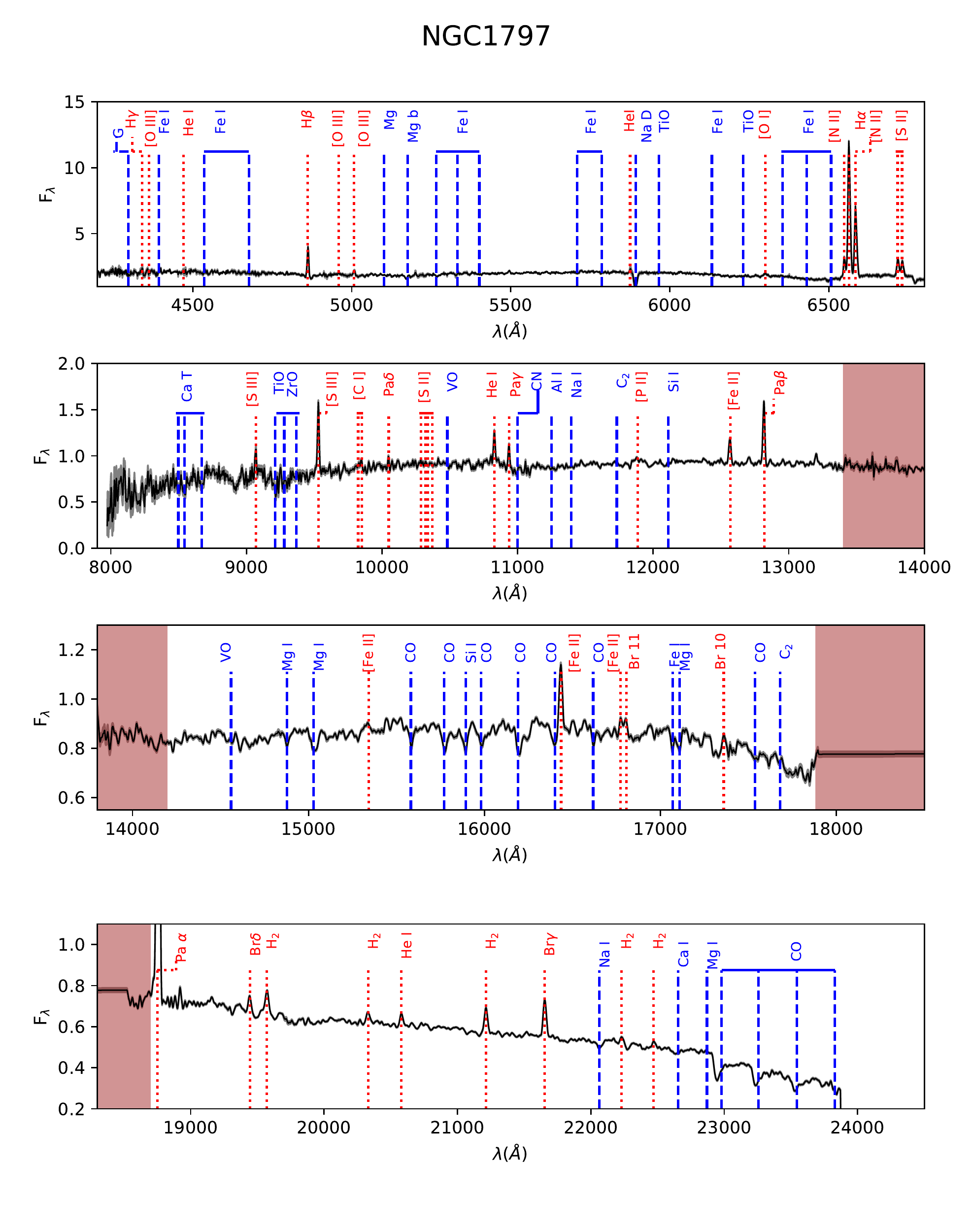}
 \end{minipage}\hfill
 \begin{minipage}[b]{0.5\linewidth}
 \includegraphics[width=\textwidth]{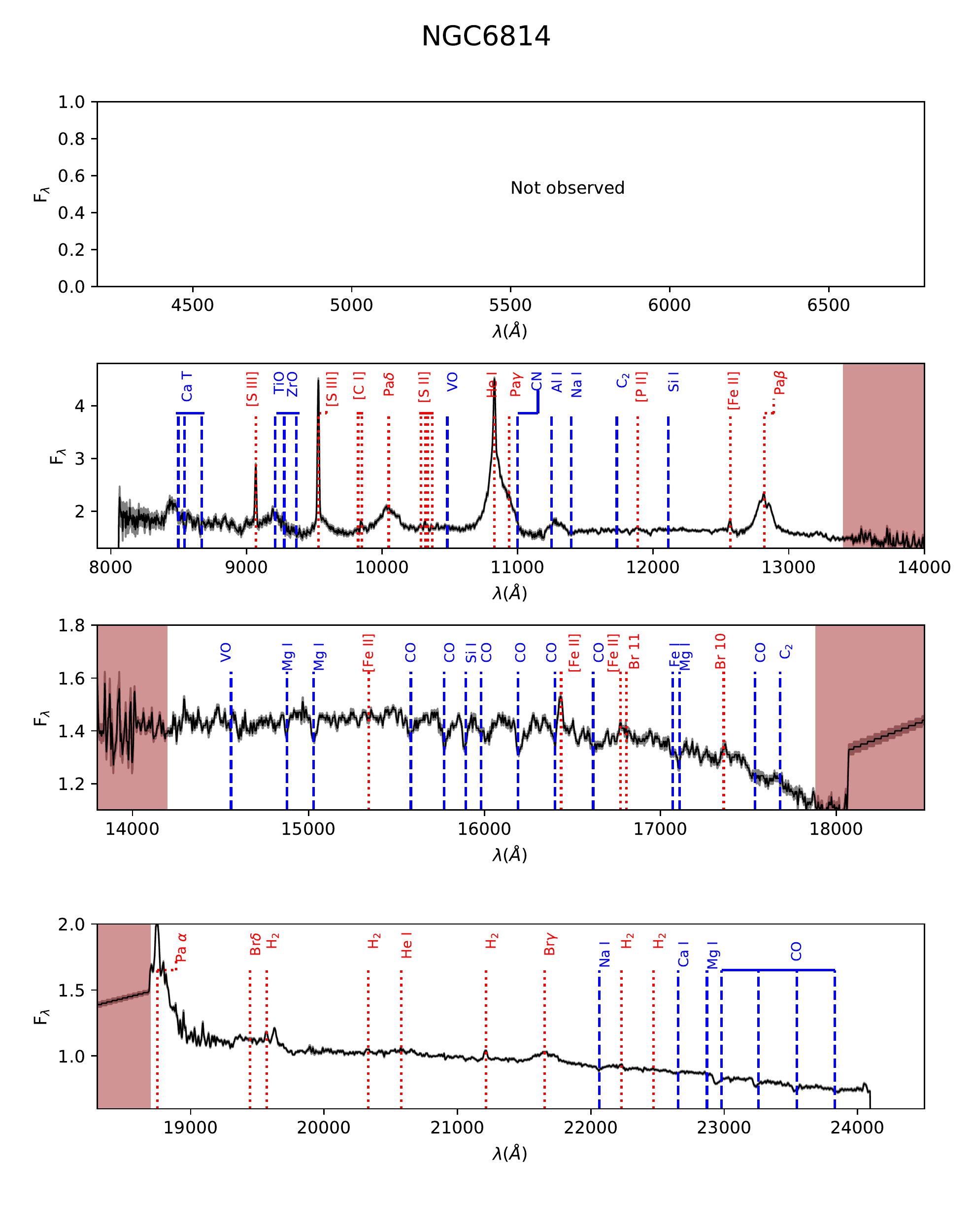}
 \end{minipage}\hfill
 \begin{minipage}[b]{0.5\linewidth}
 \includegraphics[width=\textwidth]{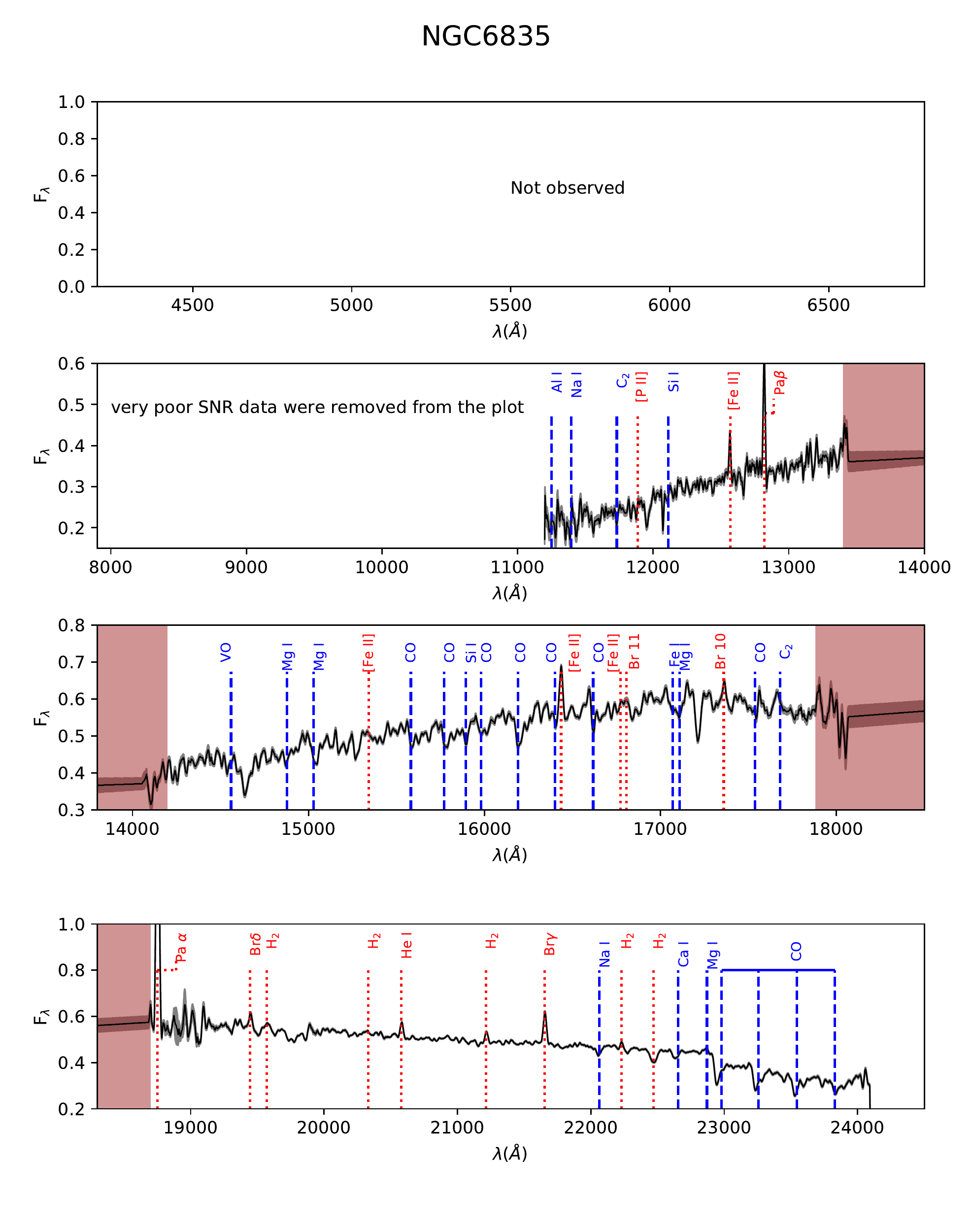}
 \end{minipage}\hfill
 \caption{Final reduced spectra in the Earth's velocity frame.  The sources are labeled. For each galaxy we show from top to bottom the optical, $z+J$ , $H$ , and $K$ bands, respectively. The flux is in units of  $\rm 10^{-15}~ erg ~ cm^{-2} ~ s^{-1}$. The shaded grey area represents the uncertainties and the brown area shows the poor transmission region between different bands.}
 \label{spectraapend2}
 \end{figure*}

 \begin{figure*}
 \begin{minipage}[b]{0.5\linewidth}
 \includegraphics[width=\textwidth]{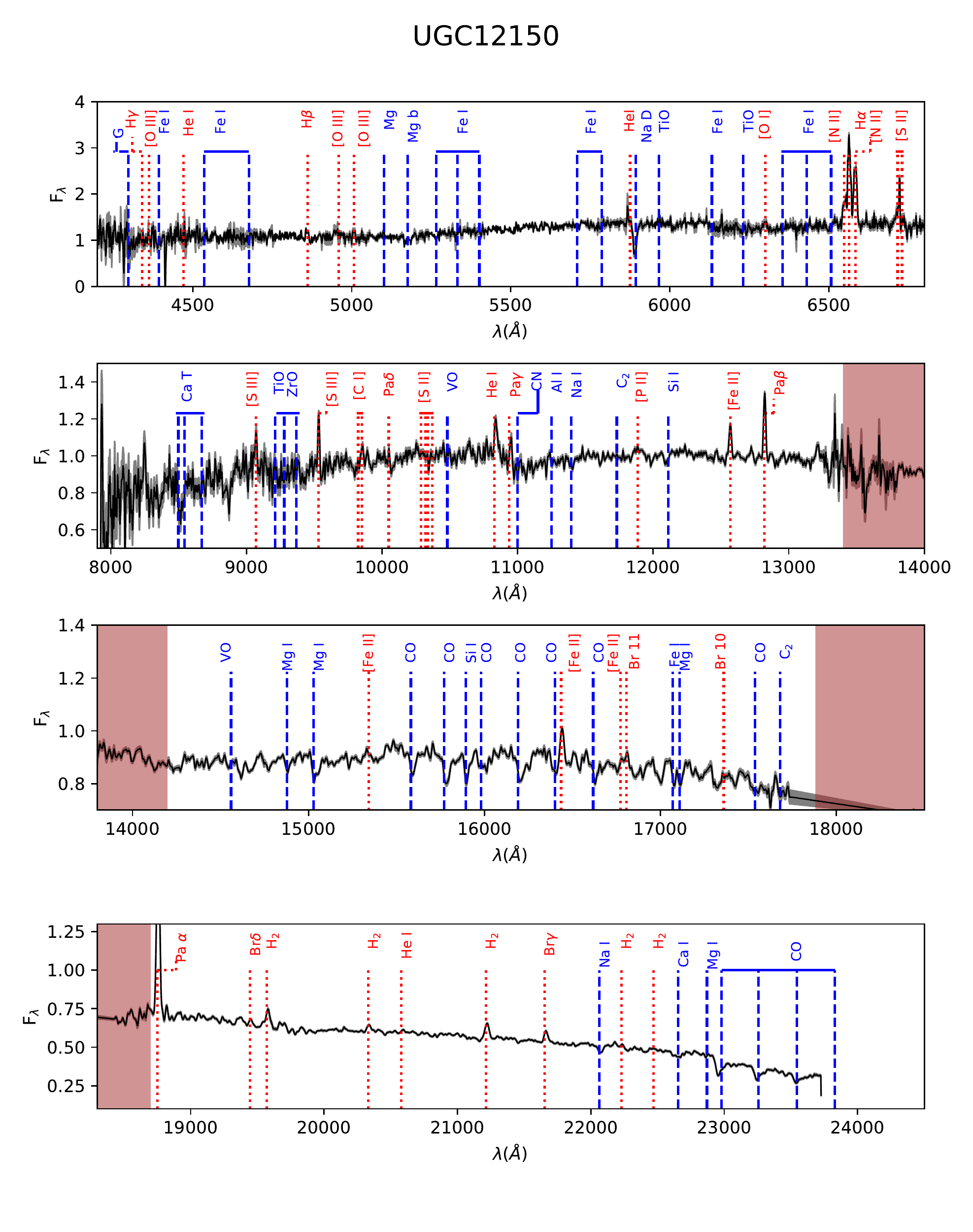}
 \end{minipage}\hfill
 \begin{minipage}[b]{0.5\linewidth}
 \includegraphics[width=\textwidth]{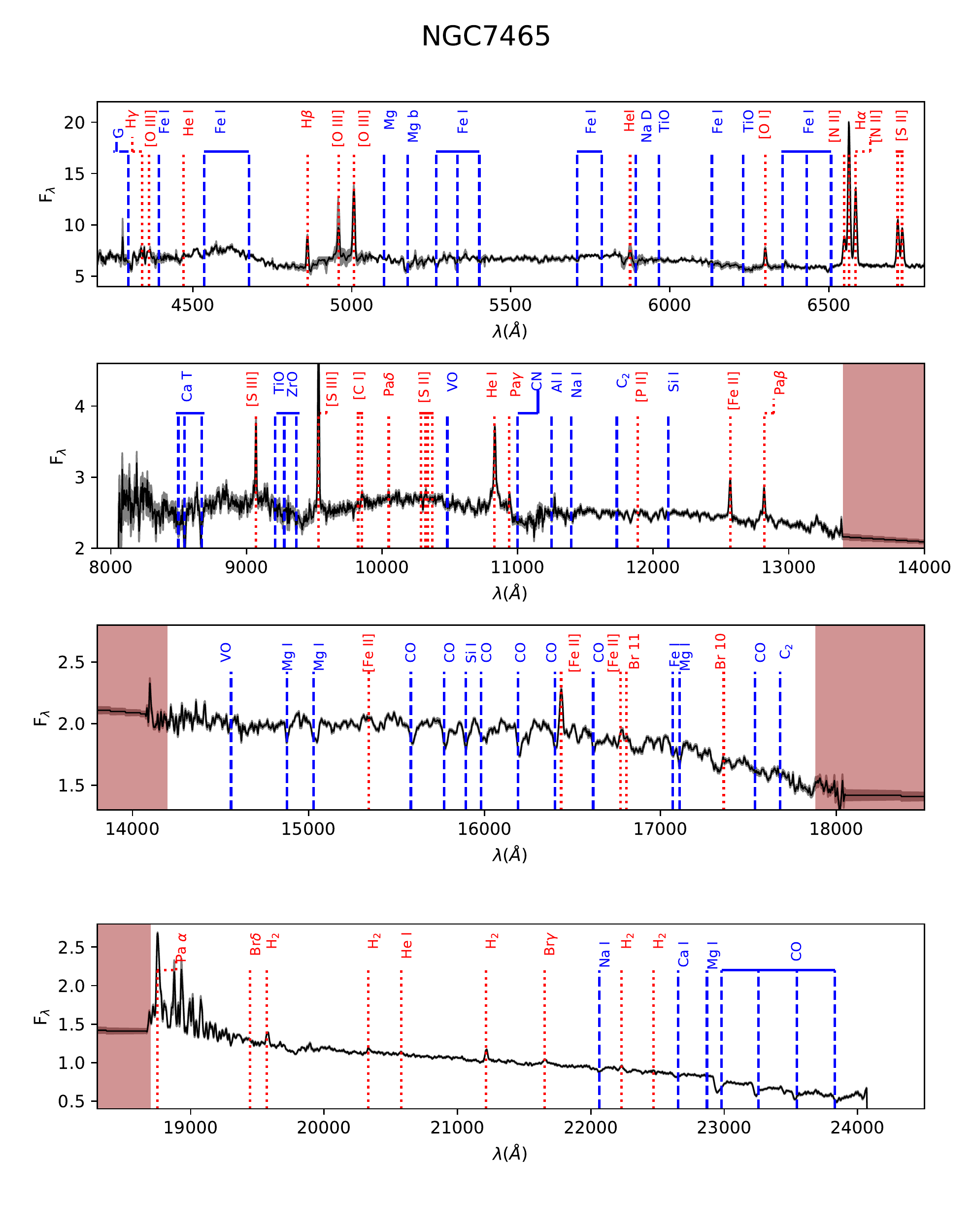}
 \end{minipage}\hfill
 \begin{minipage}[b]{0.5\linewidth}
 \includegraphics[width=\textwidth]{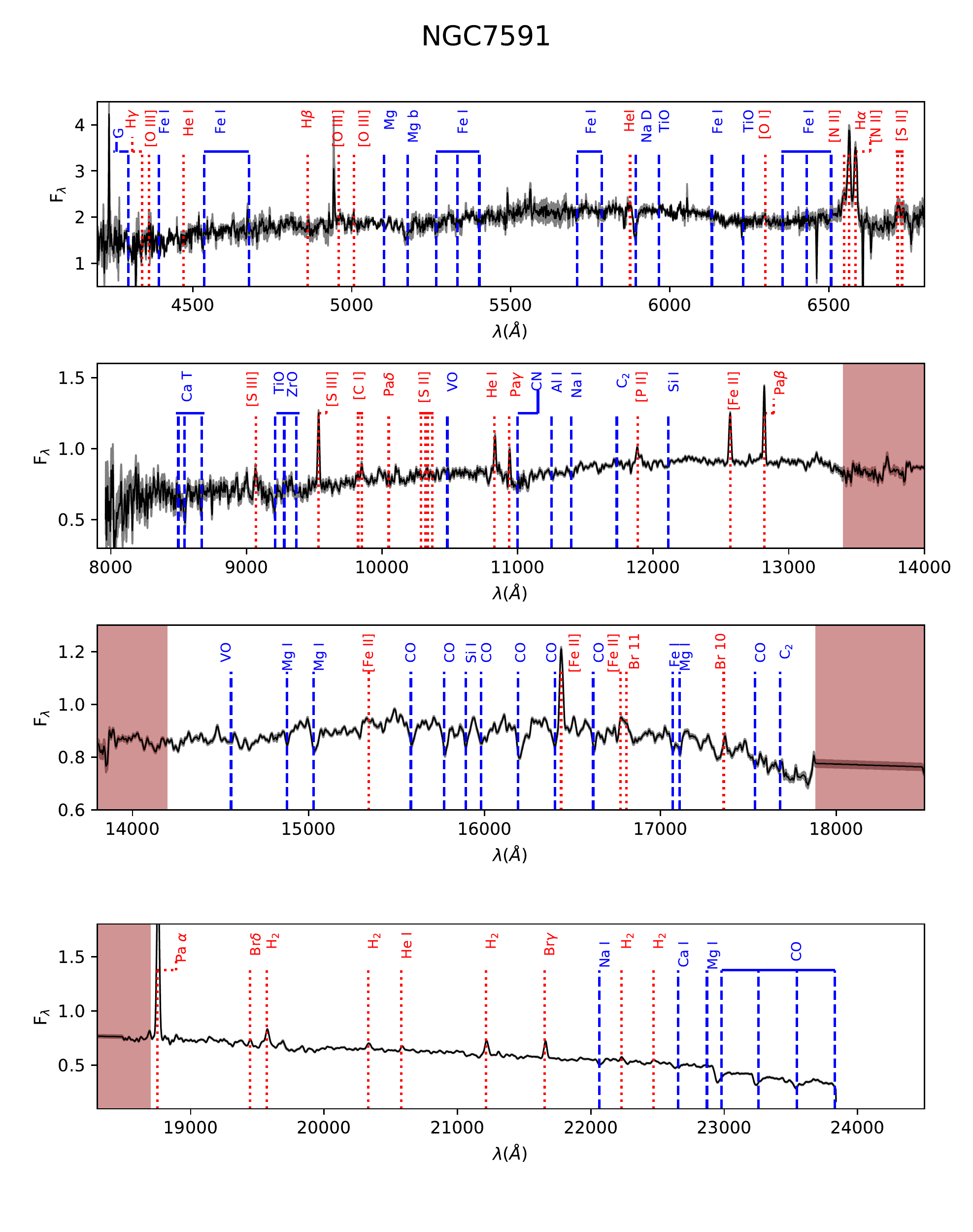}
 \end{minipage}\hfill
 \begin{minipage}[b]{0.5\linewidth}
 \includegraphics[width=\textwidth]{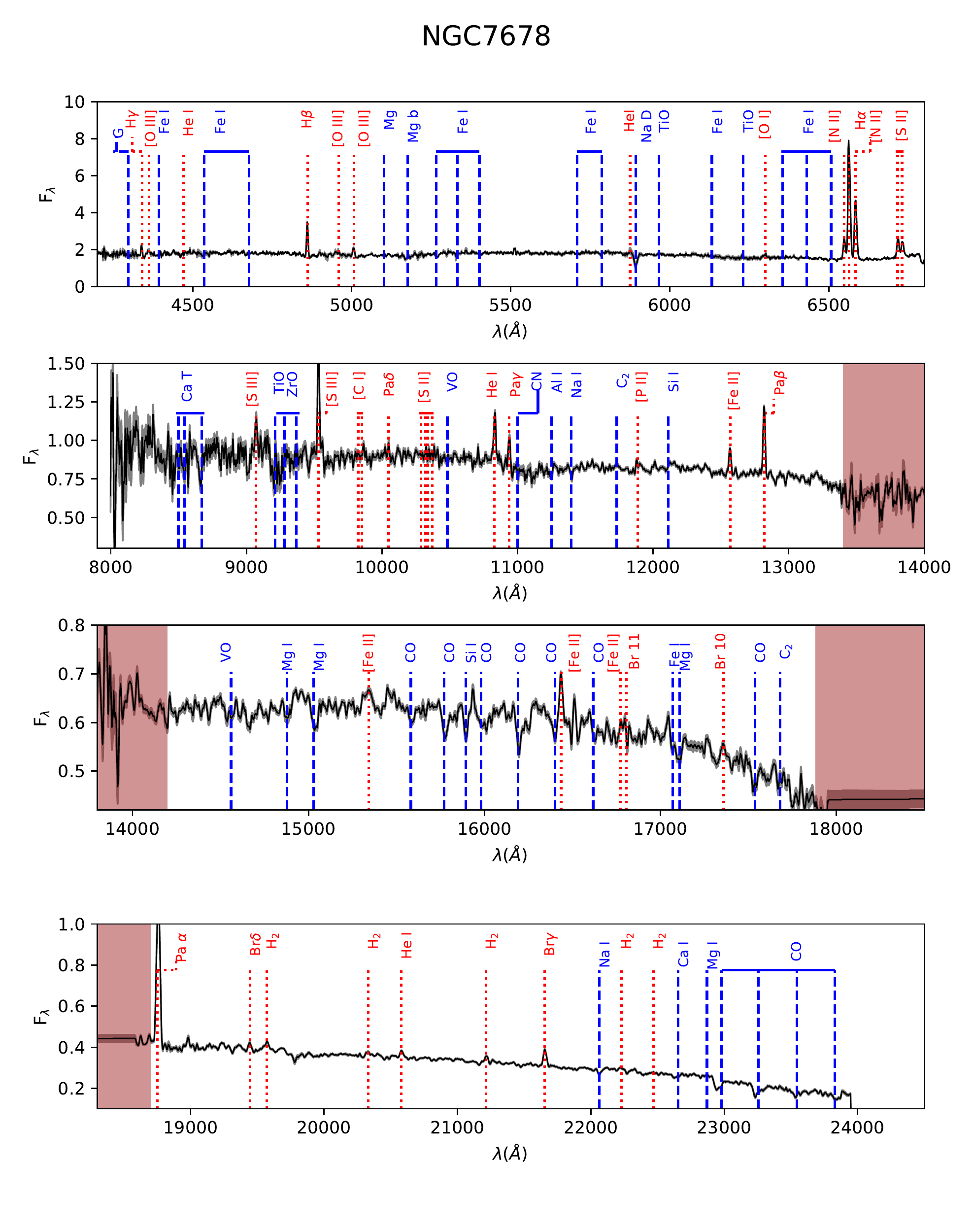}
 \end{minipage}\hfill
 \caption{Final reduced spectra in the Earth's velocity frame.  The sources are labeled. For each galaxy we show from top to bottom the optical, $z+J$ , $H$ , and $K$ bands, respectively. The flux is in units of  $\rm 10^{-15}~ erg ~ cm^{-2} ~ s^{-1}$. The shaded grey area represents the uncertainties and the brown area shows the poor transmission region between different bands.}
 \label{spectraapend3}
 \end{figure*}

 \section{Literature data}

 Here we present the measurements using the index definitions listed in Tab.~\ref{ewdefs} for the literature data. The data used here are those of \citet{Dahmer-Hahn+18} and \citet{Baldwin+17}. For the latter we found optical Sloan Digital Sky Survey data \citep{Ahn+14} for four sources.  For the remaining objects we collected the values of Fe5015, Mg$_b$ and Fe5270  from \citet{McDermid+15}, while for the sources of  \citet{Dahmer-Hahn+18} the optical data were taken from the Calar Alto Legacy Integral Field Area Survey \citep[{\sc califa}][]{Sanchez+16} and we measured the EW of the optical lines.

\begin{table*}
\renewcommand{\tabcolsep}{0.70mm}
\centering
\caption{ Absorption feature Equivalent Widths (in \AA) from the sample of \citet{Baldwin+17}. \label{Ewbald1}} 
\begin{tabular}{llccccccccccccccccc}
\hline\hline
\noalign{\smallskip}
Line           &    IC0719       &  NGC3032        &  NGC3098        &  NGC3156        &  NGC3182        &  NGC3301           \\
\noalign{\smallskip}
\hline \noalign{\smallskip}
Ca4227         &  1.68$\pm$0.01  &   --            &   --            &  0.56$\pm$0.00  &  1.41$\pm$0.00  &   --             \\
G4300          &  6.23$\pm$0.04  &   --            &   --            &  1.94$\pm$0.00  &  5.57$\pm$0.01  &   --             \\
Fe4383         &  5.99$\pm$0.05  &   --            &   --            &  1.89$\pm$0.00  &  5.33$\pm$0.01  &   --             \\
Ca4455         &  2.07$\pm$0.02  &   --            &   --            &  1.08$\pm$0.00  &  1.68$\pm$0.00  &   --             \\
Fe4531         &  3.67$\pm$0.04  &   --            &   --            &  2.99$\pm$0.00  &  3.37$\pm$0.01  &   --             \\
$C_2$4668      &  7.04$\pm$0.07  &   --            &   --            &  2.79$\pm$0.01  &  7.86$\pm$0.01  &   --             \\
Fe5015         &  5.68$\pm$0.06  &  4.44$\pm$0.14  &  4.64$\pm$0.14  &  4.07$\pm$0.01  &  3.57$\pm$0.01  &  4.95$\pm$0.14   \\
Mg$_1$         &  5.40$\pm$0.05  &   --            &   --            &  1.88$\pm$0.01  &  6.46$\pm$0.01  &   --             \\
Mg$_2$         &  8.06$\pm$0.03  &   --            &   --            &  4.10$\pm$0.00  &  8.82$\pm$0.01  &   --             \\
Mg$_b$         &  3.83$\pm$0.03  &  1.70$\pm$0.12  & 3.17$\pm$0.12   &  1.58$\pm$0.00  &  4.18$\pm$0.01  & 2.93$\pm$0.12    \\
Fe5270         &  3.02$\pm$0.03  &  2.06$\pm$0.11  & 2.56$\pm$0.11   &  1.93$\pm$0.01  &  2.97$\pm$0.01  & 2.56$\pm$0.11    \\
Fe5335         &  2.93$\pm$0.02  &   --            &   --            &  1.98$\pm$0.00  &  3.04$\pm$0.01  &   --             \\
Fe5406         &  1.51$\pm$0.02  &   --            &   --            &  1.06$\pm$0.00  &  1.87$\pm$0.00  &   --             \\
Fe5709         &  1.21$\pm$0.01  &   --            &   --            &  0.68$\pm$0.00  &  1.09$\pm$0.00  &   --             \\
Fe5782         &  0.97$\pm$0.01  &   --            &   --            &  0.51$\pm$0.00  &  0.91$\pm$0.00  &   --             \\
NaD            &  3.13$\pm$0.03  &   --            &   --            &  1.19$\pm$0.01  &  3.63$\pm$0.01  &   --             \\
TiO$_1$        &  2.20$\pm$0.04  &   --            &   --            &  0.60$\pm$0.01  &  1.66$\pm$0.01  &   --             \\
TiO$_2$        &  5.17$\pm$0.05  &   --            &   --            &  1.87$\pm$0.02  &  5.26$\pm$0.02  &   --             \\
CaT1           &   --            &   --            &   --            &   --            &   --            &   --             \\
CaT2           &   --            &   --            &   --            &   --            &   --            &   --             \\
CaT3           &   --            &   --            &   --            &   --            &   --            &   --             \\
ZrO            &  8.55$\pm$2.34  &  15.87$\pm$1.08 &  8.92$\pm$1.41  &  22.28$\pm$1.91 &  14.75$\pm$1.68 &  15.61$\pm$0.92  \\
VO             &  2.63$\pm$0.55  &   --            &  0.32$\pm$0.51  &  1.87$\pm$0.59  &  0.28$\pm$0.38  &  1.45$\pm$0.39   \\
CN11           &  5.97$\pm$0.79  &  10.58$\pm$0.37 &  11.35$\pm$0.39 &  7.31$\pm$0.64  &  7.93$\pm$0.58  &  8.47$\pm$0.36   \\
NaI1.14        &  1.82$\pm$0.21  &  3.14$\pm$0.17  &  1.48$\pm$0.14  &  1.23$\pm$0.26  &  2.37$\pm$0.17  &  1.42$\pm$0.14   \\
FeI1.16        &  1.23$\pm$0.26  &   --            &  0.61$\pm$0.10  &  0.45$\pm$0.18  &  0.39$\pm$0.22  &  0.03$\pm$0.05   \\
MgI1.24        &  1.00$\pm$0.12  &  1.04$\pm$0.09  &  0.36$\pm$0.08  &  0.81$\pm$0.15  &  1.43$\pm$0.11  &  0.81$\pm$0.07   \\
MnI1.29        &  0.85$\pm$0.19  &  1.06$\pm$0.09  &  1.20$\pm$0.13  &  1.19$\pm$0.20  &  1.20$\pm$0.20  &  1.26$\pm$0.18   \\
AlI1.31        &  1.27$\pm$0.16  &  1.87$\pm$0.10  &  1.32$\pm$0.18  &  0.74$\pm$0.15  &  2.27$\pm$0.11  &  1.27$\pm$0.11   \\
MgI1.48        &  2.05$\pm$0.13  &  1.62$\pm$0.11  &  1.71$\pm$0.08  &  1.12$\pm$0.13  &  0.85$\pm$0.27  &  1.55$\pm$0.09   \\
MgI1.50        &  4.29$\pm$0.23  &  3.58$\pm$0.17  &  3.60$\pm$0.21  &  3.23$\pm$0.22  &  1.42$\pm$0.31  &  3.37$\pm$0.18   \\
CO1.5a         &  3.72$\pm$0.14  &  4.74$\pm$0.16  &  4.05$\pm$0.28  &  4.40$\pm$0.19  &  2.75$\pm$0.27  &  4.38$\pm$0.18   \\
CO1.5b         &  4.13$\pm$0.20  &  4.09$\pm$0.14  &  4.96$\pm$0.20  &  3.61$\pm$0.16  &  2.61$\pm$0.44  &  5.56$\pm$0.10   \\
FeI1.58        &  1.37$\pm$0.13  &  0.85$\pm$0.11  &  1.55$\pm$0.11  &  1.78$\pm$0.11  &  0.62$\pm$0.24  &  2.14$\pm$0.07   \\
SiI1.58        &  2.91$\pm$0.22  &  3.44$\pm$0.24  &  3.24$\pm$0.23  &  4.14$\pm$0.23  &  2.89$\pm$0.49  &  4.85$\pm$0.14   \\
CO1.5c         &  3.97$\pm$0.17  &  3.56$\pm$0.24  &  3.20$\pm$0.16  &  2.83$\pm$0.22  &  3.35$\pm$0.29  &  3.98$\pm$0.16   \\
CO1.6a         &  5.33$\pm$0.35  &  4.73$\pm$0.45  &  6.67$\pm$0.30  &  6.48$\pm$0.41  &  5.81$\pm$0.48  &  6.30$\pm$0.37   \\
CO1.6b         &  1.27$\pm$0.16  &  0.89$\pm$0.14  &  1.51$\pm$0.13  &  1.28$\pm$0.16  &  1.37$\pm$0.21  &  1.31$\pm$0.15   \\
MgI1.7         &  1.99$\pm$0.08  &  2.30$\pm$0.08  &  2.12$\pm$0.09  &  1.85$\pm$0.10  &  2.41$\pm$0.09  &  2.14$\pm$0.10   \\
NaI2.20        &  3.41$\pm$0.08  &  5.15$\pm$0.14  &  2.84$\pm$0.13  &  2.52$\pm$0.20  &  3.74$\pm$0.08  &  2.77$\pm$0.19   \\
CaI2.26        &  1.85$\pm$0.17  &  3.66$\pm$0.26  &  2.49$\pm$0.34  &  2.55$\pm$0.29  &  2.57$\pm$0.22  &  3.09$\pm$0.14   \\
MgI2.28        &  0.76$\pm$0.11  &  0.94$\pm$0.12  &  1.05$\pm$0.09  &   --            &  1.22$\pm$0.16  &  0.94$\pm$0.04   \\
CO2.2          &  15.76$\pm$0.83 &  20.43$\pm$0.68 &  14.80$\pm$0.70 &  19.55$\pm$0.78 &  17.57$\pm$0.77 &  19.47$\pm$0.58  \\
CO2.3a         &  17.74$\pm$1.01 &  18.16$\pm$0.65 &  14.12$\pm$0.86 &  19.90$\pm$1.21 &  16.44$\pm$0.78 &  19.23$\pm$0.70  \\
CO2.3b         &  20.23$\pm$1.27 &  17.06$\pm$0.78 &  17.40$\pm$1.13 &  23.39$\pm$1.85 &  20.04$\pm$1.15 &  21.83$\pm$0.98  \\
\noalign{\smallskip}
\hline
\end{tabular}
\begin{list}{Table Notes:}
\item The values of Fe5015, Mg$_b$ and Fe5270 for NGC3032, NGC3098 and NGC3301 were taken from \citet{McDermid+15} for Re/8.
\end{list}
\end{table*}

\begin{table*}
\renewcommand{\tabcolsep}{0.70mm}
\centering
\caption{Absorption feature Equivalent Widths (in \AA) from the sample of \citet{Baldwin+17}.\label{Ewbald2}} 
\begin{tabular}{llccccccccccccccccc}
\hline\hline
\noalign{\smallskip}
Line            &   NGC3489        &  NGC4379        &  NGC4578        &  NGC4608        &  NGC4710        &  NGC5475                \\
\noalign{\smallskip}
\hline \noalign{\smallskip}
Ca4227          &   --  	  &   --	    &	--	      &   --		&   --  	  &  1.32$\pm$0.00    \\
G4300           &   --  	  &   --	    &	--	      &   --		&   --  	  &  5.76$\pm$0.01    \\
Fe4383          &   --  	  &   --	    &	--	      &   --		&   --  	  &  5.43$\pm$0.01    \\
Ca4455          &   --  	  &   --	    &	--	      &   --		&   --  	  &  1.87$\pm$0.00    \\
Fe4531          &   --  	  &   --	    &	--	      &   --		&   --  	  &  3.79$\pm$0.01    \\
$C_2$4668       &   --  	  &   --	    &	--	      &   --		&   --  	  &  7.12$\pm$0.01    \\
Fe5015          & 4.84$\pm$0.14   & 5.41$\pm$0.14   &	5.24$\pm$0.14 &  4.64$\pm$0.14  &  4.46$\pm$0.15  &  5.09$\pm$0.01    \\
Mg$_1$          &   --  	  &   --	    &	--	      &   --		&   --  	  &  5.33$\pm$0.01    \\
Mg$_2$          &   --  	  &   --	    &	--	      &   --		&   --  	  &  8.28$\pm$0.00    \\
Mg$_b$          & 2.80$\pm$0.12   & 3.66$\pm$0.12   &  3.88$\pm$0.12  & 3.78$\pm$0.12   & 3.05$\pm$0.1 2  &  4.22$\pm$0.00    \\
Fe5270          & 2.58$\pm$0.11   & 2.85$\pm$0.12   &  2.90$\pm$0.11  &	2.59$\pm$0.11   & 2.44$\pm$0.15   &  3.10$\pm$0.00    \\
Fe5335          &   --  	  &   --	    &	--	      &   --		&   --  	  &  3.24$\pm$0.00    \\
Fe5406          &   --  	  &   --	    &	--	      &   --		&   --  	  &  1.96$\pm$0.00    \\
Fe5709          &   --  	  &   --	    &	--	      &   --		&   --  	  &  1.09$\pm$0.00    \\
Fe5782          &   --  	  &   --	    &	--	      &   --		&   --  	  &  1.07$\pm$0.00    \\
NaD             &   --  	  &   --	    &	--	      &   --		&   --  	  &  3.58$\pm$0.00    \\
TiO$_1$         &   --  	  &   --	    &	--	      &   --		&   --  	  &  1.26$\pm$0.01    \\
TiO$_2$         &   --  	  &   --	    &	--	      &   --		&   --  	  &  4.73$\pm$0.01    \\
CaT1            &   --  	  &   --	    &	--	      &   --		&   --  	  &   --	      \\
CaT2            &   --  	  &   --	    &	--	      &   --		&   --  	  &   --	      \\
CaT3            &   --  	  &   --	    &	--	      &   --		&   --  	  &   --	      \\
ZrO             &  12.10$\pm$0.68 &  25.45$\pm$0.64 &  22.91$\pm$0.98 &  19.96$\pm$1.62 &  6.11$\pm$1.51  &  6.40$\pm$1.26    \\
VO              &  1.34$\pm$0.44  &  1.52$\pm$0.58  &  1.87$\pm$0.36  &   --		&  3.32$\pm$1.07  &  2.22$\pm$0.57    \\
CN11            &  7.14$\pm$0.40  &  4.53$\pm$0.49  &  5.49$\pm$0.37  &  7.03$\pm$0.43  &  9.21$\pm$1.01  &  10.28$\pm$0.47   \\
NaI1.14         &  1.35$\pm$0.20  &  1.47$\pm$0.19  &  1.63$\pm$0.16  &  1.69$\pm$0.21  &  1.22$\pm$0.25  &  1.61$\pm$0.21    \\
FeI1.16         &  0.48$\pm$0.07  &  0.38$\pm$0.10  &  0.49$\pm$0.10  &  0.44$\pm$0.09  &  0.73$\pm$0.12  &  0.33$\pm$0.11    \\
MgI1.24         &  0.60$\pm$0.04  &  0.88$\pm$0.06  &  0.99$\pm$0.07  &  0.94$\pm$0.07  &  0.71$\pm$0.15  &  0.80$\pm$0.08    \\
MnI1.29         &  0.91$\pm$0.11  &  1.06$\pm$0.35  &  0.98$\pm$0.26  &  1.10$\pm$0.21  &  0.70$\pm$0.28  &  1.71$\pm$0.25    \\
AlI1.31         &  1.19$\pm$0.10  &  1.35$\pm$0.18  &  1.65$\pm$0.12  &  1.65$\pm$0.09  &  1.38$\pm$0.15  &  2.21$\pm$0.10    \\
MgI1.48         &  1.49$\pm$0.09  &  1.62$\pm$0.11  &  2.29$\pm$0.22  &  1.33$\pm$0.11  &  1.70$\pm$0.13  &  1.72$\pm$0.09    \\
MgI1.50         &  3.13$\pm$0.13  &  4.07$\pm$0.18  &  5.10$\pm$0.35  &  3.86$\pm$0.15  &  4.02$\pm$0.19  &  4.54$\pm$0.21    \\
CO1.5a          &  4.19$\pm$0.14  &  3.36$\pm$0.17  &  5.15$\pm$0.26  &  3.43$\pm$0.16  &  3.41$\pm$0.19  &  4.02$\pm$0.17    \\
CO1.5b          &  4.46$\pm$0.13  &  3.10$\pm$0.19  &  5.95$\pm$0.47  &  3.86$\pm$0.16  &  3.44$\pm$0.21  &  4.55$\pm$0.15    \\
FeI1.58         &  2.28$\pm$0.10  &  1.40$\pm$0.11  &  2.66$\pm$0.23  &  1.97$\pm$0.10  &  1.98$\pm$0.15  &  1.84$\pm$0.10    \\
SiI1.58         &  3.75$\pm$0.18  &  2.97$\pm$0.18  &  4.43$\pm$0.42  &  3.98$\pm$0.21  &  3.70$\pm$0.22  &  4.03$\pm$0.20    \\
CO1.5c          &  3.72$\pm$0.17  &  2.68$\pm$0.11  &  3.71$\pm$0.36  &  3.37$\pm$0.25  &  3.59$\pm$0.20  &  4.26$\pm$0.20    \\
CO1.6a          &  6.74$\pm$0.29  &  5.51$\pm$0.29  &  6.64$\pm$0.54  &  6.78$\pm$0.37  &  5.59$\pm$0.34  &  6.07$\pm$0.37    \\
CO1.6b          &  0.85$\pm$0.15  &  0.95$\pm$0.09  &  1.47$\pm$0.22  &  2.12$\pm$0.17  &  0.92$\pm$0.12  &  1.48$\pm$0.17    \\
MgI1.7          &  1.94$\pm$0.06  &  2.25$\pm$0.06  &  2.27$\pm$0.10  &  2.33$\pm$0.08  &  2.13$\pm$0.06  &  2.32$\pm$0.11    \\
NaI2.20         &  3.22$\pm$0.04  &  3.91$\pm$0.22  &  4.59$\pm$0.12  &  3.11$\pm$0.11  &  3.56$\pm$0.06  &  3.96$\pm$0.08    \\
CaI2.26         &  2.38$\pm$0.09  &  0.88$\pm$0.42  &  2.10$\pm$0.15  &  2.09$\pm$0.30  &  2.41$\pm$0.28  &  3.40$\pm$0.30    \\
MgI2.28         &  0.56$\pm$0.02  &  0.48$\pm$0.18  &  0.88$\pm$0.10  &  1.04$\pm$0.10  &  0.86$\pm$0.14  &  1.57$\pm$0.20    \\
CO2.2           &  18.62$\pm$0.44 &  14.95$\pm$0.78 &  18.82$\pm$0.49 &  17.66$\pm$0.62 &  18.84$\pm$0.61 &  14.87$\pm$0.50   \\
CO2.3a          &  17.39$\pm$0.53 &  14.74$\pm$0.81 &  19.96$\pm$0.67 &  16.66$\pm$0.75 &  18.11$\pm$0.63 &  16.69$\pm$0.68   \\
CO2.3b          &  18.45$\pm$0.66 &  16.78$\pm$1.03 &  23.74$\pm$0.91 &  19.23$\pm$1.08 &  20.31$\pm$0.87 &  19.88$\pm$0.89    \\
\noalign{\smallskip}
\hline
\end{tabular}
\end{table*}

\begin{table*}
\renewcommand{\tabcolsep}{0.70mm}
\centering
\caption{Absorption feature Equivalent Widths (in \AA) from the sample of \citet{Dahmer-Hahn+18}.\label{Ewlgdh}} 
\begin{tabular}{llcccccccccccccccccc}
\hline\hline
\noalign{\smallskip}
Line &   NGC4636        &  NGC5905        &  NGC5966        &  NGC6081        &  NGC6146        &  NGC6338        &  UGC08234        \\
\noalign{\smallskip}
\hline \noalign{\smallskip}
Ca4227         &   --		 &   -- 	   &  1.23$\pm$0.10  &  1.21$\pm$0.07  &  1.00$\pm$0.03  &  1.16$\pm$0.06  &  0.76$\pm$0.05    \\
G4300          &   --		 &   -- 	   &  5.54$\pm$0.28  &  5.59$\pm$0.30  &  5.30$\pm$0.16  &  5.82$\pm$0.23  &  3.01$\pm$0.21    \\
Fe4383         &   --		 &   -- 	   &  4.46$\pm$0.35  &  4.56$\pm$0.29  &  4.47$\pm$0.24  &  4.33$\pm$0.29  &  2.63$\pm$0.25    \\
Ca4455         &   --		 &   -- 	   &  1.36$\pm$0.16  &  1.14$\pm$0.15  &  1.04$\pm$0.13  &  1.30$\pm$0.17  &  0.92$\pm$0.15    \\
Fe4531         &   --		 &   -- 	   &  3.32$\pm$0.14  &  3.04$\pm$0.21  &  3.15$\pm$0.14  &  3.24$\pm$0.12  &  2.73$\pm$0.08    \\
$C_2$4668      &   --		 &   -- 	   &  6.42$\pm$0.13  &  6.63$\pm$0.16  &  6.82$\pm$0.18  &  7.78$\pm$0.23  &  4.56$\pm$0.13    \\
Fe5015         &   --		 &   -- 	   &  4.01$\pm$0.29  &  3.52$\pm$0.37  &  3.75$\pm$0.25  &  3.98$\pm$0.24  &  4.14$\pm$0.22    \\
Mg$_1$         &   --		 &   -- 	   &  6.28$\pm$0.27  &  6.84$\pm$0.28  &  6.58$\pm$0.22  &  8.06$\pm$0.22  &  3.26$\pm$0.20    \\
Mg$_2$         &   --		 &   -- 	   &  8.70$\pm$0.15  &  8.90$\pm$0.15  &  9.08$\pm$0.12  &  10.34$\pm$0.13 &  5.96$\pm$0.13    \\
Mg$_b$         &   --		 &   -- 	   &  3.95$\pm$0.15  &  4.09$\pm$0.15  &  3.89$\pm$0.13  &  4.67$\pm$0.12  &  2.60$\pm$0.09    \\
Fe5270         &   --		 &   -- 	   &  2.51$\pm$0.15  &  2.53$\pm$0.14  &  2.47$\pm$0.10  &  2.40$\pm$0.12  &  2.30$\pm$0.15    \\
Fe5335         &   --		 &   -- 	   &  2.19$\pm$0.13  &  2.19$\pm$0.13  &  1.97$\pm$0.08  &  2.15$\pm$0.11  &  1.92$\pm$0.10    \\
Fe5406         &   --		 &   -- 	   &  1.62$\pm$0.06  &  1.36$\pm$0.05  &  1.34$\pm$0.08  &  1.46$\pm$0.05  &  1.27$\pm$0.04    \\
Fe5709         &   --		 &   -- 	   &  0.85$\pm$0.04  &  0.71$\pm$0.05  &  0.73$\pm$0.04  &  0.26$\pm$0.07  &  0.57$\pm$0.12    \\
Fe5782         &   --		 &   -- 	   &  0.50$\pm$0.03  &  0.82$\pm$0.03  &  0.67$\pm$0.03  &  0.73$\pm$0.03  &  0.60$\pm$0.03    \\
NaD            &   --		 &   -- 	   &  3.45$\pm$0.06  &  4.02$\pm$0.04  &  4.00$\pm$0.03  &  4.82$\pm$0.02  &  3.18$\pm$0.04    \\
TiO$_1$        &   --		 &   -- 	   &  1.78$\pm$0.06  &  1.82$\pm$0.08  &  1.79$\pm$0.11  &  2.16$\pm$0.07  &  1.51$\pm$0.08    \\
TiO$_2$        &   --		 &   -- 	   &  5.80$\pm$0.17  &  5.96$\pm$0.09  &  5.52$\pm$0.15  &  5.45$\pm$0.15  &  3.56$\pm$0.14    \\
CaT1           &   --		 &   -- 	   &   --	     &   --	       &   --		 &   -- 	   &   --	       \\
CaT2           &   --		 &   -- 	   &   --	     &   --	       &   --		 &   -- 	   &   --	       \\
CaT3           &   --		 &   -- 	   &   --	     &   --	       &   --		 &   -- 	   &   --	       \\
ZrO            &   --		 &   -- 	   &   --	     &   --	       &   --		 &   -- 	   &   --	       \\
VO             &  2.61$\pm$0.20  &  1.86$\pm$0.46  &  2.98$\pm$0.34  &  1.37$\pm$0.35  &  2.11$\pm$0.28  &  0.97$\pm$0.20  &  2.60$\pm$0.25    \\
CN11           &  10.82$\pm$0.68 &  0.68$\pm$0.57  &  0.31$\pm$0.35  &  0.34$\pm$0.17  &   --		 &  5.39$\pm$0.33  &  2.71$\pm$0.67    \\
NaI1.14        &  1.57$\pm$0.47  &  0.94$\pm$0.24  &   --	     &  1.18$\pm$0.09  &  2.21$\pm$0.17  &  1.15$\pm$0.09  &  1.70$\pm$0.26    \\
FeI1.16        &   --		 &  1.02$\pm$0.07  &  1.43$\pm$0.09  &  0.40$\pm$0.08  &  0.56$\pm$0.03  &  0.29$\pm$0.06  &   --	       \\
MgI1.24        &  0.56$\pm$0.06  &  1.18$\pm$0.13  &  1.12$\pm$0.06  &  0.93$\pm$0.17  &  0.47$\pm$0.05  &  2.06$\pm$0.14  &  0.76$\pm$0.26    \\
MnI1.29        &  1.15$\pm$0.14  &   -- 	   &  3.89$\pm$0.32  &  0.99$\pm$0.32  &  1.40$\pm$0.11  &  2.98$\pm$0.46  &  1.74$\pm$0.24    \\
AlI1.31        &  3.54$\pm$0.19  &  1.61$\pm$0.06  &  2.58$\pm$0.16  &  2.98$\pm$0.14  &  0.02$\pm$0.04  &  0.56$\pm$0.07  &   --	       \\
MgI1.48        &  0.92$\pm$0.14  &  2.60$\pm$0.07  &  1.68$\pm$0.08  &  1.38$\pm$0.05  &  1.33$\pm$0.06  &  1.69$\pm$0.10  &  1.31$\pm$0.10    \\
MgI1.50        &  3.88$\pm$0.12  &  3.92$\pm$0.12  &  3.50$\pm$0.17  &  4.03$\pm$0.05  &  3.03$\pm$0.09  &  3.24$\pm$0.17  &  3.23$\pm$0.17    \\
CO1.5a         &  4.59$\pm$0.06  &  3.63$\pm$0.14  &  2.93$\pm$0.10  &  1.96$\pm$0.08  &  3.89$\pm$0.02  &  5.39$\pm$0.15  &  3.07$\pm$0.13    \\
CO1.5b         &  3.89$\pm$0.08  &  4.73$\pm$0.18  &  4.91$\pm$0.09  &  3.61$\pm$0.11  &  3.95$\pm$0.03  &  4.52$\pm$0.26  &  4.56$\pm$0.09    \\
FeI1.58        &  1.10$\pm$0.05  &  1.99$\pm$0.09  &  1.83$\pm$0.06  &  1.54$\pm$0.09  &  1.80$\pm$0.03  &  1.70$\pm$0.15  &  1.77$\pm$0.06    \\
SiI1.58        &  2.79$\pm$0.07  &  3.49$\pm$0.18  &  4.90$\pm$0.11  &  3.20$\pm$0.15  &  3.89$\pm$0.06  &  4.20$\pm$0.25  &  4.17$\pm$0.12    \\
CO1.5c         &  3.24$\pm$0.04  &  3.27$\pm$0.15  &  3.02$\pm$0.13  &  3.53$\pm$0.15  &  3.03$\pm$0.07  &  4.32$\pm$0.12  &  4.26$\pm$0.11    \\
CO1.6a         &  7.43$\pm$0.17  &  5.99$\pm$0.30  &  8.47$\pm$0.24  &  6.18$\pm$0.26  &  5.60$\pm$0.14  &  6.30$\pm$0.12  &  5.85$\pm$0.20    \\
CO1.6b         &  1.22$\pm$0.04  &  1.21$\pm$0.10  &  0.96$\pm$0.07  &  1.28$\pm$0.07  &  1.18$\pm$0.08  &  1.37$\pm$0.10  &  0.34$\pm$0.10    \\
MgI1.7         &  2.12$\pm$0.02  &  1.81$\pm$0.06  &  2.14$\pm$0.03  &  1.81$\pm$0.05  &  2.18$\pm$0.05  &  1.76$\pm$0.06  &  1.57$\pm$0.08    \\
NaI2.20        &  3.73$\pm$0.06  &  3.96$\pm$0.07  &  4.36$\pm$0.11  &  2.99$\pm$0.16  &  4.00$\pm$0.18  &  4.07$\pm$0.14  &  1.79$\pm$0.12    \\
CaI2.26        &  3.83$\pm$0.04  &  2.16$\pm$0.46  &  0.81$\pm$0.39  &  2.07$\pm$0.13  &  2.51$\pm$0.29  &  5.05$\pm$0.07  &  5.31$\pm$0.41    \\
MgI2.28        &  0.92$\pm$0.08  &  0.83$\pm$0.12  &  0.86$\pm$0.09  &  0.91$\pm$0.24  &  0.42$\pm$0.15  &  1.46$\pm$0.13  &  1.26$\pm$0.25    \\
CO2.2          &  16.65$\pm$0.38 &  17.53$\pm$0.66 &  8.39$\pm$0.74  &  19.84$\pm$0.79 &  13.78$\pm$0.50 &  14.18$\pm$0.60 &  17.74$\pm$0.98   \\
CO2.3a         &  19.00$\pm$0.74 &  6.86$\pm$0.90  &  4.91$\pm$0.74  &  24.38$\pm$1.20 &  8.70$\pm$0.32  &  14.02$\pm$0.90 &  25.90$\pm$0.83   \\
CO2.3b         &  17.96$\pm$0.93 &  8.97$\pm$1.09  &  6.80$\pm$1.03  &  27.70$\pm$1.68 &   --		 &  8.52$\pm$1.12  &  35.37$\pm$1.07    \\
\noalign{\smallskip}
\hline
\end{tabular}
\begin{list}{Table Notes:}
\item The optical data where taken from {\sc califa} survey \citep{Sanchez+16}.
\end{list}
\end{table*}

\bsp	
\label{lastpage}
\end{document}